\documentclass[twocolumn,english,aps, prx, superscriptaddress,longbibliography]{revtex4-1}

\usepackage{amsmath,amssymb,amsthm}
\usepackage{hyperref}
\usepackage{amsthm}
\usepackage{graphicx}
\usepackage{subfigure}
\usepackage{color,bbm}
\usepackage{amsfonts}
\usepackage{mathrsfs}
\usepackage{float}
\usepackage{mathtools}
\usepackage{times}

\usepackage{booktabs} 

\usepackage{multirow}

\begin{document}
\date{\today}
\title{\bf  Mean-field theory of vector spin models on networks with arbitrary degree distributions}

\author{Fernando L. Metz}
\email[]{fmetzfmetz@gmail.com}
 \affiliation{Physics Institute, Federal University of Rio Grande do Sul, 91501-970 Porto Alegre, Brazil}
  \affiliation{London Mathematical Laboratory, 8 Margravine Gardens, London W6 8RH, United Kingdom}
\author{Thomas Peron} 
\affiliation{Institute of Mathematics and Computer Science, University of S\~ao Paulo, S\~ao Carlos 13566-590, S\~ao Paulo, Brazil }
\begin{abstract}
  Understanding the  relationship between the heterogeneous structure of complex networks and cooperative
  phenomena occurring on them remains
  a key problem in network science. Mean-field theories of spin models on networks constitute a fundamental
  tool to tackle this problem and a cornerstone of statistical physics, with an impressive number of applications in condensed
  matter, biology, and computer science. In this work we derive the mean-field equations for the equilibrium
  behavior of vector spin  models on high-connectivity random networks with an arbitrary degree distribution and with
  randomly weighted links. We demonstrate that the high-connectivity limit of spin models on networks
is not universal in that it depends on the full degree distribution. 
Such nonuniversal behavior is akin to a remarkable  mechanism that leads
to the breakdown of the central limit theorem  when applied to the distribution of effective local fields.
  Traditional mean-field theories on fully-connected models, such as the Curie-Weiss, the Kuramoto, and the Sherrington-Kirkpatrick
  model, are only valid if the network degree distribution is highly concentrated around its mean degree.
  We obtain a series of results that highlight the importance of degree
  fluctuations to the phase diagram of mean-field spin models
  by focusing on the Kuramoto model of synchronization and on the Sherrington-Kirkpatrick
  model of spin-glasses.
  Numerical simulations corroborate our theoretical findings and provide compelling evidence that the present mean-field theory
  describes an intermediate regime of connectivity, in which the  average degree $c$ scales as a power $c \propto N^{b}$ ($b < 1$) of the total number $N \gg 1$ of spins.
  Our findings put forward a novel class of spin models that incorporate the effects of degree fluctuations and, at the same time, are amenable
  to exact analytic solutions. 
\end{abstract}

\maketitle

\section{Introduction} \label{introdu}

Several man-made and natural complex systems are represented by networks of nodes joined by links \cite{NewmanBook}. The
study of the interplay between the structure of complex networks and dynamical
processes on top of them
has grown into a major research field \cite{BarratBook,Doro2008}, with
many applications in physics, biology, information theory, and technology.
As opposing to the homogeneous structures typically studied in solid state physics, such as Bravais and Bethe
lattices \cite{BaxterBook}, the striking feature of complex networks is the existence of strong local fluctuations
in their structure. This heterogeneous character is responsible for most of the nontrivial
dynamical properties of networked systems \cite{BarratBook}.

Spin models on networks describe systems formed by a large number of state variables, represented by scalars or continuous vectors, which
are coupled through the links of networks \cite{Doro2008,Izaakthesis,Dommersthesis}. 
The study of such models is of utmost importance
for at least two main reasons. First, they are minimal
models to address the impact of heterogeneous structures on the cooperative behavior of a large number of interacting degrees of freedom.
Second, seemingly unrelated problems across disciplines can be cast in terms of the unifying framework of spin models on random networks \cite{Mezard1987,MezardBook}. 
Models of scalar spins on networks have a vast number of applications
in a variety of research fields, such as 
opinion dynamics \cite{Castellano2009,Baumann2021}, models of socio-economic
phenomena \cite{Bouchaud2013}, artificial neural networks \cite{Hopfield82,Wemmenhove2003,Castillo2004}, agent-based models of the
market behavior \cite{Challet1997,Challet2005,Seoane2021}, dynamics of biological neural
networks \cite{Brunel2000,Ostojic2014,Schuessler2020}, information theory and computer science \cite{MezardBook}, sparse
random-matrix theory \cite{Rogers2008,Metz2019}, and the stability of large dynamical systems \cite{Neri2020,Krumbeck2021}.
Models of vector spins on networks are relevant for the study of synchronization phenomena \cite{Kuramoto1975,strogatz2000kuramoto,Rodrigues2016,fonseca2018kuramoto}, random
lasers \cite{Antenucci2015,Marruzzo2015,Antenucci2021}, vector spin-glasses \cite{Cosimothesis,Lupo2018,Lupo2019}, and the collective dynamics of swarms \cite{Olfati2006,lohe2009non,Zhu2013,Chandra2019}.

Mean-field theories stand out as one of the most celebrated tools in physics and the natural starting point to address the collective behavior
of many interacting spins. The
heart of the mean-field approach is the assumption that all spins are statistically
equivalent, in the sense that each spin experiences an effective random field drawn from the same distribution. By virtue of that, the original problem
of many interacting elements is replaced by a problem of a single spin coupled to an effective field.
Paradigmatic examples of mean-field theories are derived from spin models on fully-connected networks, such as the Curie-Weiss model of ferromagnetism \cite{Kochma2013}, the Kuramoto model
of synchronization \cite{Kuramoto1975,Rodrigues2016}, and the Sherrington-Kirkpatrick model of spin-glasses \cite{Sherrington1975,Sherrington1978}. 

Fully-connected mean-field theories
are expected to provide an universal description of spin models
on high-connectivity networks, for which the mean degree is infinitely large. In fact, it seems
sensible to argue that spin models on networks gradually flow to their fully-connected
behavior as the mean degree increases, since the
detailed structure of a network should become irrelevant in the high-connectivity limit.
Although this intuitive argument has been confirmed
for a few network models \cite{Hatchett2008,Hatchett2009,Mezard2001}, understanding the impact of degree fluctuations
on the high-connectivity limit of spin models remains a key open question.
In this work we provide a comprehensive solution to this problem.
Surprisingly, we find that the high-connectivity limit of spin models on networks is not universal, since it depends on the
full degree distribution. It follows that traditional, fully-connected mean-field theories do not generally predict the macroscopic behavior
of spin models on high-connectivity networks.

\begin{figure}[t!]
	\centering
	\includegraphics[width=1.0\columnwidth]{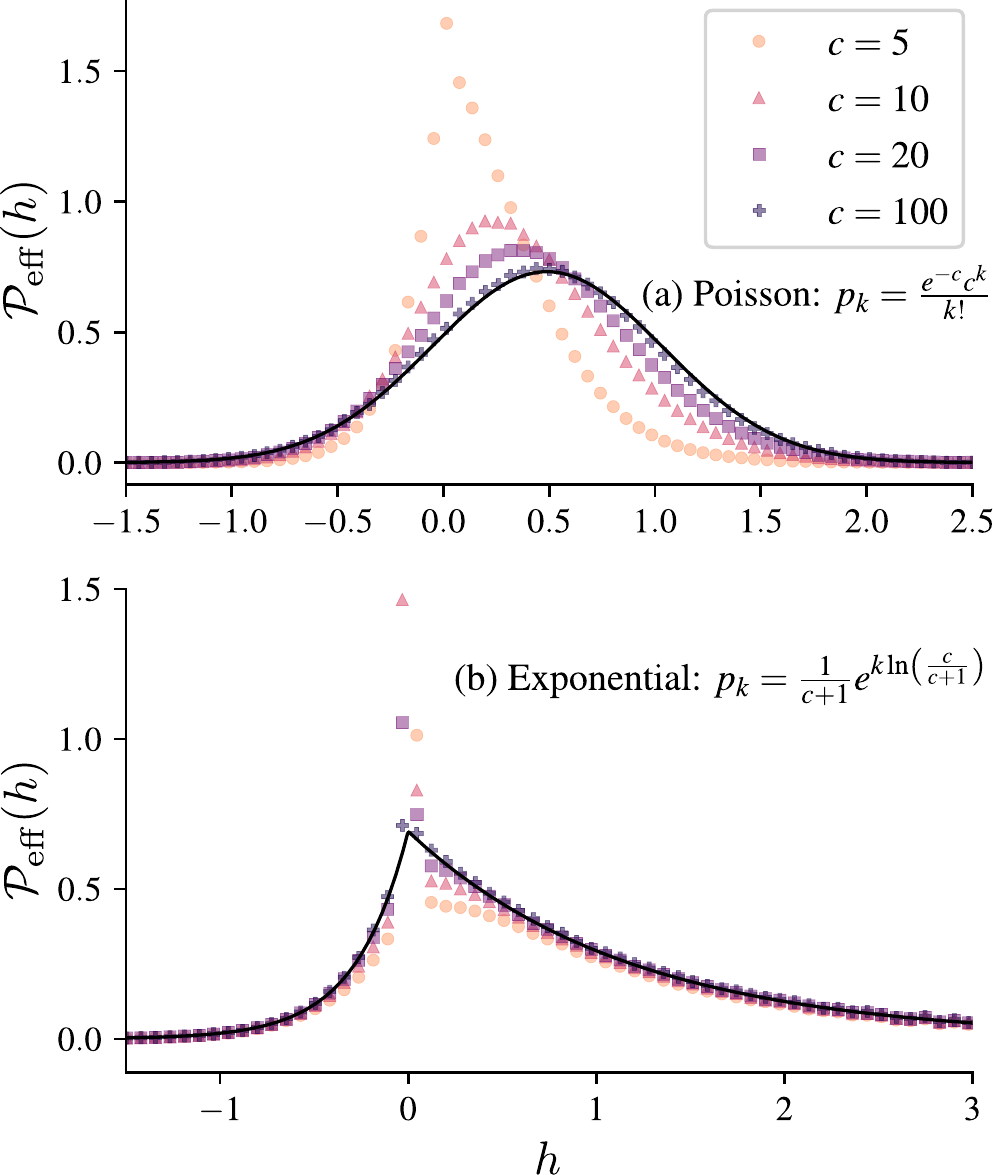}
	\caption{Distribution of effective local fields $h_{i,{\rm eff}}$ [Eq.~\eqref{retua}] for Ising spins in equilibrium at temperature $T=1$. The spins are
          coupled through networks with (a) an exponential and (b) a Poisson degree distribution $p_k$. The symbols
          are obtained from numerical simulations of networks with $N = 10^5$ nodes and different values of the average degree $c$. The coupling strengths follow a Gaussian
          distribution with mean $1.3/c$ and standard deviation $1/\sqrt{c}$.
          The solid lines are the analytic expressions for $N \rightarrow \infty$ and $c \rightarrow \infty$ derived
          in this work [see Eqs.~(\ref{ufopoas}) and \eqref{jutwe}] .}
	\label{localfields}
\end{figure}

Such nonuniversal character is intimately related to the breakdown
of the central limit theorem as applied to the distribution of effective local fields.
To illustrate this point, let us consider the equilibrium behavior of Ising spins on random networks at inverse temperature $\beta$ \cite{Mezard1987,Mezard2001,Mezard2003}. The local
magnetization $m_i = \tanh(\beta h_{i,{\rm eff}})$
at node $i$ is determined by the effective field
\begin{equation}
  h_{i,{\rm eff}} = \sum_{j \in \partial_i} J_{ij} m_j^{(i)},
  \label{retua}
\end{equation}  
where $\partial_i$ is the set of neighbors of $i$, $J_{ij}$ is the random coupling strength between spins at nodes $i$ and $j$, and
$m_j^{(i)}$ is the local magnetization at node $j$ in the absence of $i$. The network degrees are random variables with average $c$.
Figure \ref{localfields} depicts simulation
results for two different degree distributions. For random graphs with a Poisson degree distribution, the central limit theorem holds and the distribution of effective fields
converges to a Gaussian distribution when $c \rightarrow \infty$, which corresponds to the fully-connected mean-field
behavior \cite{Sherrington1978}.
For networks with an exponential degree distribution, the central limit theorem fails and the effective fields are no longer
Gaussian in the high-connectivity limit.
The  breakdown of the central limit theorem is caused by the strong fluctuations of the random number of summands
in Eq.~(\ref{retua}), which is nothing more than the degree of node $i$. This compelling mechanism for the failure
of the central limit theorem
has been studied for more than seventy years in probability theory \cite{Wald1944,Robbins1948,GnedenkoBook}, but
its evident importance for spin models on networks has so far eluded a careful analysis.
In this paper we fill this gap and derive a novel family of mean-field theories that emerge from such breakdown
of the central limit theorem.

\subsection{Main results}

The central result of this work is a set of equations for the equilibrium behavior of spin models on high-connectivity networks with
an arbitrary degree distribution. The spins are Ising variables or continuous vectors with finite dimension, while the random pairwise interactions between
spins follow an arbitrary distribution.
The high-connectivity limit of the model is cast in terms of an effective problem of a single
spin, whose configurations follow the Boltzmann distribution with an effective energy given by Eq.~(\ref{cortusa}).
The analytic expression for the distribution of the effective energy, Eq.~(\ref{tuasu1}), is one
of the main outcomes of this work.

The remarkable consequence of Eq.~(\ref{tuasu1}) is that, even in the high-connectivity limit, the behavior of spin models on graphs
is not universal, but strongly dependent on the degree distribution.
Networks for which the high-connectivity limit depends on the degree
distribution are called {\it heterogeneous networks}. In contrast, the behavior of spin
models on the so-called {\it homogeneous networks} is universal, i.e., independent of the degree
distribution and consistent with the behavior of fully-connected models.

The analytic results of section \ref{secHet} are very general and they can be applied to
a variety of specific models. We illustrate the effects of degree fluctuations on the mean-field
behavior of spin models 
by focusing on two examples: the Kuramoto model of synchronization and the Ising model of spin-glasses.
We obtain the complete phase diagrams
of both models [see figures \ref{pdferro} and \ref{pdr}].

Degree fluctuations dramatically affect the distribution of effective fields
and the magnetization inside the ferromagnetic phase. 
We show that the distribution of effective local fields of Ising spin models exhibits a long tail for large fields and a divergence at zero field [see Eq.~(\ref{porart})], in contrast
to the Gaussian \cite{Sherrington1978} or the Dirac-$\delta$ \cite{Kochma2013} distributions of effective fields in fully-connected models. 
In the case of the Kuramoto model, the magnetization or phase coherence displays a singular point that separates two distinct regions 
in the ferromagnetic phase [see figure \ref{colapsado}], each one identified by a different behavior
of the magnetization for strong couplings.
In addition, we show that degree fluctuations have an ambiguous effect in the ordered phase.
Although the ferromagnetic phase of the Kuramoto model expands over the entire phase diagram
when the variance of the degree distribution diverges, the magnetization becomes
arbitrarily small. This result sheds light on a key paradigm of network science, namely the idea that collective
  behavior is improved as the critical point vanishes with the increase of degree fluctuations~\cite{Leone2002,BarratBook}.
  Our results provide compelling evidence that in the thermodynamic limit of networks with diverging degree variance the only possible
  emerging phase is ultimately the paramagnetic (asynchronous) one. Last but not least, we find that the shape of the phase distribution
  of the Kuramoto model on heterogeneous networks fluctuates from site to site [see figure \ref{fluct}], which
exposes a lack of correspondence between local and global ensemble averages, in sharp contrast to the behavior on fully-connected networks.

The results of section \ref{rr} generalize the replica-symmetric mean-field theory and the phase diagram of
Ising spin-glasses \cite{Sherrington1975,Sherrington1978} to the case of heterogeneous networks. We explain how to derive the Almeida-Thouless
line \cite{Almeida1978}, which bounds the region of the phase diagram where the replica-symmetric theory is unstable and our results are no longer
exact. When the variance of the degree distribution diverges, the spin-glass phase as well as the replica-symmetry breaking
region are  confined to an arbitrarily small sector of the phase diagram.
Interestingly, the low-temperature portion of the Almeida-Thouless
line exhibits a non-monotonic behavior as a function of the degree fluctuations.

Throughout the paper, we present simulation results for spin models on random networks with a finite number $N$ of nodes.
Besides confirming our theoretical findings, the simulations reveal that the heterogeneous mean-field theory derived in this work is valid
in the limit $N \rightarrow \infty$ when the mean degree scales as $c \propto N^b$ ($0 < b < 1$). This intermediate regime  of connectivity
lies between the sparse ($b=0$) and the fully-connected ($b=1$) regimes.
Our work uncovers the non-universal behavior of spin models on heterogeneous networks in such intermediate scaling of $c$.

The paper is organized as follows. In the next section we define the Hamiltonian of the vector spin model on an ensemble
of random networks with an arbitrary degree distribution. Section \ref{cav11} presents the cavity or message-passing equations
for the equilibrium behavior of spin models on networks with finite  mean degree, as derived in previous works \cite{Skantzos2005,Coolen2005}. This section
also features the distributional version of the cavity equations. Section \ref{secHet} is the core of the paper.
Initially, we thoroughly explain how to calculate the high-connectivity limit of the distributional
cavity equations, from which the heterogeneous mean-field theory emerges. The explicit results
for the effect of degree fluctuations are presented in two subsections. Subsection \ref{sub1} is focused on the
Kuramoto model with ferromagnetic couplings, while subsection \ref{rr} presents results for Ising
spin models with random pairwise interactions. We further discuss our findings and main conclusions in section \ref{concl}. The paper
contains two appendices. Appendix \ref{ATline00}
explains in detail how to derive the Almeida-Thouless line for Ising spin models
on heterogeneous networks, while appendix \ref{sec:simulations} presents some details of the numerical simulations.

\section{The model set-up}

We consider $D$-dimensional spins $\boldsymbol{\sigma}_1, \dots,\boldsymbol{\sigma}_N$, with
$\boldsymbol{\sigma}_i = (\sigma_{i,1} \,\, \sigma_{i,2} \, \dots \, \sigma_{i,D})^{T}$, placed
on the nodes of a simple random graph \cite{BolloBook}. Each spin $\boldsymbol{\sigma}_i$ is a continuous
vector that identifies a point on the surface of the $D$-dimensional hypersphere $\mathcal{R}_D$ with unit radius.
The probability to observe
a global spin configuration $\{ \boldsymbol{\sigma} \}=(\boldsymbol{\sigma}_1,\dots,\boldsymbol{\sigma}_N)$ in thermal equilibrium follows the Boltzmann distribution
\begin{equation}
p_{\rm B}(\{ \boldsymbol{\sigma} \}) = \frac{1}{\mathcal{Z}} e^{- \beta \mathcal{H}(\{ \boldsymbol{\sigma} \}  )},
\end{equation}  
where $\beta = 1/T$ is the inverse temperature, $\mathcal{H}(\boldsymbol{\sigma})$ is the Hamiltonian, and $\mathcal{Z}$ is the partition function
\begin{equation}
\mathcal{Z} = \int_{\mathcal{R}_D} \left( \prod_{i=1}^N d \boldsymbol{\sigma}_i  \right) e^{- \beta \mathcal{H}\left( \{ \boldsymbol{\sigma} \} \right)}.
\end{equation}  
The shorthand notation $\int_{\mathcal{R}_D} d \boldsymbol{\sigma}_i$ denotes an integral over all possible configurations of $\boldsymbol{\sigma}_i$ such
that $\boldsymbol{\sigma}_i^2 = 1$.

We study a generic family of spin models invariant under orthogonal transformations and defined by the Hamiltonian \cite{Stanley1968}
\begin{equation}
  \mathcal{H}(\{ \boldsymbol{\sigma} \}) = - \frac{1}{2} \sum_{ij=1}^N c_{ij} J_{ij} \boldsymbol{\sigma}_i^{T}  \boldsymbol{\sigma}_j,
  \label{tuacu}
\end{equation}  
where $\{ c_{ij} \}_{i,j=1}^N$ are the elements of the graph adjacency matrix $\boldsymbol{C}$, and
$\{ J_{ij} \}_{i,j=1}^N$ are the coupling strengths between the spins.
The set-up specified by Eq.~(\ref{tuacu}) comprises a broad class of traditional models of spins randomly coupled through
the links of a network. The Ising model, the XY or Kuramoto model with identical oscillators, and the classical Heisenberg model on a random network
are obtained by setting, respectively, $D=1$, $D=2$, and $D=3$.

The binary matrix elements $c_{ij} \in \{ 0,1 \}$, with $c_{ii}=0 \, \forall \, i$,  specify the network
topology: if $c_{ij} = c_{ji} =  1$, there is an undirected edge between
nodes $i$ and $j$, whereas $c_{ij}=0$ otherwise. The coordination number or degree  $k_i = \sum_{j=1}^N c_{ij}$ gives the number of nodes connected to $i$.
The adjacency random matrix $\boldsymbol{C}$ is generated according to the configuration model of networks \cite{Molloy95,Newman2001,Fosdick2018}, in which
a graph instance is uniformly drawn  from the set of all
possible simple graphs with a prescribed degree distribution \cite{Molloy95,Newman2001,Fosdick2018}
\begin{equation}
p_k = \lim_{N \rightarrow \infty} \sum_{i=1}^N \delta_{k,\sum_{j=1}^N c_{ij}  }.
\end{equation}
The average degree $c$, defined as
\begin{equation}
c  = \sum_{k=0}^\infty k p_k   ,
\end{equation}
is independent of the total number of nodes $N$ in the graph. The configuration model of networks, in which $p_k$ is specified at the outset, is the
ideal setting to explore how $p_k$ impacts the macroscopic behavior of the spin model of Eq.~(\ref{tuacu}) in the high-connectivity
limit $c \rightarrow \infty$. We assume that $p_k$ is an arbitrary degree distribution with finite moments as long as $c < \infty$.

Although the main results presented below are valid for any $p_k$, we will be particularly interested on random graphs 
with a negative binomial degree distribution \cite{EvansBook}
\begin{equation}
  p_k^{(b)} = \frac{\Gamma(\alpha + k)}{k! \, \Gamma(\alpha)} \left( \frac{c}{\alpha} \right)^k \left( \frac{\alpha}{ \alpha + c} \right)^{\alpha + k},
  \label{utaop}
\end{equation}  
parametrized by $c$ and $\alpha > 0$. The parameter $\alpha$ is related to the variance $\sigma_{b}^2$ of $p_k^{(b)}$ as follows
\begin{equation}
\sigma_{b}^2 = c + \frac{c^2}{\alpha}.
\end{equation}  
The distribution $p_k^{(b)}$ becomes identical to the exponential distribution for $\alpha=1$, and it converges
to the Poisson distribution when $\alpha \rightarrow \infty$.
In the limit $c \rightarrow \infty$, the relative
variance $\sigma_{b}^2/c^2$ is solely governed by $\alpha$ and, as we will show below, the mean-field theory depends
on the degree distribution only through $\alpha$, which
enables to study the role of the degree fluctuations in a clear-cut way.
In the context of complex networks, the negative binomial distribution finds
applications in models of weighted random graphs \cite{Garlaschelli2009}, in studies of the spread
of infectious diseases on networks \cite{Volz2011,Miller2013}, and as the empirical degree distribution
of real-world contact networks \cite{Aleta2020}.

The variable $J_{ij} \in \mathbb{R}$ in Eq.~(\ref{tuacu}) denotes the strength of the mutual interaction between $\boldsymbol{\sigma}_i$ and $\boldsymbol{\sigma}_j$. The coupling
strengths $\{ J_{ij} \}_{i,j=1}^N$ are independently and identically distributed random variables
drawn from an arbitrary distribution $p_J$ with mean $K_0/c$ and standard deviation $K_1/\sqrt{c}$. Note that the ensemble
of random graphs is fully specified by $p_k$ and $p_J$.


\section{The distributional version of the message-passing equations} \label{cav11}

In this section we obtain a set of distributional equations for the single-site marginals of the spins when $c$ is finite.
These distributional equations, originally derived in references
\cite{Skantzos2005,Coolen2005} and further studied in \cite{Hatchett2008,Hatchett2009}, build on the message-passing or cavity equations for network models with a local
tree-like structure, such as the configuration model \cite{Bordenave2010}. The mean-field equations for heterogeneous
networks follow from the limit $c \rightarrow \infty$ of the distributional equations.

\subsection{The cavity equations for single-site marginals}

The probability density to observe a configuration $\boldsymbol{\sigma}_i$ at an arbitrary node $i$ follows from
the Boltzmann distribution
\begin{equation}
p_i (\boldsymbol{\sigma}_i) = \left( \prod_{j=1 (\neq i)}^N  \int_{\mathcal{R}_D} d \boldsymbol{\sigma}_j   \right) p_{\rm B} \left( \{ \boldsymbol{\sigma} \}  \right), \quad i=1,\dots,N.
\end{equation}  
Thanks to the approximate local tree-like structure
of the graph when $N \gg 1$ \cite{Newman2001,Bordenave2010}, the local marginals $\{ p_i (\boldsymbol{\sigma}_i) \}_{i=1,\dots,N}$ on a
single graph instance fulfill \cite{Skantzos2005}
\begin{equation}
  p_i(\boldsymbol{\sigma}_i) = \frac{1}{\mathcal{Z}_i} \prod_{l \in \partial_i}
   \int_{\mathcal{R}_D} d \boldsymbol{\sigma}_l \,     e^{\beta  J_{il}  \boldsymbol{\sigma}_{i}^{T} \boldsymbol{\sigma}_l  }  p^{(i)}_{l} (\boldsymbol{\sigma}_l)   ,
  \label{gugu1}
\end{equation}  
where $\partial_i$ is the set of nodes adjacent to node $i$, $\mathcal{Z}_i$ is the normalization factor of $p_i(\boldsymbol{\sigma}_i)$, and
$p^{(i)}_{l} (\boldsymbol{\sigma}_l)  $ is the distribution of $\boldsymbol{\sigma}_l$ on the so-called cavity graph $\mathcal{G}^{(i)}$, obtained from
the original graph $\mathcal{G}$ by deleting node $i$ and all its edges.
The distributions $p^{(i)}_{j} (\boldsymbol{\sigma}_j)$ solve the following self-consistency equations
\begin{equation}
  p^{(i)}_{j} (\boldsymbol{\sigma}_j)  = \frac{1}{\mathcal{Z}_{j}^{(i)}} \prod_{l \in \partial_j \setminus i}     \int_{\mathcal{R}_D} d \boldsymbol{\sigma}_l \, 
  e^{\beta  J_{jl}     \boldsymbol{\sigma}_{j}^{T}  \boldsymbol{\sigma}_l     }  p^{(j)}_{l} (\boldsymbol{\sigma}_l)   ,
\label{cav}
\end{equation}  
in which $ \partial_j \setminus i$ is the set of nodes adjacent to $j$ except for $i \in \partial_j$, and $\mathcal{Z}_{j}^{(i)}$ is the normalization
factor of $p^{(i)}_{j} (\boldsymbol{\sigma}_j)$. After solving the message-passing Eqs.~(\ref{cav}), the marginals on the original
graph are reconstructed from Eqs.~(\ref{gugu1}).

\subsection{The distributional equations for an ensemble of networks}

Equations (\ref{gugu1}) and (\ref{cav}) yield an approximation for the local marginals of an ensemble of spin models defined
on a {\it single graph instance} with $N \gg 1$ nodes. These equations become
asymptotically exact for $K_1=0$  when $N \rightarrow \infty$ \cite{Mezard2001,Mezard2003}, due to the existence of a vanishing fraction of short
cycles in the graph. Nevertheless, if the coupling strengths are random ($K_1 > 0$), Eqs.~(\ref{gugu1}) and (\ref{cav}) exhibit a large
number of solutions at low temperatures \cite{Mezard2001,Mezard2003}, which reflects the existence of many local minima
in the free-energy
\cite{Mezard1987,Monasson1998}. In this regime of parameters,  Eqs.~(\ref{gugu1})
and (\ref{cav})  only provide an approximate description of the system.

Inasmuch as $p_i(\boldsymbol{\sigma}_i)$ and $p^{(i)}_{j} (\boldsymbol{\sigma}_j)$ are (positive) random functions whose shape fluctuates along
the nodes of a network, it is instrumental to adopt an ensemble viewpoint and work with the functional probability densities
of $p_i(\boldsymbol{\sigma}_i)$ and $p^{(i)}_{j} (\boldsymbol{\sigma}_j)$, defined respectively as
\begin{equation}
W[p] = \lim_{N \rightarrow \infty} \frac{1}{N} \sum_{i=1}^N    \delta_{F} \left[ p(\boldsymbol{\sigma})   -  p_i(\boldsymbol{\sigma}) \right]
\end{equation}  
and
\begin{equation}
  R[p] = \lim_{N \rightarrow \infty}  \frac{\sum_{j=1}^N \sum_{i \in \partial_j}  \delta_F \left[   p(\boldsymbol{\sigma}) - p_{j}^{(i)}(\boldsymbol{\sigma})     \right] }{ \sum_{j=1}^N K_j },
\end{equation}  
with $\delta_F[f(\boldsymbol{\sigma})] $ representing the functional Dirac-$\delta$ defined over the space
of all possible functions $f(\boldsymbol{\sigma})$.

Let us obtain the distributional equations for the functionals $W[p]$ and $R[p]$.
Equations (\ref{gugu1}) and (\ref{cav}) can be intuitively rewritten as 
\begin{align}
  p_i(\boldsymbol{\sigma}_i) &= \frac{e^{\beta H_i(\boldsymbol{\sigma}_i) }}{  \int_{\mathcal{R}_D} d \boldsymbol{\sigma}^{\prime} \,  e^{\beta H_i (\boldsymbol{\sigma}^{\prime} ) }             },  \label{zigui1}  \\
   p_{i}^{(j)}(\boldsymbol{\sigma}_i) &= \frac{e^{\beta H_i^{(j)}(\boldsymbol{\sigma}_i) }}{   \int_{\mathcal{R}_D} d \boldsymbol{\sigma}^{\prime} \,  e^{\beta H_i^{(j)} (\boldsymbol{\sigma}^{\prime} ) }      },  \label{zigui2}
\end{align} 
where
\begin{equation}
  H_i(\boldsymbol{\sigma}_i) = \frac{1}{\beta} \sum_{l \in \partial_i}
  \ln{\left(   \int_{\mathcal{R}_D} d \boldsymbol{\sigma}_{l} \, e^{\beta  J_{il}  \boldsymbol{\sigma}_{i}^{T} \boldsymbol{\sigma}_l  }  p^{(i)}_{l} (\boldsymbol{\sigma}_l)    \right)}
  \label{curti}
\end{equation}  
and
\begin{equation}
  H_i^{(j)} (\boldsymbol{\sigma}_i) = \frac{1}{\beta} \sum_{l \in \partial_i \setminus j}  \ln{\left( \int_{\mathcal{R}_D} d \boldsymbol{\sigma}_{l} \,
    e^{\beta  J_{il}  \boldsymbol{\sigma}_{i}^{T}  \boldsymbol{\sigma}_l  }  p^{(i)}_{l} (\boldsymbol{\sigma}_l)    \right)} .
\end{equation}  
Clearly, the functional distribution of $H_i(\boldsymbol{\sigma}_i)$ ($H_i^{(j)} (\boldsymbol{\sigma}_i)$) determines the distribution
$W[p]$ ($R[p]$) of $p_i(\boldsymbol{\sigma}_i)$ ($p_{i}^{(j)}(\boldsymbol{\sigma}_i)$). Since the left and right hand
sides  of Eqs.~(\ref{zigui1}) and (\ref{zigui2}) are equal in the distributional sense, $W[p]$ and $R[p]$
fulfill
\begin{align}
  W[p] &=   \int \mathcal{D}h \,  \mathcal{F}_0 [h]  \, \delta_F \left[ p(\boldsymbol{\sigma}) -     \frac{e^{\beta h(\boldsymbol{\sigma})}  }
    {  \int_{\mathcal{R}_D} d \boldsymbol{\sigma}^{\prime} \,  e^{\beta h(\boldsymbol{\sigma}^{\prime} ) }          }      \right], \label{tu1} \\
  R[p] &= \int \mathcal{D}h \,  \mathcal{F}_1 [h]  \, \delta_F \left[ p(\boldsymbol{\sigma}) -     \frac{e^{\beta h(\boldsymbol{\sigma})}  }
    {     \int_{\mathcal{R}_D} d \boldsymbol{\sigma}^{\prime} \,  e^{\beta h(\boldsymbol{\sigma}^{\prime} ) }        }      \right], \label{tu2}  
\end{align}  
where $\mathcal{D} h$ is a functional integration measure, while
\begin{align}
 \mathcal{F}_n [h] &= \sum_{k=0}^\infty \frac{k^n p_k}{c^n} \int \left( \prod_{r=1}^{k-n} d J_r \mathcal{D}q_r  p_J(J_r)   R[q_r] \right) \nonumber \\
 &\times  \delta_F \left[h(\boldsymbol{\sigma}) - \frac{1}{\beta} \sum_{r=1}^{k-n} \ln{\left(  \int_{\mathcal{R}_D} d \boldsymbol{\sigma}^{\prime}
     e^{\beta J_r    \boldsymbol{\sigma}^{T} \boldsymbol{\sigma}^{\prime} }       q_r(\boldsymbol{\sigma}^{\prime})      \right) }    \right] \label{hopa1} 
\end{align}  
yields the functional distribution of $H_i(\boldsymbol{\sigma}_i)$ or $H_i^{(j)} (\boldsymbol{\sigma}_i)$ by setting $n=0$ or $n=1$, respectively.
Equations (\ref{tu1}) and (\ref{tu2}) form a closed system of equations for $W[p]$ and $R[p]$. The central quantity
$\mathcal{F}_n [h]$ is determined, in the
limit $N \rightarrow \infty$, by the degree distribution $p_k$, the distribution of coupling
strengths $p_J$, and the functional distribution $R[p]$. As we will show below, $\mathcal{F}_n [h]$ simplifies in the high-connectivity limit $c \rightarrow \infty$.


\subsection{Local and global ensemble averages}

The distributions $W[p]$ and $R[p]$ are key quantities in the study of the macroscopic behavior of spin models
on networks as they allow to compute ensemble averages of the spins and derive equations for the order-parameters \cite{Skantzos2005,Coolen2005}.
Due to the fluctuations in the local structure of a random network, it is important to clearly distinguish
between local and global ensemble averages. Let $O( \boldsymbol{\sigma}_i)$ be an observable defined in terms of
$\boldsymbol{\sigma}_i$.
The local ensemble average of $O( \boldsymbol{\sigma}_i)$ on the
original graph  is defined as
%
\begin{equation}
  \langle O \left( \boldsymbol{\sigma} \right) \rangle_{p_{i}} =    \int_{\mathcal{R}_D} d \boldsymbol{\sigma} \, p_i(  \boldsymbol{\sigma} ) O(\boldsymbol{\sigma}),
  \label{huawe}
\end{equation}  
while the local average on the cavity graph reads
\begin{equation}
  \langle O \left( \boldsymbol{\sigma} \right) \rangle_{p_{i}^{(j)}} =    \int_{\mathcal{R}_D} d \boldsymbol{\sigma} \, p_i^{(j)}(  \boldsymbol{\sigma} ) O(\boldsymbol{\sigma}).
  \label{huawe1}
\end{equation}  
Local averages only depend on the random functions $p_i(  \boldsymbol{\sigma} )$ and $p_i^{(j)}(  \boldsymbol{\sigma} )$ defined
at node $i$.
By taking the averages of Eqs.~(\ref{huawe}) and (\ref{huawe1}) with respect to the functional distributions
of $p_i(  \boldsymbol{\sigma} )$ and $p_i^{(j)}(  \boldsymbol{\sigma} )$, we obtain the global ensemble averages
\begin{align}
 \langle \langle  O  \left( \boldsymbol{\sigma}  \right)\rangle_{p} \rangle_W   &= \int \mathcal{D} p \, W[p]    \langle  O  \left( \boldsymbol{\sigma}  \right) \rangle_{p}, \label{gutas} \\
  \langle \langle  O  \left( \boldsymbol{\sigma}  \right) \rangle_{p} \rangle_Q  &= \int \mathcal{D} p \, R[p]    \langle  O  \left( \boldsymbol{\sigma}  \right) \rangle_{p}. \label{gutas1}
\end{align}  
For instance, the choice $O( \boldsymbol{\sigma}) = \boldsymbol{\sigma}$ yields the global magnetization $\boldsymbol{m}$ on the original graph,
\begin{equation}
  \boldsymbol{m} \equiv  \langle \langle \boldsymbol{\sigma}  \rangle_{p} \rangle_W = \int \mathcal{D} p \, W[p]   \langle \boldsymbol{\sigma}  \rangle_{p},
  \label{defini}
\end{equation}  
and the global magnetization $\boldsymbol{M}$ on the cavity graph,
\begin{equation}
  \boldsymbol{M} \equiv \langle \langle \boldsymbol{\sigma}  \rangle_{p} \rangle_Q = \int \mathcal{D} p \, R[p]   \langle \boldsymbol{\sigma}  \rangle_{p}.
  \label{defini1}
\end{equation}  

Notably, the present formalism also gives access to the fluctuations of the random function $p_{i}( \boldsymbol{\sigma})$ along the different nodes.
Instead of working directly with the functional distribution $W[p]$, whose domain is infinite dimensional, it is sensible
to study the moments of $p_{i}( \boldsymbol{\sigma}) $  for
fixed $\boldsymbol{\sigma}$
\begin{equation}
  \langle \left[ p (\boldsymbol{\sigma}) \right]^r \rangle_W =   \int \mathcal{D} p \, W[p]  \left[ p (\boldsymbol{\sigma}) \right]^r, \,\,\, r \geq 0.
  \label{hua1}
\end{equation}  
In particular, the variance
\begin{equation}
\sigma^{2}_{W} = \langle \left[ p (\boldsymbol{\sigma}) \right]^2  \rangle_W - \langle   p (\boldsymbol{\sigma})  \rangle_W
\end{equation}
quantifies the spread of the functional shape of the local marginals $p_{i}( \boldsymbol{\sigma})$ around their average $\langle p (\boldsymbol{\sigma})  \rangle_W$.


\section{The heterogeneous mean-field theory} \label{secHet}

The heterogeneous mean-field theory describes spin models on
high-connectivity random graphs characterized by strong degree
fluctuations, in contrast to the standard, homogeneous mean-field
theory, whose predictions are limited to random graphs with
vanishing degree fluctuations.

The heterogeneous mean-field equations are derived by taking the $c \rightarrow \infty$ limit of Eqs.~(\ref{tu1}) and (\ref{tu2}) for arbitrary $p_k$.
The core of the calculation lies in the high-connectivity limit of the distribution $\mathcal{F}_n [h]$.
Before performing this calculation in full generality, let us consider the behavior
of the random variable $H_i( \boldsymbol{\sigma}_i)$ for large $c$ in the particular case of Ising spins, in order to gain insight
into the physical meaning of $H_i(\boldsymbol{\sigma}_i)$ and its distribution $\mathcal{F}_0[h]$.
Thanks to the scaling of the moments of $J_{il}$ with $c$, the coupling strengths become very small for $c \gg 1$ and
it suffices to expand Eq.~(\ref{curti}) in powers of  $J_{il}$  up to $\mathcal{O}(J_{il}^2)$, leading to
the following expression 
\begin{equation}
  H_i( \sigma_i) = \sigma_i \sum_{l \in \partial_i } J_{il} \langle \sigma  \rangle_{p_{l}^{(i)}} + \frac{\beta}{2} \sum_{l \in \partial_i }   J_{il}^2 \left(1 - \langle \sigma  \rangle_{p_{l}^{(i)}}^2   \right),
  \label{sumat}
\end{equation}  
with $\langle \sigma  \rangle_{p_{l}^{(i)}}$ the local magnetization at node $l$ in the absence of $i \in \partial_l$.
Equation (\ref{sumat}) reveals that $H_i( \sigma_i)$ can be regarded as the interaction energy of the
spin $\sigma_i$ with an effective random field, specified by the local magnetizations of the neighboring spins in the absence of $\sigma_i$.
The physical meaning of the random variable $H_i^{(j)}( \sigma_i)$ is completely analogous.

Equation (\ref{sumat}) involves two sums over a large number of independent random
variables. Hence it is tempting to invoke the central limit theorem and argue that $H_i( \sigma_i)$ follows
a Gaussian distribution, which is in fact correct for Erd\"os-R\'enyi and regular random graphs \cite{Hatchett2008,Hatchett2009}.
However, the total number of terms in the sums of Eq.~(\ref{sumat}) is itself a random variable, whose fluctuations are controlled
by the degree distribution $p_k$. As we will show below, the central limit theorem breaks down when $p_k$ is not sufficiently concentrated around its
mean value $c$ \cite{GnedenkoBook}, which leads to a family of heterogeneous mean-field equations that explicitly depend on $p_k$.

With the purpose of calculating the $c \rightarrow \infty$ limit of $\mathcal{F}_n [h]$ for arbitrary $p_k$, it is convenient to introduce the characteristic
functional 
\begin{eqnarray}
  \mathcal{G}_n [t] = \int \mathcal{D} h \, e^{- i   \int_{\mathcal{R}_D} d \boldsymbol{\sigma}  \, t(\boldsymbol{\sigma}) h(\boldsymbol{\sigma})     }  \mathcal{F}_n [h],
\end{eqnarray}  
from which $\mathcal{F}_n [h]$ is determined via the inverse Fourier transform
\begin{eqnarray}
  \mathcal{F}_n [h] = \int   \frac{\mathcal{D} t}{(2 \pi)^{S_D}}       
  \, e^{i  \int_{\mathcal{R}_D} d \boldsymbol{\sigma} \, t(\boldsymbol{\sigma}) h(\boldsymbol{\sigma})     }  \mathcal{G}_n [t], \label{erda1}
\end{eqnarray}  
where $S_D$ is the total number of single-spin states
\begin{equation}
S_D = \int_{\mathcal{R}_D} d \boldsymbol{\sigma} = \frac{2 \pi^{\frac{D}{2} } }{\Gamma \left( \frac{D}{2} \right)}.
\end{equation}  
By inserting Eq. (\ref{hopa1}) in $\mathcal{G}_n [t]$,
\begin{align}
& \mathcal{G}_n [t] = \sum_{k=0}^\infty \frac{k^n p_k}{c^n} \int \left( \prod_{r=1}^{k-n} d J_r \mathcal{D} q_r  p_{J}(J_r)   R[q_r] \right) \nonumber \\
 &\times
 \exp{\left[  - \frac{i}{\beta}   \int_{\mathcal{R}_D} d \boldsymbol{\sigma}  t(\boldsymbol{\sigma}) \sum_{r=1}^{k-n} \ln{\left(  \int_{\mathcal{R}_D} d \boldsymbol{\sigma}^{\prime}
        e^{\beta J_r  \boldsymbol{\sigma}^{T} \boldsymbol{\sigma}^{\prime}  } q_r(\boldsymbol{\sigma}^{\prime})      \right) }  \right]}, \nonumber
\end{align}  
and noting that the above expression factorizes in terms of a product over the
neighborhood of a single node, it is straightforward to show that $\mathcal{G}_n [t]$ assumes
the compact form
\begin{align}
  \mathcal{G}_n [t] &= \int_{0}^{\infty} d g \, g^n \, \nu(g) \left( \eta[t] \right)^g  \label{poutaq1}
\end{align}  
in the limit $c \rightarrow \infty$. The functional $\eta[t]$ is given by
\begin{align}
  &\eta[t]  = \lim_{c \rightarrow \infty} {\rm exp} {\Bigg \{}   c \, {\rm \ln}  \Bigg[ \int  d J \mathcal{D}q \,  p_{J}(J)   R[q]  \nonumber \\
    &\times
      \exp{\left(  - \frac{i}{\beta}  \int_{\mathcal{R}_D} d \boldsymbol{\sigma}    t(\boldsymbol{\sigma}) \ln{\left[  \int_{\mathcal{R}_D} d \boldsymbol{\sigma}^{\prime} 
       e^{\beta J \boldsymbol{\sigma}^{T} \boldsymbol{\sigma}^{\prime}      }
       q(\boldsymbol{\sigma}^{\prime})      \right] }  \right)}  \Bigg]   {\Bigg \}} ,
\label{paqu}
\end{align}  
and $\nu(g)$ is the high-connectivity limit of the distribution of rescaled degrees $k_1/c,\dots,k_N/c$ 
\begin{equation}
  \nu(g) = \lim_{c \rightarrow \infty} \sum_{k=0}^{\infty} p_k \delta{\left(g - \frac{k}{c}   \right)}.
  \label{forte}
\end{equation}  
The assumption that the distribution of rescaled degrees
attains a well-defined limit $\nu(g)$ for $c \rightarrow \infty$ underlies the validity of Eq.~(\ref{poutaq1}).

The next step is to calculate $\eta[t]$ from Eq. (\ref{paqu}).
Since the coupling strengths become very weak for $c \gg 1$, we expand Eq.~(\ref{paqu}) in powers of $J_r$ and in turn compute
the limit $c \rightarrow \infty$, finding
\begin{align}
  \eta[t] &= \exp{ \left[  - i   \int_{\mathcal{R}_D} d \boldsymbol{\sigma}      t(\boldsymbol{\sigma}) \left( K_0 \boldsymbol{ \sigma}^T \boldsymbol{M} 
      + \frac{1}{2} \beta K_1^2 \boldsymbol{ \sigma}^T \boldsymbol{\chi}  \boldsymbol{ \sigma} \right) \right]} \nonumber \\
  &\times \exp{\left( - \frac{ K_{1}^2 }{2 }  \int_{\mathcal{R}_D} d \boldsymbol{\sigma} d \boldsymbol{\sigma}^{\prime}     t(\boldsymbol{ \sigma})
    t( \boldsymbol{ \sigma}^{\prime})  \boldsymbol{ \sigma}^T \boldsymbol{\mathcal{C}} \boldsymbol{\sigma}^{\prime}     \right)},
  \label{popoli}
\end{align}  
in which we have introduced the $D$-dimensional magnetization $\boldsymbol{M}$
\begin{equation}
 M_{\alpha} = \int \mathcal{D}p  \, R[p] \langle \sigma_{\alpha} \rangle_{p}, \label{order1} 
\end{equation}  
the $D \times D$ connected correlation matrix $\boldsymbol{\chi}$ between the components of a spin
\begin{equation}
  \chi_{\alpha \beta} = \int \mathcal{D}p   \, R[p] \left( \langle \sigma_{\alpha} \sigma_{\beta}  \rangle_{p }
  - \langle \sigma_{\alpha}  \rangle_{ p}  \langle \sigma_{\beta}  \rangle_{p }   \right)  ,  \label{order2}
\end{equation}  
and the $D \times D$ correlation matrix $\boldsymbol{\mathcal{C}}$ between the components of the local magnetization
\begin{equation}
\mathcal{C}_{\alpha \beta} =  \int \mathcal{D}p  \, R[p]  \langle \sigma_{\alpha}  \rangle_{p }  \langle \sigma_{\beta}  \rangle_{p }    . \label{order3}
\end{equation}  
The order-parameters of Eqs.~(\ref{order1}-\ref{order3}) are defined in terms of the distribution
$R[p]$ of the single-site marginals on the cavity graph.

The final step is to perform the inverse Fourier transform in Eq.~(\ref{erda1}). Nonetheless, the exponential in $\mathcal{G}_n [t] $
has a quadratic term in $t(\boldsymbol{\sigma})$, as can be noticed by inserting
Eq.~(\ref{popoli}) in Eq.~(\ref{poutaq1}). This quadratic term, given by
\begin{equation}
\Omega[t]  =  \exp{\left( - \frac{ K_{1}^2 g }{2 }  \int_{\mathcal{R}_D} d \boldsymbol{\sigma} d \boldsymbol{\sigma}^{\prime}  t(\boldsymbol{ \sigma})
    t( \boldsymbol{ \sigma}^{\prime})  \boldsymbol{ \sigma}^T \boldsymbol{\mathcal{C}} \boldsymbol{\sigma}^{\prime}     \right)},
\end{equation}  
is linearized by a multivariate Gaussian integral over $\boldsymbol{z} = (z_1 \,\, z_2 \, \dots \, z_D)^T$
\begin{align}
  \Omega[t]   &=  \int_{-\infty}^{\infty}  {\rm D} \boldsymbol{z}
  \exp{\left[  - i K_1 \sqrt{g}  \boldsymbol{z}^T \int_{\mathcal{R}_D} d \boldsymbol{\sigma} \, t(\boldsymbol{\sigma} )  \boldsymbol{\sigma}  \right]}, \nonumber
\end{align}  
with the  Gaussian measure
\begin{equation}
{\rm D} \boldsymbol{z} =  \frac{d \boldsymbol{z}}{\sqrt{  ( 2 \pi)^D   \det \boldsymbol{\mathcal{C}}}}
   \exp{\left(- \frac{1}{2 } \boldsymbol{z}^{T}   \boldsymbol{\mathcal{C}}^{-1} \boldsymbol{z}  \right)},
\end{equation}
and $d \boldsymbol{z} = \prod_{\alpha=1}^D d z_{\alpha}$. By carrying out the above linearization, we arrive at the appealing expression
for 
\begin{align}
 \mathcal{G}_n [h] &= \int_{0}^{\infty} d g  g^n  \nu(g)
  \int_{-\infty}^{\infty} {\rm D} \boldsymbol{z}  \nonumber \\
  &\times \exp{\left[ - i \int_{\mathcal{R}_D} d \boldsymbol{\sigma} \,  t( \boldsymbol{\sigma}) \mathcal{H}_{\rm eff} \left( \boldsymbol{\sigma} |g, \boldsymbol{z} \right) \right] },  \label{tuasug}
\end{align}  
where $\mathcal{H}_{\rm eff} \left( \boldsymbol{\sigma} |g, \boldsymbol{z} \right)$ is the effective Hamiltonian
of a single spin
\begin{align}
  \mathcal{H}_{\rm eff} \left( \boldsymbol{\sigma} |g, \boldsymbol{z} \right) &=
  g  K_0 \boldsymbol{\sigma}^{T} \boldsymbol{M} +  K_1 \sqrt{g} \boldsymbol{\sigma}^{T} \boldsymbol{z}   \nonumber \\
  &+ \frac{1}{2}g \beta K_1^2 \boldsymbol{\sigma}^{T}  \boldsymbol{\chi}  \boldsymbol{\sigma}.
  \label{cortusa}
\end{align}  
Inserting Eq.~(\ref{tuasug}) in Eq.~(\ref{erda1}), the functional integral over $t( \boldsymbol{\sigma}) $ is immediately  performed, leading to
\begin{align}
 \mathcal{F}_n [h] &= \int_{0}^{\infty} d g  g^n  \nu(g)
 \int_{-\infty}^{\infty} {\rm D} \boldsymbol{z} \, \delta_F \left[ h(\boldsymbol{\sigma}) -  \mathcal{H}_{\rm eff} \left( \boldsymbol{\sigma} |g, \boldsymbol{z} \right)  \right].  \label{tuasu1}
\end{align}  

Equations (\ref{cortusa}) and (\ref{tuasu1}) constitute the central result of this paper, as they provide the analytic expression
for the high-connectivity limit of the distribution of the coupling energy between a single spin
and the effective field coming from its neighborhood. These equations are valid for the generic spin model
defined by Eq.~(\ref{tuacu}).
Remarkably, the distribution of $\mathcal{H}_{\rm eff}$ is generally
not Gaussian, but it explicitly depends  on the distribution $\nu(g)$ of rescaled degrees, which means
the model retains information about the graph structure even when $c \rightarrow \infty$.
The non-Gaussian nature of $\mathcal{H}_{\rm eff}$ is a direct consequence of the breakdown of the  central limit theorem, as discussed
in the beginning of this section.
For random graph models in which $\nu(g) = \delta(g-1)$, such
as regular and Erd\H{o}s-R\'enyi graphs \cite{NewmanBook}, the central limit theorem holds and $\mathcal{H}_{\rm eff}$ follows a Gaussian
distribution. The strength of the degree fluctuations for $c \rightarrow \infty$ is
quantified by the variance of $\nu(g)$
\begin{equation}
  \Delta_{\nu}^2 = \int_{0}^{\infty} d g \, g^2 \, \nu(g) - 1 .
  \label{furtade}
\end{equation}  
In the present context, we say that a network model is {\it homogeneous} if $\Delta_{\nu}^2 = 0$, whereas an {\it heterogeneous} network
is characterized by $\Delta_{\nu}^2 > 0$. 

Plugging Eq~(\ref{tuasu1}) back into Eqs.~(\ref{tu1})
and (\ref{tu2}), we find
\begin{align}
  W[p] &=   \int_{0}^{\infty} d g  \,  \nu(g)
  \int_{-\infty}^{\infty} {\rm D } \boldsymbol{z}    \nonumber \\
  &\times   \delta_F \left[ p(\boldsymbol{\sigma}) -   \frac{e^{\beta   \mathcal{H}_{\rm eff} \left( \boldsymbol{\sigma} |g, \boldsymbol{z} \right) }  }
    {  \int_{\mathcal{R}_D} d \boldsymbol{\sigma}^{\prime}   e^{\beta \mathcal{H}_{\rm eff} \left( \boldsymbol{\sigma}^{\prime} |g, \boldsymbol{z} \right)}  } \right], \label{turwe1} \\
   R[p] &=   \int_{0}^{\infty} d g  \, g \,  \nu(g)
  \int_{-\infty}^{\infty}   {\rm D } \boldsymbol{z}   \nonumber \\
  &\times   \delta_F \left[ p(\boldsymbol{\sigma}) -   \frac{e^{\beta   \mathcal{H}_{\rm eff} \left( \boldsymbol{\sigma} |g, \boldsymbol{z} \right) }  }
    {   \int_{\mathcal{R}_D} d \boldsymbol{\sigma}^{\prime}   e^{\beta \mathcal{H}_{\rm eff} \left( \boldsymbol{\sigma}^{\prime} |g, \boldsymbol{z} \right)  }  } \right], \label{turwe2}
\end{align}  
from which a closed set of equations for the order-parameters is derived from Eqs.~(\ref{order1}-\ref{order3}).
Equation (\ref{turwe1}) summarizes the gist of the
mean-field approach: the macroscopic behavior of an infinitely large system is reduced to an effective problem of a single spin, in which the
single-site states are sampled from a Boltzmann distribution with effective energy $\mathcal{H}_{\rm eff} \left( \boldsymbol{\sigma} |g, \boldsymbol{z} \right)$.
Equations (\ref{turwe1}) and (\ref{turwe2}), together with the order-parameter Eqs. (\ref{order1}-\ref{order3}), define
what we call the {\it heterogeneous mean-field theory}.
It is unfeasible to solve the generic order-parameter equations that follow from Eqs.~(\ref{turwe1}) and (\ref{turwe2}). Thus, in the sequel we explore
the role of degree fluctuations in specific models.


\subsection{Ferromagnetic couplings} \label{sub1}

In this section we address the impact of heterogeneous degrees on the behavior of ferromagnetic spin models. For $K_1 = 0$, the
coupling energy is given by
\begin{equation}
  \mathcal{H}_{\rm eff} \left( \boldsymbol{\sigma} |g \right) = g  K_0 \boldsymbol{\sigma}^{T}\boldsymbol{M},
  \label{eff}
\end{equation}
and the distribution of the effective field $g K_0 \boldsymbol{M}$ on a spin $\boldsymbol{\sigma}$ follows
from $\nu(g)$. Due to the orthogonal invariance of the Hamiltonian, Eq.~(\ref{tuacu}), the
magnetization  $\boldsymbol{m}$
is determined by its absolute value $|\boldsymbol{m}|$. For heterogeneous networks we get
\begin{equation}
  |\boldsymbol{m}| =  \int_{0}^{\infty} d g  \,  \nu(g)  \frac{I_{\frac{D}{2}} \left(  \beta g K_0 |\boldsymbol{M}|  \right) }{I_{\frac{D}{2}-1 } \left(  \beta g K_0 |\boldsymbol{M}|  \right) },
  \label{tuda1}
\end{equation}  
where $I_{a}(x)$ is the modified Bessel function of the first kind. The order-parameter $\boldsymbol{M}$ is the magnetization on the ensemble of cavity graphs, and its
absolute value solves the self-consistency equation
\begin{equation}
  |\boldsymbol{M}| =  \int_{0}^{\infty} d g  \, g \, \nu(g)  \frac{I_{\frac{D}{2}} \left(  \beta g K_0 |\boldsymbol{M}|  \right) }{I_{\frac{D}{2}-1 } \left(  \beta g K_0 |\boldsymbol{M}|  \right) }.
  \label{tuda2}
\end{equation}  
Equations (\ref{tuda1}) and (\ref{tuda2}) generalize the fully-connected  mean-field equations for ferromagnetic models with $D$-dimensional spins \cite{Cosimothesis} to the case
of heterogeneous networks. In particular, for Ising spins ($D=1$) we have $I_{\frac{1}{2}}(x)/I_{-\frac{1}{2}}(x) = \tanh(x)$, and the scalar magnetization $m$ is determined from 
\begin{align}
  m = \int_{0}^{\infty} d g \, \nu(g) \, \tanh{\left(\beta K_0 g M   \right)} \label{suad1} , \\
  M = \int_{0}^{\infty} d g \, g \, \nu(g) \, \tanh{\left(\beta K_0 g M   \right)}, \label{suad2} 
\end{align}  
which generalizes the Curie-Weiss mean-field equations \cite{Kochma2013} to heterogeneous random graphs.

Both parameters, $|\boldsymbol{M}|$ and $|\boldsymbol{m}|$, quantify the coherence of vector spins, but in slightly
different ways: $|\boldsymbol{m}|$
is the centroid of the spin configuration $\{\boldsymbol{\sigma}\}$ on the surface of the $D$-dimensional
hypersphere, while  $|\boldsymbol{M}|$ is the average of the spins weighted by the rescaled degree $g$. For homogeneous
random graphs, identified by $\nu(g) = \delta(g-1)$, Eqs.~(\ref{tuda1}) and (\ref{tuda2}) reduce to a single equation
for $|\boldsymbol{m}|$, normally derived from fully-connected graphs. For
heterogeneous networks, however, these parameters do not coincide, since highly
connected nodes (hubs) contribute more to the integral in Eq.~(\ref{tuda2}).
Theoretical approaches for the Kuramoto model ($D=2$) on heterogeneous networks have often characterized the
synchronization phase transition solely in terms of $|\boldsymbol{M}|$~\cite{Rodrigues2016}. While this is appropriate for homogeneous
networks, it leads to discrepancies between theory and simulations for the behavior of $|\boldsymbol{m}|$ when the network has strongly
fluctuating degrees, such as in
scale-free networks~\cite{yook2018two}.
Here we do not face this problem of choosing suitable observables beforehand, since both quantities, $|\boldsymbol{m}|$ and $|\boldsymbol{M}|$,
and the relationship between them
emerge naturally from the high-connectivity limit of local tree-like networks, without any additional assumption on the network topology.

\begin{figure}[t!]
	\centering
	\includegraphics[width=0.9\columnwidth]{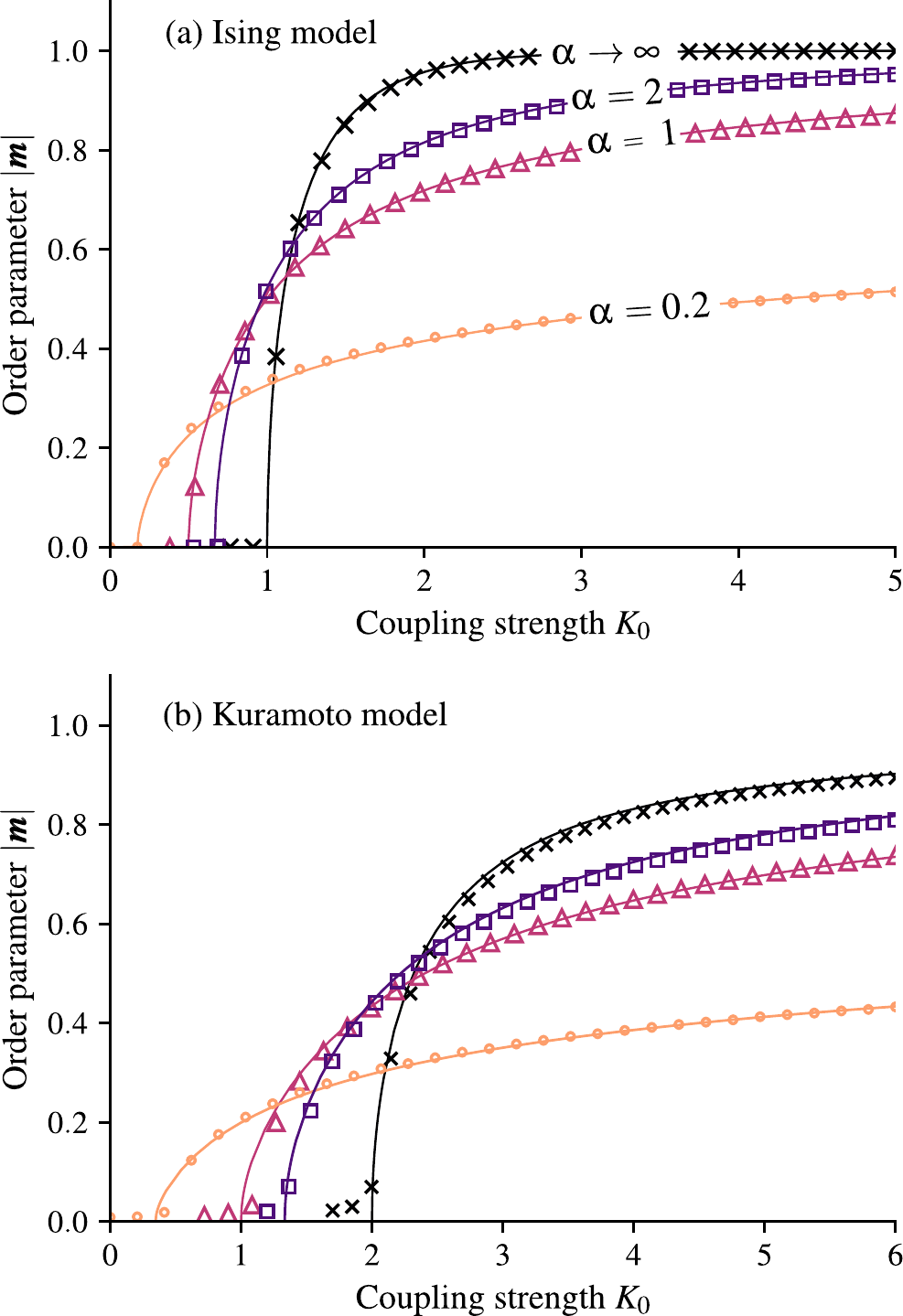}
	\caption{The magnetization $|\boldsymbol{m}|$ as a function of the coupling strength $K_0$ for the  Ising [panel (a)] and the
          Kuramoto model [panel (b)] on heterogeneous networks with an infinitely large average degree and temperature $T=1$. The heterogeneity parameter $\alpha$ controls
          the variance of the negative binomial degree distribution [Eqs.~(\ref{roar}) and (\ref{pocas1})].
          The solid lines are obtained by solving Eqs.~(\ref{tuda1}) and (\ref{tuda2}), while the symbols are results from numerical simulations. The simulated
          networks have $N = 10^4$ nodes and average degree $c = 100$, and they were generated according to the configuration model with
          the negative binomial distribution [Eq.~(\ref{utaop})]. See Appendix~\ref{sec:simulations} for further details on the simulations.}
	\label{orderpar}
\end{figure}

The solutions of Eqs.~(\ref{tuda1}) and (\ref{tuda2}) allow to study the effect of degree fluctuations on the
phase diagram of the ferromagnetic spin model.
In the limit $T \rightarrow \infty$, we find the solution $\boldsymbol{M}=\boldsymbol{m}=0$ and the system lies in the paramagnetic phase.
By expanding Eq.~(\ref{tuda2}) up to $O(|\boldsymbol{M}|)$, we conclude that $|\boldsymbol{M}|$ has a nontrivial solution provided
\begin{equation}
\beta K_0 = \frac{D}{1 + \Delta_{\nu}^2 }  ,
  \label{critic}
\end{equation}  
where $\Delta_{\nu}^2$ is the {\it heterogeneity parameter}, given by Eq.~(\ref{furtade}). The above expression defines a line in the parameter
space at which the model undergoes a continuous phase transition between the paramagnetic and the ferromagnetic phase.
Equation (\ref{critic}) shows that the size of the ferromagnetic phase increases as a function of the
heterogeneity parameter. Indeed, the critical value of $\beta K_0$ vanishes as $\Delta_{\nu}^2 \rightarrow \infty$ for
any finite $D$, which is in line with previous results for the critical temperature of the Ising model on scale-free networks \cite{Leone2002,Doro2002}.

\begin{figure}[t!]
	\centering
	\includegraphics[width=1.0\columnwidth]{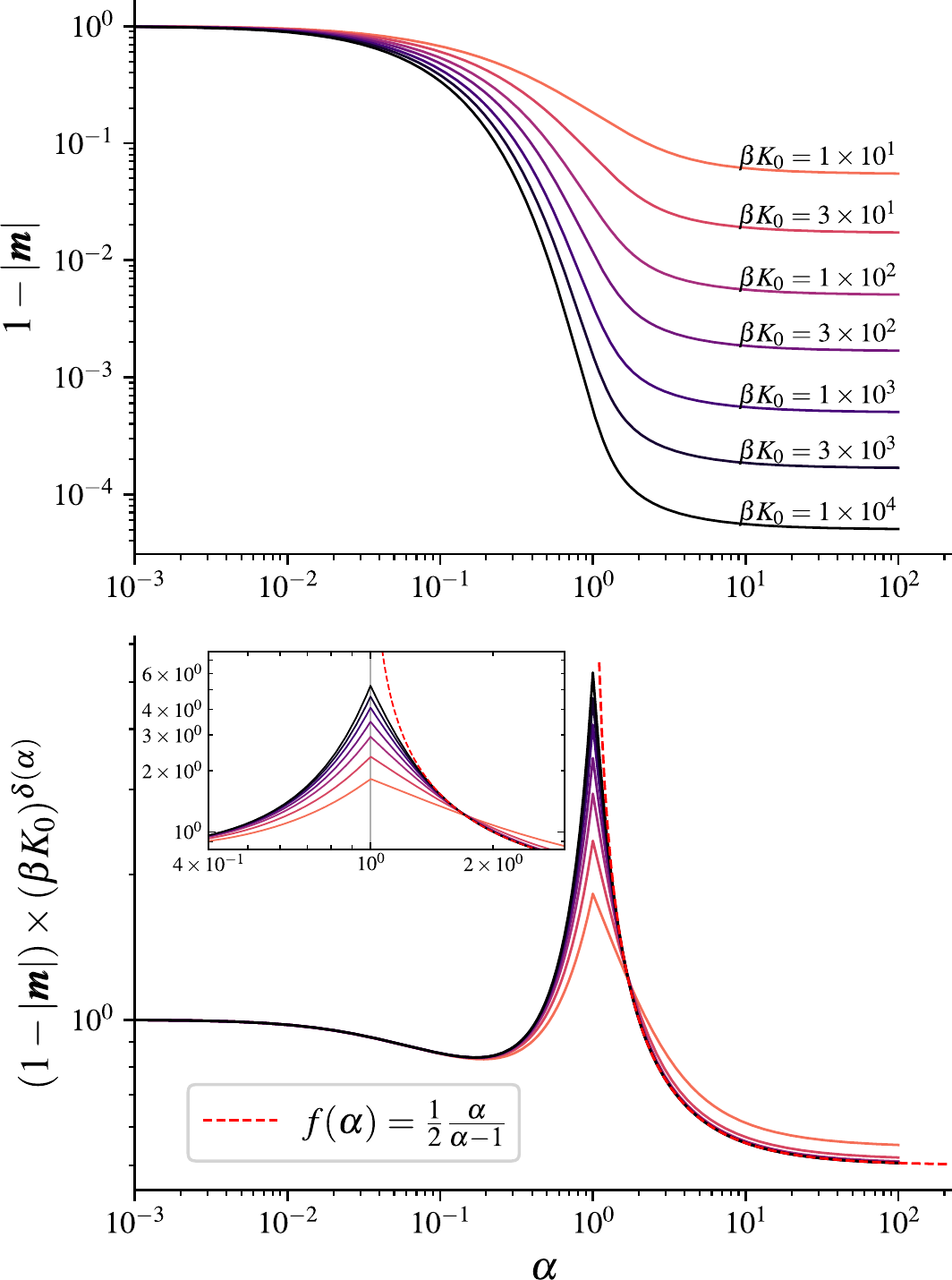}
	\caption{(a) The magnetization $|\boldsymbol{m}|$ as a function of the heterogeneity parameter $\alpha$ for the
          Kuramoto model on heterogeneous networks with an infinitely large average degree, temperature $T=1$, and several values of the coupling
          strength $\beta K_0$. The results follow from the numerical solutions of Eqs.~(\ref{tuda1}) and (\ref{tuda2}). (b) The same results of panel (a), but
          rescaled by the factor $( \beta K_0 )^{\delta(\alpha)}$, where $\delta(\alpha) =\alpha$ if $0 < \alpha \leq 1$, and  $\delta(\alpha) =1$ otherwise. The collapse of the
          data for $\beta K_0 \gg 1$ onto a single curve corroborates Eq.~(\ref{gugua}) for $|\boldsymbol{m}|$.
          The red dashed curve is the analytic expression for $f(\alpha) = (1 - |\boldsymbol{m}|)  ( \beta K_0 )^{\delta(\alpha)}$ ($\beta K_0 \gg 1$) in
          the regime $\alpha > 1$ (see the main text). }
	\label{colapsado}
\end{figure}

In the rest of this section, we present explicit results for random networks with the negative binomial degree distribution. Substituting Eq.~(\ref{utaop}) in
Eq.~(\ref{forte}), we calculate the corresponding distribution of rescaled degrees
\begin{equation}
  \nu_{\rm nb} (g) = \frac{\alpha^{\alpha}}{\Gamma(\alpha)} g^{\alpha-1} e^{-\alpha g},
  \label{roar}
\end{equation}  
where $\alpha > 0$ is related to $\Delta_{\nu}^2 $ as follows
\begin{equation}
  \Delta_{\nu}^2 = \frac{1}{\alpha}.
  \label{pocas1}
\end{equation}
Equation (\ref{roar}) allows us to interpolate continuously between homogeneous networks ($\alpha \rightarrow \infty$)
and strongly heterogeneous networks ($\alpha \rightarrow 0$) by varying a single parameter $\alpha$ that controls
the degree fluctuations in the high-connectivity limit. 

Figure \ref{orderpar} illustrates the effect of $\alpha$ on the behavior of $|\boldsymbol{m}|$ as a function of
the coupling strength $K_0$ for the Ising ($D=1$) and the Kuramoto model ($D=2$). The ferromagnetic or synchronized
phase, identified by $|\boldsymbol{m}| > 0$, appears through a continuous phase transition at the critical
coupling strength predicted by Eq. (\ref{critic}). Moreover, figure \ref{orderpar} compares the theoretical results, derived 
from the numerical solutions of Eqs.~(\ref{tuda1}) and (\ref{tuda2}), with data from numerical simulations of the dynamics
of each model. The agreement between theory and simulations is excellent in both cases, confirming the exactness of Eqs.~(\ref{tuda1}) and (\ref{tuda2}).

Even though the critical value of $\beta K_0$ vanishes for $\alpha \rightarrow 0$, which could suggest the absence of a paramagnetic phase
in the strongly heterogeneous regime, we note that $|\boldsymbol{m}|$ becomes gradually smaller inside the ordered
phase as $\alpha$ is reduced. To better understand how the heterogeneity parameter $\alpha$ impacts the ordered phase of spin models, we present
in figure \ref{colapsado} the behavior of $|\boldsymbol{m}|$ as a function
of $\alpha$ for the Kuramoto model. The main outcome is that, for $\beta K_0 \gg 1$, deeply in the synchronized phase, $|\boldsymbol{m}|$ has the functional form
\begin{equation}
  |\boldsymbol{m}| = 1 -  \frac{ f(\alpha) }{( \beta K_0 )^{\delta(\alpha)}} \,\, (\beta K_0 \gg 1).
  \label{gugua}
\end{equation}  
In the sector $\alpha > 1$, we are able to show analytically that $\delta(\alpha)= 1$ and $f(\alpha) = \frac{\alpha}{2 (\alpha-1)}$, by
expanding Eqs.~(\ref{tuda1}) and (\ref{tuda2}) for $\beta K_0 \gg 1$. In the range $0 < \alpha \leq 1$, there are strong degree
fluctuations and the perturbative expansion fails, leading to a divergence in $|\boldsymbol{m}|$. In spite of that, figure \ref{colapsado}
shows that the results for different $\beta K_0$ collapse onto a single curve by choosing $\delta(\alpha) = \alpha$
in the sector $0 < \alpha \leq 1$, confirming the functional form put forward in Eq.~(\ref{gugua}). This numerical procedure does
not give the explicit form of $f(\alpha)$, but it clearly shows that  $\lim_{\alpha \rightarrow 0} f(\alpha)=1$.

\begin{figure}[t!]
	\centering
	\includegraphics[width=1.0\columnwidth]{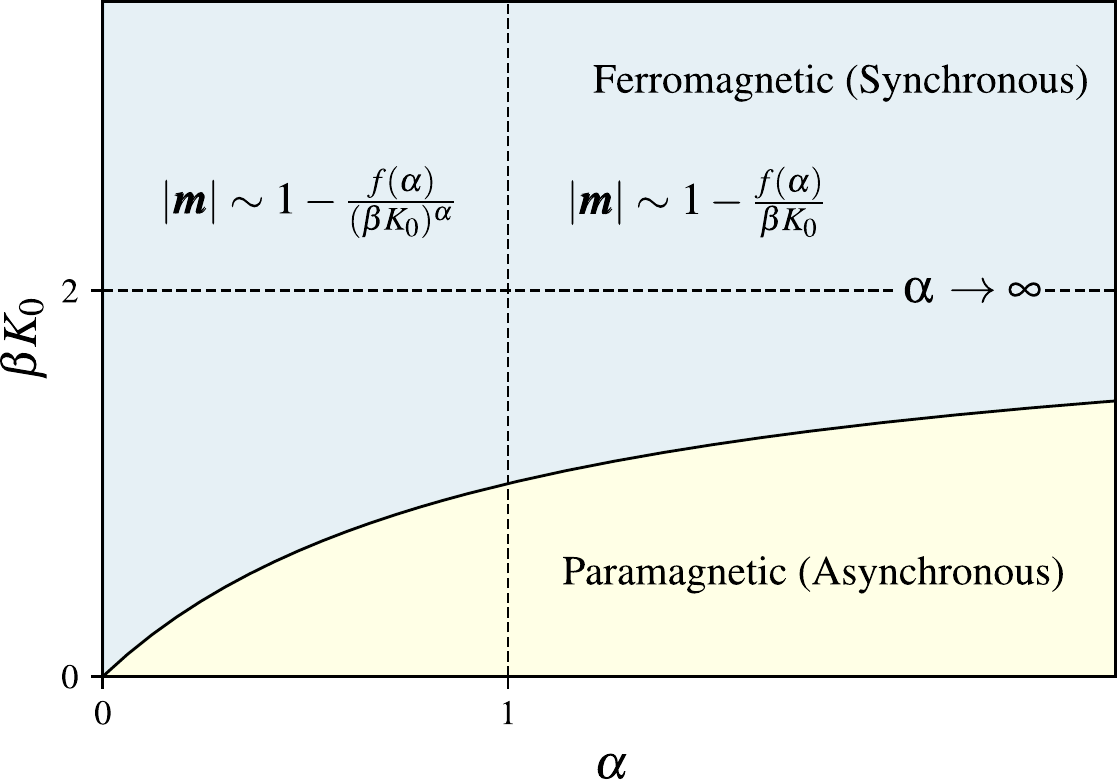}
	\caption{Phase diagram of the Kuramoto model ($D=2$) on heterogeneous networks with an infinitely large average degree. The heterogeneity
          parameter $\alpha$ measures the variance of the negative binomial degree distribution (see Eq.~(\ref{roar})), $\beta =1/T$ is the inverse temperature, and $K_0 > 0$ is the coupling
          strength between the spins. The solid line identifies the continuous transition between the ferromagnetic (synchronous) and the
          paramagnetic (asynchronous) phases (see Eq.~(\ref{critic})).
          The vertical dashed line at $\alpha=1$ separates two distinct asymptotic regimes of the magnetization $|\boldsymbol{m}|$
          for $\beta K_0 \gg 1$, each one displayed on the figure. The horizontal dashed line at $\beta K_0 =2$ is the critical
          coupling strength of the Kuramoto model on fully-connected networks. }
	\label{pdferro}
\end{figure}

On the whole, Eq.~(\ref{gugua}) leads to two interesting conclusions. First, the
magnetization $|\boldsymbol{m}|$ vanishes inside the ordered phase as $\alpha \rightarrow 0$ ($\Delta_{\nu} \rightarrow \infty$).
This is a surprising finding given
 that highly heterogeneous networks have been regarded as optimal structures for the emergence of synchronization.
 Indeed, the vanishing of the critical coupling for the Kuramoto model on heterogeneous networks has been reported in previous
 studies~\cite{moreno2004synchronization,ichinomiya2004frequency,restrepo2005onset,arenas2008synchronization,Rodrigues2016,peron2019onset}, but it
 has remained unclear how $|\boldsymbol{m}|$ behaves when the variance of the degree distribution diverges.
 Our results confirm that, in the limit $c \rightarrow \infty$, the critical coupling does vanish as $\Delta_{\nu} \rightarrow \infty$, but
 they also reveal that $|\boldsymbol{m}|$ goes to zero, suggesting that highly heterogeneous networks do not sustain
 any level of synchronous oscillation in such limit.
The second interesting conclusion is that the point $\alpha=1$ separates two different regions inside the synchronous phase, each one marked by a particular
behavior of the magnetization $|\boldsymbol{m}|$ for $\beta K_0 \gg 1$. For $0 < \alpha \leq 1$, degree fluctuations have a significant
impact in the ordered
phase and the magnetization is given by $1 - |\boldsymbol{m}| \sim (\beta K_0)^{-\alpha}$. For $\alpha > 1$, degree fluctuations
are immaterial and the magnetization behaves as $1 - |\boldsymbol{m}| \sim (\beta K_0)^{-1}$, regardless of
the heterogeneity parameter $\alpha$. Interestingly, the function $f(\alpha)$ displays a cusp at
$\alpha=1$, reflecting the non-analytic behavior of $|\boldsymbol{m}|$ as a function of $\alpha$.
Although Eq.~(\ref{gugua}) and figure \ref{colapsado} are results for the Kuramoto model ($D=2$), we expect that the above
two conclusions  hold for arbitrary $D < \infty$.

The main results discussed up to now are summarized in
figure \ref{pdferro}, in which we present the phase diagram of the Kuramoto model on heterogeneous networks. Notice
that for $\alpha \rightarrow \infty$ the transition line approaches $\beta K_0 = 2$, which is the critical point
of fully connected networks with identical stochastic oscillators ~\cite{sakaguchi1988cooperative,sonnenschein2013approximate}. 

We end this section by discussing another consequence of heterogeneous degrees, namely the lack
of correspondence between local and global ensemble averages. We present
results for the Kuramoto model ($D=2$), but the main conclusions should be once more valid for any $D < \infty$.
The state of a spin $\boldsymbol{\sigma}_i$ in $D=2$ is fully specified by
the phase $\phi_i \in (-\pi,\pi]$, distributed according to the local marginal $P_i (\phi_i)$.   
Combining Eqs.~(\ref{hua1}), (\ref{turwe1}), and (\ref{eff}), it is straightforward to write the analytic expression
for the moments of the random function $P_i (\phi_i)$ 
\begin{equation}
  \langle \left[ P (\phi) \right]^r \rangle_W = \int_{0}^{\infty} d g \, \nu(g) \frac{e^{r \beta g K_0 |\boldsymbol{M}| \cos{\phi}  } }
          {\left[ 2 \pi I_0 (  \beta g K_0 |\boldsymbol{M}|  )  \right]^r  },
          \label{jocuq}
\end{equation}  
where we have set $\boldsymbol{M} = (|\boldsymbol{M}| \,\, 0)^T$ without loosing any generality.
The above equation
yields all moments of the functional distribution $W[P]$ of the single-site marginals $P_i (\phi_i)$ over the entire network. In particular, we are interested 
in the mean $\mu_P(\phi)$ and the variance $\sigma_{P}^{2}(\phi)$ 
\begin{eqnarray}
\mu_P(\phi) &=& \langle P (\phi) \rangle_W , \nonumber \\
\sigma_{P}^{2}(\phi) &=&  \langle \left[ P (\phi) \right]^2  \rangle_W - \langle P (\phi)  \rangle_{W}^2.
\end{eqnarray}  

\begin{figure*}[t!]
	\centering
	\includegraphics[width=2.0\columnwidth]{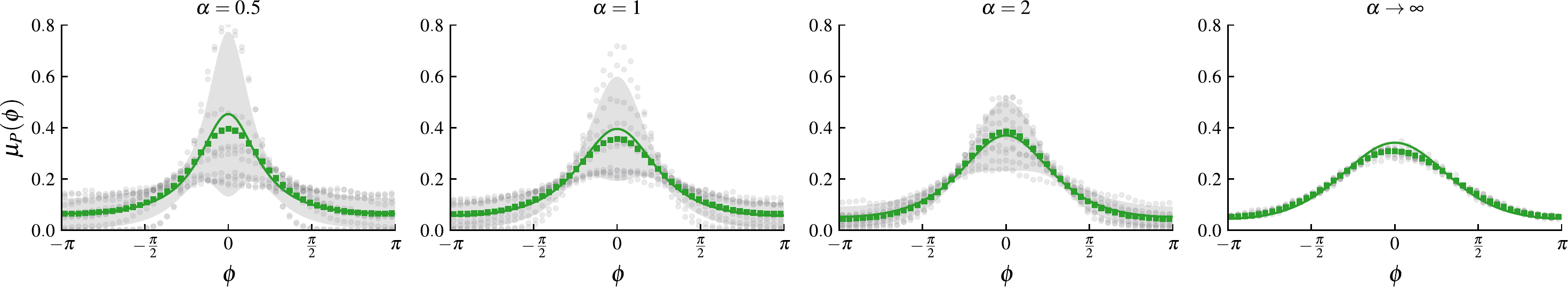}
	\caption{The mean value $\mu_P(\phi)$ of the single-site marginal $P_i(\phi_i)$ over all nodes $i=1,\dots,N$ of
          the stochastic Kuramoto model on heterogeneous networks with a negative binomial degree distribution (see Appendix~\ref{sec:simulations}).
          The panels show $\mu_P(\phi)$ for temperature $T=1$ and different values of the heterogeneity parameter $\alpha$ [see Eq.~(\ref{pocas1})]. In all panels, the coupling
          strength $K_0$ is adjusted such that $|\boldsymbol{m}| \simeq 0.45$. Solid lines are calculated from Eq.~(\ref{jocuq}) while the shaded
          area highlights the dispersion $[\mu_P(\phi) - \sigma_{P}(\phi),\mu_P(\phi) + \sigma_{P}(\phi)]$ around  $\mu_P(\phi)$, with $\sigma_{P}(\phi)$ denoting the
          standard deviation of $P_i(\phi_i)$, also computed from Eq.~\eqref{jocuq}. The symbols are histograms obtained from numerical simulations of the dynamics of the model. Each
            individual histogram (gray circles) is constructed by monitoring the time evolution of a single phase $\phi_i(t)$ selected at random, while the green squares represent
            the average over all individual histograms. All numerical simulations were performed with $N = 10^4$ and $c = 100$.}
	\label{fluct}
\end{figure*}
Figure \ref{fluct} illustrates the effect of the heterogeneity parameter $\alpha$ on the mean $\mu_P(\phi)$  inside
the synchronized phase of the Kuramoto model. The different panels are for the same value of the magnetization $|\boldsymbol{m}|$. More
importantly, the shaded area in figure \ref{fluct} represents the dispersion
$[\mu_P(\phi) - \sigma_{P}(\phi),\mu_P(\phi) + \sigma_{P}(\phi)]$ around the mean $\mu_P(\phi)$, quantifying the
  fluctuations of the random function  $P_i (\phi_i)$ from site to site. Each set of gray points in figure \ref{fluct} is the histogram
  of a single-site phase, which weights the amount of time that a randomly chosen spin $\boldsymbol{\sigma}_i$
  has a certain orientation $\phi_i$. Each one of these single-site histograms is constructed  from a single run of the numerical simulation by storing the values of
  a given $\phi_i$ over a long time interval  after the system has reached equilibrium. 

  For homogeneous networks ($\alpha \rightarrow \infty$), the variance $\sigma_{P}^{2}(\phi)$ is zero and the distribution $W[P]$ is
  peaked at the mean value $\mu_P(\phi)$. Accordingly, each single-site histogram coincides with $\mu_P(\phi)$, confirming that
  all spins are identical, in the sense that the fluctuations of the individual phases are described by the same distribution $\langle P (\phi) \rangle_W$.
  Thus, local and global ensemble averages become equivalent in the high-connectivity limit of ferromagnetic spin models on homogeneous networks, such as regular and
  Erd\H{o}s-R\'enyi random graphs. In contrast, we observe that in the case of heterogeneous networks ($\alpha < \infty$), the size of the shaded
  area in figure \ref{fluct} becomes gradually larger as $\alpha$ decreases, confirming that the functional form of $P_i (\phi_i)$ fluctuates from site to site,
  and the statistical properties of the spins remain different from each other, even in the high-connectivity limit.

  Although the stochastic
    Kuramoto model on heterogeneous networks has been extensively studied in previous works
    (see, e.g.~\cite{sonnenschein2012onset,sonnenschein2013approximate,Rodrigues2016}), little attention has been paid to the characterization
    of site to site fluctuations. It is thus noteworthy that the theory presented in Sec.~\ref{secHet} allows us to study not only global
    observables of the system, but also fluctuations at the level of individual nodes.


\subsection{Random couplings} \label{rr}

In this subsection we discuss the effect of heterogeneous degrees in mean-field spin models
with random coupling strengths. We restrict ourselves to the simplest
case of Ising spins ($D=1$), with $\sigma_i \in \{ -1, 1 \}$, in which
the weights  $p_i(\sigma)$ of a single-site marginal are encoded in the two-dimensional
vector $\vec{p}_i = (p_i(1) \,\,\, p_i(-1))^T$.

The mean-field equations for the high-connectivity limit of Ising spins coupled through the random links
of heterogeneous networks follow from Eqs.~(\ref{cortusa}), (\ref{turwe1}), and (\ref{turwe2}). The distributions
of single-site marginals read
\begin{align}
  W(\vec{p}) &= \int_{-\infty}^{\infty} d h \, \omega_0(h) \prod_{\sigma \in \{ 1 , -1 \} }  \delta \left[ p(\sigma) -   \frac{e^{\beta   \sigma h     }  } { 2 \cosh{\left( \beta h \right)}  } \right], \label{tur1} \\
  R(\vec{p})  &=
  \int_{-\infty}^{\infty} d h \, \omega_1(h) \prod_{\sigma \in \{ 1 , -1 \}  }  \delta \left[ p(\sigma) -   \frac{e^{\beta   \sigma h     }  } {  2 \cosh{\left( \beta h \right)}   } \right], \label{tur2}
\end{align}  
where the distributions $\omega_0(h)$ and $\omega_1(h)$ follow from
\begin{align}
\omega_n (h) =   \int_{0}^{\infty} d g  \, g^n \,  \nu(g) \int_{-\infty}^{\infty} d z P_{\rm g} (z) \nonumber \\
\delta \left( h -   g  K_0  M -  K_1 \sqrt{g Q    }  z \right) \label{efffie}
\end{align}
by setting $n=0,1$. The function $P_{\rm g} (z)$ is the normal probability density
\begin{equation}
  P_{\rm g} (z) = \frac{1}{\sqrt{2 \pi }}   e^{- \frac{z^2}{2   }}.
  \label{uffu}
\end{equation}  
The order-parameters $M$ and $Q$, appearing in Eq.~(\ref{efffie}), are defined as
\begin{align}
M = \int d \vec{p} \, R(\vec{p}) \langle \sigma \rangle_{\vec{p}}, \label{tuya1} \\
Q = \int d \vec{p} \, R(\vec{p}) \langle \sigma \rangle_{\vec{p}}^2, \label{tuya2}
\end{align}  
with $d \vec{p} = dp(1) dp(-1)$. The local magnetization $\langle \sigma \rangle_{\vec{p}}$ is computed 
by replacing the integral in Eq.~(\ref{huawe1}) by a discrete sum over $\sigma \in \{ 1, -1 \}$.
The quantities $M$ and $Q$ are, respectively, the global magnetization and the Edwards-Anderson order-parameter \cite{Sherrington1975}
on the ensemble of cavity graphs. Inserting Eq.~(\ref{tur2}) in the above definitions and integrating over $\vec{p}$, we arrive at the self-consistency
equations
\begin{align}
M = \int_{-\infty}^{\infty} d h \, \omega_1(h) \tanh{\left(\beta h    \right)}, \label{jotur1} \\
Q = \int_{-\infty}^{\infty} d h \, \omega_1(h) \tanh^2{\left(\beta h    \right)}.\label{jotur2}
\end{align} 
The global magnetization 
\begin{equation}
m = \int d \vec{p} \, W(\vec{p}) \langle \sigma \rangle_{\vec{p}}
\end{equation}
and the Edwards-Anderson order-parameter
\begin{equation}
q_{\rm EA} = \int d \vec{p} \, W(\vec{p}) \langle \sigma \rangle_{\vec{p}}^2
\end{equation}
follow from the distribution $W(\vec{p})$ of marginals on the original graph.
Substituting Eq.~(\ref{tur1}) in the above definitions and integrating over $\vec{p}$, we get
\begin{align}
m = \int_{-\infty}^{\infty} d h \, \omega_0(h) \tanh{\left(\beta h    \right)}, \label{kotu1} \\
q_{\rm EA} = \int_{-\infty}^{\infty} d h \, \omega_0(h) \tanh^2{\left(\beta h    \right)}.\label{kotu2}
\end{align} 
Equations (\ref{jotur1}), (\ref{jotur2}), (\ref{kotu1}) and (\ref{kotu2}) generalize the replica-symmetric mean-field equations of Ising spin-glass models with random
interactions to the case of heterogeneous networks. For homogeneous networks, characterized by $\nu(g) = \delta(g-1)$, we
obtain $\omega_0(h) = \omega_1 (h)$ and the above equations reduce to the replica-symmetric equations of the Sherrington-Kirkpatrick model \cite{Sherrington1975,Sherrington1978}.
The solutions of the self-consistency Eqs.~(\ref{jotur1}) and (\ref{jotur2}) allow us to address the impact of heterogeneous degrees
on the phase diagram of the system.

According to Eq.~(\ref{tur1}), the local marginal $p_i(\sigma)$ at node $i$ is a Boltzmann factor parametrized by the scalar
effective field $h_{i,{\rm eff}}$  \cite{Monasson1998}, 
given in terms of the local magnetization $\langle \sigma \rangle_{\vec{p}_i}$ as follows
\begin{equation}
  h_{i,{\rm eff}} = \frac{1}{\beta}\tanh^{-1}{\langle \sigma \rangle_{\vec{p}_i} } ,
  \label{tutar}
\end{equation}  
where $\vec{p}_i$ is drawn from $W(\vec{p})$. Combining Eqs.~(\ref{sumat}) and (\ref{zigui1}), one can show that $h_{i,{\rm eff}}$ is also
given by Eq.~(\ref{retua}).
The object $\omega_0 (h)$ is nothing more
than the 
empirical distribution of effective fields
\begin{equation}
  \omega_0(h) \equiv \mathcal{P}_{\rm eff} (h) = \lim_{N \rightarrow \infty} \frac{1}{N} \sum_{i=1}^N \delta \left( h - h_{i,{\rm eff}}  \right),
  \label{ufopoas}
\end{equation}  
which also gives information about the fluctuations of the weights $\vec{p}_i$ throughout the network.
The random variable $h_{i,{\rm eff}}$  is defined for a single
realization of the random graph and its distribution $\mathcal{P}_{\rm eff}(h)$ is determined by the fluctuations in the graph structure.

According to Eq.~(\ref{efffie}), $\omega_n (h)$ is determined only by
the probability densities $P_{\rm g} (z)$ and $\nu(g)$, which enables to derive the general formula
\begin{equation}
  \omega_n (h) = \frac{1}{K_1 \sqrt{Q}} \int_{0}^{\infty} d g \, g^{n-\frac{1}{2}}  \nu(g) P_{\rm g} \left( \frac{h - g K_0 M}{K_1 \sqrt{Q g }  }  \right)
  \label{omigaw}
\end{equation}  
for arbitrary $n$.
Note that Eq.~(\ref{omigaw}) is valid for an arbitrary distribution $\nu(g)$ of rescaled degrees. If the network is
homogeneous ($\nu(g) = \delta(g-1)$), then $\omega_0(h)$ is a
Gaussian distribution with mean $K_0 m$ and standard deviation $K_1 \sqrt{q_{EA}}$ \cite{Sherrington1975,Sherrington1978}.
If the network is heterogeneous, then $\omega_0(h)$ is not Gaussian, which stems from the breakdown of the central limit
theorem due to the large variance  of the random number of summands in Eq.~(\ref{retua}).
The average $\langle p(\sigma) \rangle_W$ in the present formalism
corresponds with the replica-symmetric {\it ansatz} employed in the study of spin models on random graphs \cite{Monasson1998}.

Before discussing the phase diagram for $K_1 > 0$, we point out that the heterogeneous mean-field theory is
not exact in the whole phase diagram. In the context of the replica method \cite{Mezard1987}, we say that
equations (\ref{jotur1}), (\ref{jotur2}), (\ref{kotu1}) and (\ref{kotu2}) are the replica symmetric (RS)
solution of the model. In the case of the Sherrington-Kirkpatrick model \cite{Sherrington1975,Sherrington1978}, obtained
for the choice $\nu(g)= \delta(g-1)$, the RS solution becomes unstable at low temperatures owing to the existence of an exponentially large number of metastable
states \cite{Mezard1987}. The sector of the phase diagram where the replica symmetric theory fails is
bounded by the so-called Almeida-Thouless (AT) line \cite{Almeida1978}, which is determined by the
eigenvalues of the Hessian of the free energy.

It is thus important to establish the limit of validity of the heterogeneous mean-field theory and ask how degree fluctuations impact the location of the AT line. Instead of tackling
this problem by analyzing the Hessian of the free-energy, here we follow an alternative strategy  \cite{Kwon1991,Pagnani2003,Neri2010}, based on the
number of solutions of the cavity equations (\ref{cav}). By defining $\{ p_{j,1}^{(i)} (\sigma_j ) \}$ and
$\{ p_{j,2}^{(i)} (\sigma_j ) \}$ as two fixed-point solutions of Eqs.~(\ref{cav}) for the {\it same}
realization of the random network, we introduce the correlation $\rho$ between the local magnetizations
of each solution on an ensemble of graphs
\begin{equation}
  \rho = \int d \vec{p}_1 d \vec{p}_2   R_{12} \left( \vec{p}_1,\vec{p}_2 \right)  \langle \sigma \rangle_{  \vec{p}_1  }
  \langle \sigma \rangle_{  \vec{p}_2   },
  \label{guasw}
\end{equation}  
where the joint distribution $R_{12} \left( \vec{p}_1,\vec{p}_2 \right)$ of weights is defined as follows
\begin{equation}
  R_{12} \left( \vec{p}_1,\vec{p}_2 \right)  = \! \! \! \lim_{N \rightarrow \infty}  \! \! \! \frac{\sum_{j=1}^N \sum_{i \in \partial_j}
   \!  \delta \left(  \vec{p}_1 - \vec{p}_{j,1}^{(i)} \right)  \! \delta \left(  \vec{p}_2 - \vec{p}_{j,2}^{(i)} \right)}{ \sum_{j=1}^N K_j }.
  \label{joiunuta}
\end{equation}  
When the RS theory is stable, Eqs.~(\ref{cav}) exhibit a single solution, the distribution
$R_{12} \left( \vec{p}_1,\vec{p}_2 \right)$ becomes diagonal
\begin{equation}
  R_{12} \left( \vec{p}_1,\vec{p}_2 \right) = \delta \left( \vec{p}_1  -  \vec{p}_2  \right) R\left( \vec{p}_1 \right),
  \label{jurema}
\end{equation}
and we simply have $\rho = Q$, with $Q$ given by Eq.~(\ref{jotur2}).
When the RS theory is unstable, the cavity Eqs.~(\ref{cav}) have a large number of fixed-point solutions resulting from the existence of a
large number of extrema in the free-energy. To calculate the AT line, it suffices to consider
the stability of the RS theory under a perturbation of $\rho$. By plugging $\rho = Q + \delta \rho$ in Eq.~(\ref{guasw}) and expanding
its right hand side up to $O(\delta \rho)$, we conclude that the RS theory is unstable provided
\begin{equation}
  \beta^2 K_1^2
  \int_{-\infty}^{\infty} d h \, \omega_{2}(h) {\rm sech}^{4} \left( \beta h \right) > 1,
  \label{iuts}
\end{equation}  
where $\omega_{2}(h)$ is defined through Eq.~(\ref{omigaw}). Equation (\ref{iuts}) is the condition for the breakdown of the
RS theory, generalizing the AT line of Ising
spin models with random couplings to  heterogeneous networks. For homogeneous networks with $\nu(g) = \delta(g-1)$, we recover
the AT line for the Sherrington-Kirkpatrick model \cite{Almeida1978}. The details involved in the derivation of Eq.~(\ref{iuts}) are
explained in appendix \ref{ATline00}. We remark that, although we have limited ourselves to Ising spins, the current approach
to obtain the AT line can be also applied to $D$-dimensional spins. 

Let us discuss our results for the phase diagram of the Ising model on heterogeneous networks with $K_1 > 0$. 
Equations (\ref{jotur1}) and (\ref{jotur2}) exhibit the paramagnetic solution $M=Q=0$ at
sufficiently high temperatures. The model undergoes
a second-order phase transition from the paramagnetic to the ferromagnetic phase, characterized by $M \neq 0$ and $Q \neq 0$, at the
critical temperature
\begin{equation}
  T = K_0 \left( 1 + \Delta_{\nu}^2  \right) ,
  \label{turaca1}
\end{equation}  
derived from an expansion of Eqs.~(\ref{jotur1}) and (\ref{jotur2}) for $|M| \ll 1$ and $Q \ll 1$. As long as $K_1 >0$, the model
has a spin-glass phase, in which $M=0$ and $Q \neq 0$. By setting $M=0$ and expanding Eq.~(\ref{jotur2}) for $Q \ll 1$, we find
the critical temperature 
\begin{equation}
  T = K_1 \sqrt{  1 + \Delta_{\nu}^2  } 
  \label{turaca2}
\end{equation}  
that marks the second-order transition between the paramagnetic and the spin-glass phase.

\begin{figure}[t!]
	\centering
	\includegraphics[width=1.0\columnwidth]{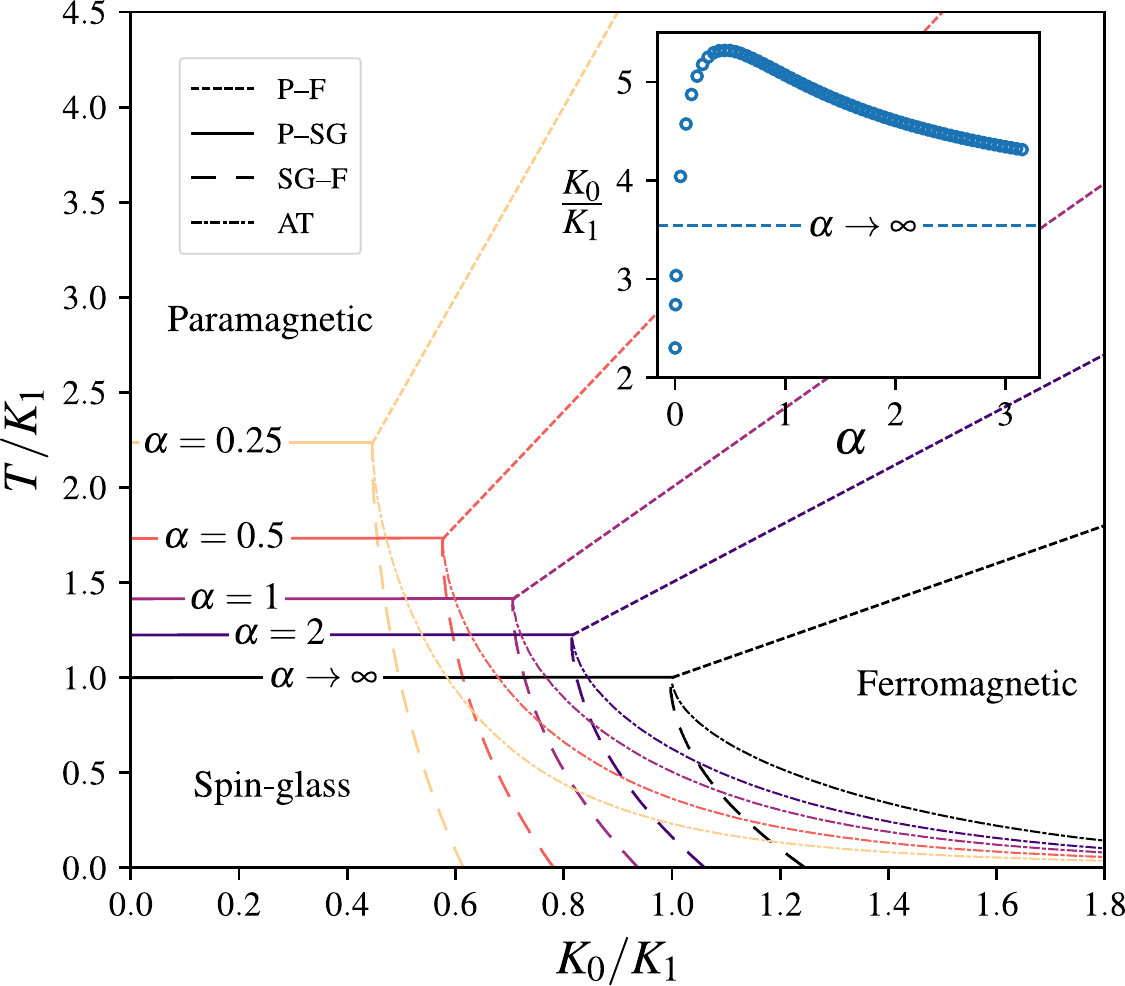}
	\caption{Phase diagram of the Ising model on heterogeneous networks with random couplings and an infinitely large average
          degree $c$. The degrees follow a binomial degree distribution with heterogeneity parameter
          $\alpha$ [see Eqs.~(\ref{utaop}) and (\ref{roar})]. The coupling strengths are drawn from a distribution with mean $K_0/c$ and
          standard deviation $K_1/\sqrt{c}$. The model exhibits a paramagnetic (P), a ferromagnetic (F), and a spin-glass (SG) phase
          (see the main text). The Almeida-Thouless (AT) line bounds the low temperature
          region within which the replica-symmetric theory is unstable. The inset shows the critical value of $K_0/K_1$ at the AT transition
          for $T/K_1=10^{-3}$ as a function of $\alpha$. The order-parameters $M$ and $Q$ [see Eqs.~(\ref{tuya1}) and (\ref{tuya2})]  change continuously across each
          one of the transitions in the phase diagram.        
        }
	\label{pdr}
\end{figure}

Equations (\ref{turaca1}) and (\ref{turaca2}) yield the continuous phase transitions for an arbitrary degree distribution $\nu(g)$. The AT line as well as the
continuous transition between the ferromagnetic and the spin-glass phase are calculated by numerically solving the order-parameter
Eqs.~(\ref{jotur1}) and (\ref{jotur2}). As before, we
present results for networks with a negative binomial degree
distribution, in which $\nu(g)$ is given by Eq.~(\ref{roar}). The function $\omega_n(h)$ ($n=0,1,2$) is the
key quantity that determines the order-parameters and the AT line. Inserting Eq.~(\ref{roar}) in Eq.~(\ref{omigaw}) and integrating
over $g$ \cite{Gradstein}, we get the analytic expression
\begin{align}
  \omega_n(h) & = \frac{1}{\mathcal{N}_n\left( M,Q \right)  } \, |h|^{n + \alpha - \frac{1}{2}} \, \exp{\left( \frac{K_0 M h  }{K_{1}^{2} Q } \right)} \nonumber \\
  &\times  K_{n + \alpha - \frac{1}{2}} \left( \frac{|h| \sqrt{K_{0}^2 M^2 + 2 \alpha K_{1}^{2} Q } }{K_{1}^{2} Q}   \right),
  \label{jutwe}
\end{align}  
where
\begin{align}
  \mathcal{N}_n\left( M,Q \right) &= \frac{\Gamma(\alpha) \sqrt{2 \pi K_{1}^{2} Q  }   }{2 \alpha^{\alpha}} \nonumber \\
        &\times     \left( \sqrt{ K_{0}^2 M^2 + 2 \alpha K_{1}^{2} Q } \right)^{ n + \alpha - \frac{1}{2}     } ,
\end{align}  
and $K_a(x)$ is the modified Bessel function of the second kind with order $a$ \cite{Gradstein}.
In the limit $\alpha \rightarrow \infty$, $\omega_n(h)$ converges to a Gaussian distribution with mean $K_0 M$ and
variance $K_1^2 Q$, independently of the index $n=0,1,2$.

Figure \ref{pdr} depicts the phase diagram of the Ising spin-glass model on networks with a negative binomial
degree distribution for different values of the heterogeneity parameter $\alpha$.
The phase diagram of the Sherrington-Kirkpatrick model
is recovered for  $\alpha \rightarrow \infty$ \cite{Sherrington1975,Sherrington1978}.
The different critical lines in figure \ref{pdr} meet at the point
\begin{equation}
  \left( \frac{K_0}{K_1},\frac{T}{K_1} \right) = \left( \sqrt{\frac{\alpha  }{1 + \alpha  }},\sqrt{\frac{1 + \alpha  }{\alpha  }} \right),
  \label{point}
\end{equation}  
which serves as a useful guide to understand the effect of $\alpha$ on the phase diagram.
From Eq.~(\ref{point}) we note that, in the limit $\alpha \rightarrow 0$, the ferromagnetic phase
essentially expands over the entire phase diagram, while the spin-glass phase is confined to an
arbitrary small region around $K_0/K_1 \simeq 0$. The critical temperature between the
paramagnetic and the spin-glass phase diverges as $1/\sqrt{\alpha}$ for $\alpha \rightarrow 0$.

According
to figure \ref{pdr}, the location of the AT line shows that the RS theory
is unstable throughout the spin-glass phase and in the low-temperature sector of the ferromagnetic phase, for any value of $\alpha$.
Although decreasing values of $\alpha$ seem to gradually stabilize the replica-symmetric ferromagnetic
phase for lower temperatures, the impact of the network heterogeneity on the AT line is fundamentally different in the regime $T \rightarrow 0$.
In fact, the inset in figure \ref{pdr} demonstrates that the
critical value of $K_0$ marking the AT transition for $T \rightarrow 0$ is a non-monotonic
function of $\alpha$, with a maximum around $\alpha \simeq 0.5$. Moreover, the inset suggests that, for strong
heterogeneous networks with $\alpha \rightarrow 0$, the RS theory may
become stable in the low-temperature sector of the ferromagnetic
phase.

We end this section by studying how the network heterogeneity impacts the distribution $\mathcal{P}_{\rm eff}(h)$
of effective fields. The full analytic expression for $\mathcal{P}_{\rm eff}(h)$ is obtained directly
from the distribution $\omega_0(h)$ (see Eq.~(\ref{ufopoas})), defined by means of  Eq.~(\ref{efffie}).
We point out that $\mathcal{P}_{\rm eff}(h)$ is always given by $\mathcal{P}_{\rm eff}(h) = \delta(h)$ in the paramagnetic phase.

Let us first consider $\mathcal{P}_{\rm eff}(h)$ for networks with ferromagnetic couplings ($K_0 >0 $ and $K_1=0$). The distribution
of effective fields for homogeneous networks has the typical form $\mathcal{P}_{\rm eff}(h) = \delta(h - K_0 m)$ of the
Curie-Weiss model, while we obtain
\begin{equation}
\mathcal{P}_{\rm eff}(h) = \frac{1}{K_0 M} \, \nu \hspace{-0.1cm} \left(  \frac{h}{K_0 M}   \right) \Theta\left( h \right)
\end{equation}  
for heterogeneous networks with arbitrary degree distribution. The above equation holds for 
the choice $M >0$, and the symbol $\Theta(h)$ represents the Heaviside step function. For a negative binomial degree
distribution, the above expression becomes
\begin{equation}
\mathcal{P}_{\rm eff}(h) = \frac{\alpha^{\alpha}}{\Gamma(\alpha) \left(K_0 M   \right)^{\alpha} } h^{\alpha-1} \exp{\left( - \frac{\alpha h }{K_0 M } \right) }\Theta\left( h \right).
\end{equation}
We see that strong degree fluctuations lead to striking modifications in $\mathcal{P}_{\rm eff}(h)$ when compared
to homogeneous networks. The first interesting aspect concerns the behavior of $\mathcal{P}_{\rm eff}(h)$ for large fields.
For $\alpha < 1$, $\mathcal{P}_{\rm eff}(h)$ exhibits a power-law regime $\mathcal{P}_{\rm eff}(h) \sim h^{\alpha-1}$ ($h \gg 1$) up to the point $h = O(K_0 M/ \alpha)$, above
which it decays exponentially fast. Therefore, the power-law decay of $\mathcal{P}_{\rm eff}(h)$ persists over an arbitrarily large range of $h$ as $\alpha \rightarrow 0$.
The second interesting effect of the network heterogeneity occurs in the behavior of $\mathcal{P}_{\rm eff}(h)$ around $h=0$. In the
limit $h \rightarrow 0$,  $\mathcal{P}_{\rm eff}(h)$ converges to a constant if $\alpha \geq 1$, whereas it
diverges as $\mathcal{P}_{\rm eff}(h) \sim h^{\alpha-1}$ if $\alpha <1$. Thus, strong degree fluctuations induce the appearance
of a substantial fraction of vanishing effective fields, which has a detrimental effect on the ordered phase.

In the sequel, we discuss how the randomness in the coupling strengths impacts $\mathcal{P}_{\rm eff}(h)$.
For homogeneous networks with $K_1 >0$, the central limit theorem can be applied to Eq.~(\ref{retua}) and $\mathcal{P}_{\rm eff}(h)$ is a Gaussian
distribution with mean $K_0 m$ and variance $K_{1}^2 q_{EA}$ \cite{Sherrington1975}. The analytic expression of $\mathcal{P}_{\rm eff}(h)$ for networks with negative
binomial degrees and $K_1 >0$ is derived by setting $n=0$ in Eq.~(\ref{jutwe}). Similarly to the
above results for $K_1=0$, $\mathcal{P}_{\rm eff}(h)$ features a crossover between a
power-law and an exponential decay for sufficiently large $|h|$. The main difference with respect to $K_1=0$ appears
around $h=0$. In the regime $|h| \rightarrow 0$, $\mathcal{P}_{\rm eff}(h)$ converges to a constant if $\alpha > 1/2$, whereas
it displays the following asymptotic forms if $\alpha \leq 1/2$
\begin{equation}
\mathcal{P}_{\rm eff}(h) =
\begin{cases} 
      -\ln{|h|} & \text{   for $\alpha = 1/2$}, \\
      |h|^{1-2\alpha} & \text{   for $0 < \alpha < 1/2$}.
\end{cases}
\label{porart}
\end{equation}
Thus, the interplay between random coupling strengths and
heterogeneous degrees leads to a logarithmic divergence in $\mathcal{P}_{\rm eff}(h)$, a feature that appears as well in the spectral density of heterogeneous networks \cite{Metz2020}.
We further note that fluctuations in the coupling strengths mitigate the divergence around $h=0$, in the sense
that the power-law exponent in Eq.~(\ref{porart}) is smaller when compared to the case $K_1=0$. The comparison between numerical simulations
for the effective fields and the analytic expressions for $\mathcal{P}_{\rm eff}(h)$ in the cases $\alpha \rightarrow \infty$ and $\alpha =1$ are shown
in figure \ref{localfields}. The agreement between the theory and simulations is excellent, confirming the exactness
of our theoretical findings for sufficiently high temperatures.

\begin{figure}[t!]
	\centering
	\includegraphics[width=1.0\columnwidth]{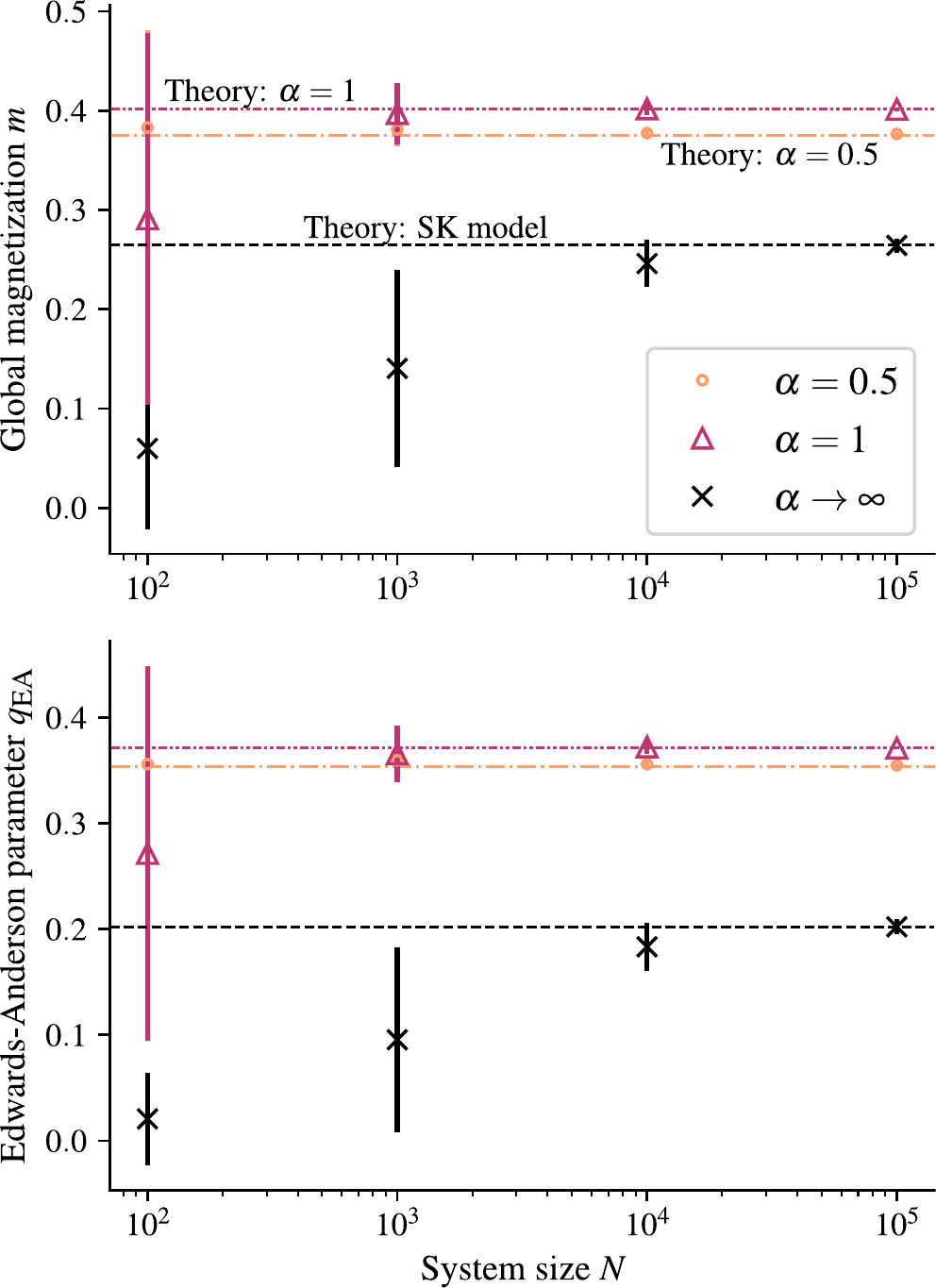}
	\caption{Numerical simulation results for the global magnetization $m$ and for the Edwards-Anderson order-parameter $q_{EA}$ of the Ising
          model on heterogeneous networks with random couplings and average degree $c=\sqrt{N}$, where $N$ is the total number of spins. The
          degrees follow a negative binomial distribution with heterogeneity parameter
          $\alpha$ [see Eq.~(\ref{utaop})], and the coupling strengths are drawn from a Gaussian distribution with mean $1.2/c$ and
          standard deviation $1/\sqrt{c}$. The symbols denote the mean values of $m$ and $q_{EA}$ over $30$ independent samples, while the vertical bars
          are the standard deviations around the mean. The horizontal lines are the theoretical results for $N \rightarrow \infty$ obtained
          from the heterogeneous mean-field theory and from the fully-connected Sherrington-Kirkpatrick model \cite{Sherrington1975}.
        }
	\label{gugartu}
\end{figure}

The heterogeneous mean-field theory is derived by taking the high-connectivity limit $c \rightarrow \infty$
{\it after} the limit $N \rightarrow \infty$. As such, one expects that our theory is valid
for high-connectivity networks in which the density  of edges $c/N$ goes to zero for $N \rightarrow \infty$.
In fact, figure \ref{gugartu} presents numerical simulation results for the global magnetization and the Edwards-Anderson order-parameter
of spin models on heterogeneous networks with average degree $c=\sqrt{N}$ and with the negative binomial degree distribution
of Eq.~(\ref{utaop}). For homogeneous networks ($\alpha \rightarrow \infty$), such as regular and Erd\H{o}s-R\'enyi random graphs, the
order-parameters flow as $N \rightarrow \infty$ to their fully-connected values, obtained from the solution of the Sherrington-Kirkpatrick model. For
heterogeneous networks, in which $\alpha$ is finite, the order-parameters
converge, in the limit $N \rightarrow \infty$, to the results of the heterogeneous mean-field theory derived in the present paper.
Figure \ref{gugartu} confirms that the mean-field theory presented
in this work describes spin models on heterogeneous networks where $c$ scales as  $c \propto N^b$, with $0 < b < 1$. This regime of connectivity lies
between sparse networks ($b=0$) and diluted networks ($b=1$) \cite{Sompolinsky1986,Canning1992,Metz2006,Boettcher2020}.

\section{Discussion and conclusions} \label{concl}

Mean-field theories are formidable tools to study the macroscopic properties of spin models
on networks. The most well-studied family of mean-field theories are realized on fully-connected architectures, in which
a given spin interacts with all others. As a network becomes more densely connected, it is natural to expect
that local fluctuations in the network structure are gradually washed out, and the macroscopic properties of the
underlying spin model converge to those of a fully-connected system. In this work we have shown that this
is generally not the case.  We have derived a novel class of exact mean-field equations that explicitly depend on the
degree distribution and that apply to the high-connectivity limit of heterogeneous networks.
Fully-connected mean-field theories, in contrast, are limited
to homogeneous networks, for which the degree distribution is peaked at its mean value.

Put differently, we have proven that the high-connectivity limit of spin models on networks is
nonuniversal, as it depends on the full degree distribution. On the other hand, the universality with respect
to the randomness in the coupling strengths is robust to degree fluctuations.
In fact, the heterogeneous mean-field equations only depend on the
first two moments of the distribution of couplings.
The nonuniversal behavior of the heterogeneous mean-field theory is accompanied by the
failure of the central limit theorem for the effective fields, caused by large fluctuations in the number
of summands in Eq.~(\ref{retua}). Apart from a few exceptions \cite{Jakeman1978,Metz2020,Godreche2021}, the consequences of this interesting mechanism for the breakdown of the
central limit theorem to statistical physics remain largely unexplored. We have illustrated its crucial role
for the equilibrium of spin models, but one can envisage the far-reaching importance of this mechanism
for a variety of problems on networks, such as the nonequilibrium dynamics of
spin models \cite{Mimura2009,Neri2009}, the storage capacity of neural networks \cite{Castillo2004}, and the stability of large dynamical systems \cite{Neri2020}.

We have presented several results that highlight the importance of degree fluctuations
to the high-connectivity limit of spin models,
with particular emphasis on the mean-field theory of synchronization and of
Ising spin-glasses.
Although the heterogeneous mean-field equations are valid for any degree distribution,
most of the  explicit results have been derived for a negative binomial degree distribution, which allows to smoothly interpolate between homogeneous and
heterogeneous networks by changing
the variance of the degree distribution.

Degree fluctuations have a conspicuous effect on the ferromagnetic phase. As the variance
of the degree distribution increases, the ferromagnetic phase gradually covers the whole phase diagram of the Kuramoto model, suggesting that the oscillators
synchronize at any coupling strength, as argued in several previous
works~\cite{moreno2004synchronization,ichinomiya2004frequency,restrepo2005onset,arenas2008synchronization,Rodrigues2016,peron2019onset}. However, we
have shown that not only the critical coupling becomes arbitrarily small as the network heterogeneity diverges, but also the magnetization in the synchronous phase.
Therefore, our results of Sec.~\ref{sub1} show that instead of facilitating synchronization, degree fluctuations actually hamper the formation of an
ordered synchronous component, and ultimately inhibit the emergence of any coherent behavior when the variance of the degrees is infinitely large.

Moreover, the magnetization of the Kuramoto model displays a cusp that separates the ferromagnetic
phase in two qualitatively distinct regions, each one characterized by a different impact
of degree fluctuations on the magnetization. Such non-analytic behavior
is not exclusive of the magnetization, but it seems to be a generic feature of the macroscopic behavior.
Indeed, the effective field distribution of Ising spin-glass models on heterogeneous networks exhibits
either a power-law or a logarithmic divergence at zero field,
which contrasts
with the Gaussian effective fields of fully-connected models. This divergence reflects the substantial fraction of spins with
a vanishing local magnetization inside the ferromagnetic phase.

As another genuine feature of strong degree fluctuations in the ferromagnetic phase, we have found that the shape of the single-site phase distribution of the
Kuramoto model fluctuates from site to site, in sharp contrast to fully-connected models, for which
the functional form of the local phase distribution is fixed. These results
follow from the mean and the variance of the single-site phase distribution.
In fact, the path-integral formalism developed in this paper gives access to all moments of the
marginal distribution of single-site configurations.
This is particularly
relevant in the case of vector spins, for which the  functional distribution of local marginals is not parametrized in terms
of a finite number of fields.
Thus, when compared to other mean-field techniques for coupled oscillators~\cite{Rodrigues2016}, the path-integral
formalism has the technical advantage of allowing the calculation of dynamical properties of vector spin models at the level of individual nodes.
It would be interesting to investigate the individual phase fluctuations for three-dimensional spins \cite{Chandra2019}, a problem that can
give insights into the synchronous dynamics of swarms and flocks in three dimensions.

We have studied the interplay between random coupling strengths and degree fluctuations in Ising
spin models. Figure \ref{pdr} generalizes the replica-symmetric phase diagram of the fully-connected Sherrington-Kirkpatrick model \cite{Sherrington1978} to
heterogeneous networks. In this case, the heterogeneous mean-field theory
is not exact on the entire phase diagram, but it becomes unstable at temperatures below the Almeida-Thouless
line \cite{Almeida1978}.
In spite of that, the results of figure \ref{pdr} 
allow to conclude that degree fluctuations promote
the ferromagnetic phase in detriment of the spin glass phase.
In addition,  the low-temperature sector of the Almeida-Thouless line exhibits a non-monotonic
behavior as the degree fluctuations increase. In view of this distinctive behavior and
given that the replica symmetry breaking theory on sparse networks is a notorious difficult problem \cite{Mezard2001,Mezard2003,Concetti2018},
it would be very interesting to employ the replica symmetry breaking machinery \cite{Parisi1979,Parisi1983}
and derive the exact version of the heterogeneous mean-field theory.
This would give an alternative, simpler route to exploit how the complex picture describing the spin-glass phase \cite{Mezard1987} is modified by the presence
of network heterogeneities.

The heterogeneous mean-field theory has been derived by taking the high-connectivity limit $c \rightarrow \infty$ after the
thermodynamic limit $N \rightarrow \infty$, where $c$ is the average degree and $N$ is the total
number of spins. An implicit assumption in this procedure is that the local tree-like structure of the network
is preserved and hence the distributional cavity equations remain valid. This can be only achieved
if the density of links $c/N$ approaches zero for $c \rightarrow \infty$, which suggests that the heterogeneous
mean-field theory should apply when $c \propto N^{b}$ ($0 < b < 1$).
We have confirmed this conjecture by means of numerical simulations.
The connectivity regime $c \propto N^{b}$ ($0 < b < 1$)
lies between sparse ($c=\mathcal{O}(1)$) and diluted ($c=\mathcal{O}(N)$) networks \cite{Sompolinsky1986,Canning1992,Metz2006,Boettcher2020}. Even though this intermediate
connectivity range, called the extreme diluted regime, has been known for a long time in the field of neural networks \cite{Derrida1987,Watkin1991}, it has been studied only in
the case of homogeneous networks, for which degree fluctuations are unimportant.
Here we have examined the extreme diluted regime of heterogeneous networks and unveiled
its nonuniversal features.
In the diluted regime, degree fluctuations are of $\mathcal{O}(N^0)$ and the macroscopic
behavior of spin models is captured  by the fully-connected mean-field equations.
The universality properties of spin models
in the three different regimes of connectivity are summarized in table \ref{tabfitIPRmin}.

\begin{table}[ht]
  \setlength{\tabcolsep}{7pt}
\centering
\begin{tabular}{cccc}
 \toprule[0.8pt] 
 & \footnotesize{Sparse}  & \footnotesize{Extreme diluted} & \footnotesize{Diluted} \\  
\midrule[0.8pt] 
\footnotesize{$p_k$} &  \footnotesize{Nonuniversal} & \footnotesize{Nonuniversal} & \footnotesize{Universal} \\  [0.2ex]
\footnotesize{$p_J$}  &  \footnotesize{Nonuniversal}  & \footnotesize{Universal} & \footnotesize{Universal} \\  [0.2ex]
\footnotesize{Effective fields}  &  \footnotesize{Non-Gaussian}   & \footnotesize{Non-Gaussian} & \footnotesize{Gaussian} \\  [0.2ex]
\bottomrule[0.8pt]
\end{tabular}
\caption{Universal properties of spin models in the three different regimes of connectivity (see the main text) with respect to the
  degree distribution $p_k$ and to the distribution $p_J$ of coupling strengths. The third row classifies
  the distribution of effective fields, Eq.~(\ref{retua}), in each regime.
 }
\label{tabfitIPRmin}
\end{table}

On the whole, our findings demonstrate that network heterogeneities, here expressed by degree fluctuations, play a surprisingly important role in the
high-connectivity limit of spin models. Other types of topological features, such as modular structure and the presence of loops \cite{Newman2009, Metz2011,Kirkley2021}, should as well
play a fundamental role in the
high-connectivity behavior. While spin models on sparse networks pose many technical challenges \cite{Mezard2001,Mezard2003,Concetti2018}, the mean-field theory of fully-connected
models has a simpler formal structure \cite{Mezard1987}, at the cost of completely neglecting the network structure.
Our work puts forward a novel family of mean-field theories that explicitly takes
into consideration the network structure and, at the same time, are simple enough that they can be
thoroughly analyzed. This framework opens the door to the development of similar mean-field theories
for other complex systems.

\acknowledgments

F. L. M. gratefully acknowledges London Mathematical Laboratory and  CNPq/Brazil for financial support. T. P. acknowledges FAPESP (Grant
No. 2016/23827-6). This research was carried out using the computational
resources of the Center for Mathematical Sciences Applied to Industry (CeMEAI) funded by FAPESP (Grant No. 2013/07375-0).


\appendix

\section{Stability of the replica-symmetric theory} \label{ATline00}

In this appendix we explain in more detail how to derive Eq.~(\ref{iuts}), which bounds the region in the phase diagram
where the replica-symmetric (RS) mean-field equations are unstable. Although we limit ourselves to Ising spin
models ($D=1$), the approach discussed here can be generalized to spins with arbitrary dimension $D$.

The key quantity, defined by Eq.~(\ref{joiunuta}), is the joint distribution $R_{12} ( \vec{p}_1,  \vec{p}_2)$ of two solutions $\{ p_{j,1}^{(i)} (\sigma_j ) \}$ and
$\{ p_{j,2}^{(i)} (\sigma_j ) \}$  of Eqs.~(\ref{cav}). Since both solutions refer to the  {\it same} realization of the random network, each solution describes
the local marginals on a copy of the original system. The RS theory is stable if Eqs.~(\ref{cav}) admit a single solution, which implies that $R_{12} ( \vec{p}_1,  \vec{p}_2)$ must
be given by Eq.~(\ref{jurema}).

The first task is to compute $R_{12} ( \vec{p}_1,  \vec{p}_2)$ in the high-connectivity limit. The steps to perform this calculation are completely
analogous to those discussed in section \ref{secHet}, with the main difference that the two copies of the system become correlated
after taking the ensemble average. The resulting expression for $R_{12} ( \vec{p}_1,  \vec{p}_2)$ reads
\begin{align}
  &R_{12} ( \vec{p}_1,  \vec{p}_2) = \int_{0}^{\infty} d g \, g  \,\nu(g) \int_{-\infty}^{\infty} D \vec{u} \nonumber \\
  & \times \! \prod_{\sigma \in \{ -1,1 \}  } \! \! \delta{  \Bigg{\{} p_1(\sigma) - \frac{e^{\beta \sigma \left( g K_0 M_1 + K_1 \sqrt{g} u_1    \right)   }   }
    {2 \cosh{\left[ \beta \left( g K_0 M_1 + K_1 \sqrt{g} u_1    \right) \right]}      }   \Bigg{\}}   }   \nonumber \\
  &\times
  \delta{  \Bigg{\{} p_2(\sigma) - \frac{e^{\beta \sigma \left( g K_0 M_2 + K_1 \sqrt{g} u_2    \right)   }   }
    {2 \cosh{\left[ \beta \left( g K_0 M_2 + K_1 \sqrt{g} u_2    \right) \right]}      }   \Bigg{\}}   },   \label{butas}
\end{align}  
where $D \vec{u}$ is the Gaussian bivariate measure of $\vec{u} = (u_1 \,\,\, u_2)^T$
\begin{equation}
D \vec{u} = \frac{d u_1 d u_2}{2 \pi \sqrt{\det{\mathcal{A}} }} \exp{\left( -\frac{1}{2} \vec{u}^{\, T} \mathcal{A}^{-1} \vec{u} \right)   },
\end{equation}  
with $\mathcal{A}$ denoting  the $2 \times 2$ matrix
\[
\mathcal{A} =
\begin{pmatrix}
    Q_1 & \rho  \\
    \rho & Q_2
\end{pmatrix}.
\]
The order-parameters $M_1$, $M_2$, $Q_1$, $Q_2$, and $\rho$ are determined from $R_{12} ( \vec{p}_1,  \vec{p}_2)$ as follows
\begin{align}
  M_1 = \int d \vec{p}_1 d \vec{p}_2 R_{12} ( \vec{p}_1,  \vec{p}_2) \langle \sigma \rangle_{\vec{p}_1}, \label{ko1} \\
  M_2 = \int d \vec{p}_1 d \vec{p}_2 R_{12} ( \vec{p}_1,  \vec{p}_2) \langle \sigma \rangle_{\vec{p}_2}, \label{ko2} \\
  Q_1 = \int d \vec{p}_1 d \vec{p}_2 R_{12} ( \vec{p}_1,  \vec{p}_2) \langle \sigma \rangle_{\vec{p}_1}^{2}, \label{ko3}  \\
   Q_2 = \int d \vec{p}_1 d \vec{p}_2 R_{12} ( \vec{p}_1,  \vec{p}_2) \langle \sigma \rangle_{\vec{p}_2}^{2}, \label{ko4} \\
  \rho = \int d \vec{p}_1 d \vec{p}_2 R_{12} ( \vec{p}_1,  \vec{p}_2) \langle \sigma \rangle_{\vec{p}_1} \langle \sigma \rangle_{\vec{p}_2}, \label{ko5}
\end{align}
with the local average $\langle \sigma \rangle_{\vec{p}}$ defined by Eq.~(\ref{huawe1}). The order-parameters $M$ and $Q$ are duplicated
for the simple reason that we are dealing with two replicas of the original system. The quantity $\rho$ measures the correlation between the local
magnetizations of each replica. If the RS solution is stable, then $R_{12} ( \vec{p}_1,  \vec{p}_2)$ simplifies to the diagonal form of Eq.~(\ref{jurema}) and
the order-parameters fulfill $M_1 = M_2 =M$, $Q_1 = Q_2 =Q$, and $\rho=Q$, where $M$ and $Q$ follow from the solutions
of Eqs.~(\ref{jotur1}) and (\ref{jotur2}). 

The correlation order-parameter $\rho$ fulfills the self-consistent equation
\begin{align}
  \rho &=  \int_{0}^{\infty} d g \, g \, \nu(g) \int_{-\infty}^{\infty} D \vec{u}  \, \tanh{\left[   \beta \left( g K_0 M_1 + K_1 \sqrt{g} u_1    \right)     \right]}\nonumber \\
  &\times \tanh{\left[   \beta \left( g K_0 M_2 + K_1 \sqrt{g} u_2    \right) \right] }, \label{fdut}
\end{align}  
derived by substituting Eq.~(\ref{butas}) in Eq.~(\ref{ko5}). In order to analyze the stability of Eq.~(\ref{fdut}) around
the RS solution, it is convenient to make an orthogonal change of integration variables.
Let $\mathbb{T}$ be the $2 \times 2$ orthogonal
matrix that diagonalizes $\mathcal{A}$
\[
\mathbb{T}^{-1} \, \mathcal{A} \, \mathbb{T} =
\begin{pmatrix}
    \lambda_{+} & 0  \\
    0 & \lambda_{-}
\end{pmatrix}.
\]
The matrix $\mathbb{T}$  and the eigenvalues of $\mathcal{A}$ have, respectively, the explicit forms
\[
\mathbb{T} =
\begin{pmatrix}
  t_{11} & t_{12} \vspace{0.1cm}  \\
   t_{21}   & t_{22}
\end{pmatrix}
=
\begin{pmatrix}
  \frac{2 \rho}{ a_{+} } & \frac{2 \rho}{ a_{-} } \vspace{0.1cm}  \\
    \frac{Q_2 - Q_1 + \Delta}{a_{+}}   & \frac{Q_2 - Q_1 - \Delta}{a_{-}}
\end{pmatrix},
\]
and
\begin{align}
\lambda_{\pm} = \frac{1}{2} \left( Q_1 + Q_2 \pm \Delta  \right),
\end{align}  
where we have defined
\begin{align}
a_{\pm} &= \sqrt{4 \rho^2 + \left( Q_2 - Q_1 \pm \Delta  \right)^2  }, \nonumber \\
\Delta &= \sqrt{4 \rho^2 + \left(Q_1 - Q_2   \right)^2  }.
\end{align}  
The dependency of some quantities with respect to the order-parameters has been omitted in order to simplify the notation. By making
the orthogonal change of integration variables $\vec{u} = \mathbb{T} \vec{z}$, with $\vec{z} = (z_1 \,\,\, z_2)^T$, we can rewrite Eq.~(\ref{fdut}) as
\begin{align}
  &\rho =  \int_{0}^{\infty} d g \, g \, \nu(g) \int_{-\infty}^{\infty} d z_1 \, d z_2 \, P_{\rm g}(z_1) \, P_{\rm g}(z_2) \nonumber \\
  &\! \times \! \tanh \! { \Bigg{\{} \! \beta \left[  g K_0 M_1 + K_1 \sqrt{g} \left( t_{11} \sqrt{\lambda_{+}} z_1  +  t_{12} \sqrt{\lambda_{-}} z_2 \right)  \right] \! \! \Bigg{\}}  } \nonumber \\
  & \! \times \! \tanh \! { \Bigg{\{} \! \beta \left[  g K_0 M_2 + K_1 \sqrt{g}  \left( t_{21} \sqrt{\lambda_{+}} z_1  +  t_{22} \sqrt{\lambda_{-}} z_2 \right)    \right] \! \! \Bigg{\}} }, \label{duraq}
\end{align}  
after rescaling the components of $\vec{z}$ as $z_1 \rightarrow \sqrt{\lambda_{+}} z_1$ and $z_2 \rightarrow \sqrt{\lambda_{-}} z_2$. The
normal probability density  $P_{\rm g} (z)$ is defined by Eq.~(\ref{uffu}).

Equation (\ref{duraq}) is the proper starting point to analyze the stability of the RS theory. The fixed-point solutions describing
the macroscopic states of the system composed of two replicas are given in terms of five order-parameters: $M_1$, $M_2$, $Q_1$, $Q_2$, and $\rho$.
In order to probe the stability of the RS solution, it suffices to consider fluctuations solely in the direction of $\rho$.
Thus, by setting $M_1=M_2=M$, $Q_1=Q_2=Q$, and $\rho = Q + \delta \rho$ in Eq.~(\ref{duraq}), and then expanding its right-hand side
up to $O(\delta \rho)$, we obtain the following equation for the perturbation $\delta \rho$
\begin{align}
  \delta \rho &= \delta \rho \, \Bigg{\{} \beta^2 K_{1}^2 \int_{0}^{\infty} d g \, \nu (g) \, g^2 \, \int_{-\infty}^{\infty} dz \, P_{\rm g}(z) \nonumber \\
  &\times {\rm sech}^{4} \left[\beta \left(   g K_0 M + K_1 \sqrt{g Q} z  \right)   \right]  \Bigg{\}}.
  \label{poita}
\end{align}  
If the coefficient of $\delta \rho$ in the above equation is larger than one, then the RS fixed-point solution is unstable, since
the iteration of the above linear equation leads
to the growth of the perturbation $\delta \rho$. After a simple change of integration variables, this condition becomes identical to Eq.~(\ref{iuts}).


\section{Simulation details}
\label{sec:simulations}

In this appendix we describe the details behind the numerical simulations of Ising spin models ($D=1$)
and of the Kuramoto model ($D=2$). The networks employed in the simulations are constructed  following the standard
configuration model: from a negative binomial degree distribution $p_k^{(b)}$ with a given $\alpha$ [see Eq.~\eqref{utaop}], we draw
the degrees $k_1,...,k_N$ independently, and we assign $k_i$ stubs of edges to each node $i$. We then choose two stubs
uniformly at random and join them to form an edge. This process is repeated until all remaining stub pairs are joined. Multiple-edges and self-loops are erased
after the network is formed. The impact of the latter step on the mean-field calculations is negligible, since the density of such edges, as well as minor discrepancies from the 
original sequence $k_1,...,k_N$, tends to zero as $N \rightarrow \infty$~\cite{NewmanBook}.

The pairwise coupling strengths $\{ J_{ij} \}_{i,j=1,\dots,N}$ between spins fulfill the symmetry condition $J_{ij} = J_{ji}$ $\forall i,j$ .
For ferromagnetic spin models, we
set the coupling strengths to the constant value $J_{ij} = K_0/c$. For Ising spin-glass models, the coupling strengths
are independently and identically distributed random variables drawn from a Gaussian distribution with mean $K_0/c$ and standard deviation $K_1/\sqrt{c}$.

\subsection{Ising spin models}

The global configuration of Ising spin models at time $t$ is defined by $\sigma_1(t),\dots,\sigma_N(t)$, where
$\sigma_i(t) \in \{ -1,1 \}$. After generating a network from the configuration
model, the spins are initialized in the ordered state $\sigma_i(t=0)=1$ $\forall i$.
The subsequent states are updated with probability \cite{coolen2001statistical}
\begin{equation}
\textrm{Prob}[\sigma_i(t+1)] = \frac{1}{2} \Big{\{}  1 + \sigma_i(t+1) \tanh \left[ \beta h_i(t) \right]   \Big{\}} ,
\label{eq:Prob_Ising}
\end{equation}
where $\beta=1/T$ is the inverse temperature, and
\begin{equation}
h_i(t) = \sum_{j \in \partial_i} J_{ij} \sigma_j(t)
\end{equation}
is the local field at node $i$ at time $t$. The symbol $\partial_i$ represents the set of nodes adjacent to node $i$. 
The global spin configuration  $\sigma_1(t),\dots,\sigma_N(t)$  evolves in time according to the probability given by Eq.~\eqref{eq:Prob_Ising}. After
the stationary state is reached, the observables are averaged over a sufficiently long time span. 

\subsection{Kuramoto model}

For $D=2$, the spins are two-dimensional vectors with unit norm that rotate on a plane. As such, we can simulate the spin system as a set of
stochastic Kuramoto oscillators~\cite{strogatz2000kuramoto,fonseca2018kuramoto}, in which the state of a node $i$ at time
$t$ is fully specified by the phase $\phi_i(t) \in ( - \pi, \pi]$. The stochastic dynamics of the model is dictated by the following set of equations~\cite{Rodrigues2016}
\begin{equation}
\frac{ d \phi_i}{d t} = \xi_i(t) + \sum_{j \in \partial_i}^N J_{ij} \sin[\phi_j(t) - \phi_i(t)]\; (i=1,...,N), 
\label{eq:KuramotoModel}
\end{equation}
where $\xi_i(t)$ is a Gaussian white noise that satisfies
\begin{equation}
\begin{aligned}
\left\langle \xi_i(t) \right\rangle &= 0,\\
\left\langle \xi_i(t) \xi_j(t') \right\rangle &= 2T\delta_{ij}\delta(t-t'),
\end{aligned}
\end{equation} 
with $T$ being the temperature. For all numerical results shown in this paper, we numerically integrate the
stochastic equations~\eqref{eq:KuramotoModel} with the Heun's method with time step $dt = 0.01$. The complex magnetization 
$m_{c}(t)$ is calculated as the average of the phasors that rotate 
in the complex unit circle, that is
\begin{equation}
m_c(t) = R(t) e^{i\psi(t)} = \frac{1}{N} \sum_{j=1}^N e^{i\phi_j(t)} , 
\label{eq:order_parameter_KM}
\end{equation}
where $0 \leq R(t) \leq 1 $  measures the level of phase-coherence in the system, and $\psi(t)$ is the average phase. In this context, the absolute value $|\boldsymbol{m}(t)|$
of the vector magnetization $\boldsymbol{m}(t)$ in section \ref{sub1} is obtained as $|\boldsymbol{m}(t)| = R(t)$.
In particular, the long-time behavior of
the magnetization $|\boldsymbol{m}(t)|$ in Figure~\ref{orderpar}(b) is obtained by averaging Eq.~\eqref{eq:order_parameter_KM} over the interval $t \in [1000, 2000]$. 

\bibliography{biblio}

\begin{thebibliography}{96}%
\makeatletter
\providecommand \@ifxundefined [1]{%
 \@ifx{#1\undefined}
}%
\providecommand \@ifnum [1]{%
 \ifnum #1\expandafter \@firstoftwo
 \else \expandafter \@secondoftwo
 \fi
}%
\providecommand \@ifx [1]{%
 \ifx #1\expandafter \@firstoftwo
 \else \expandafter \@secondoftwo
 \fi
}%
\providecommand \natexlab [1]{#1}%
\providecommand \enquote  [1]{``#1''}%
\providecommand \bibnamefont  [1]{#1}%
\providecommand \bibfnamefont [1]{#1}%
\providecommand \citenamefont [1]{#1}%
\providecommand \href@noop [0]{\@secondoftwo}%
\providecommand \href [0]{\begingroup \@sanitize@url \@href}%
\providecommand \@href[1]{\@@startlink{#1}\@@href}%
\providecommand \@@href[1]{\endgroup#1\@@endlink}%
\providecommand \@sanitize@url [0]{\catcode `\\12\catcode `\$12\catcode
  `\&12\catcode `\#12\catcode `\^12\catcode `\_12\catcode `\%12\relax}%
\providecommand \@@startlink[1]{}%
\providecommand \@@endlink[0]{}%
\providecommand \url  [0]{\begingroup\@sanitize@url \@url }%
\providecommand \@url [1]{\endgroup\@href {#1}{\urlprefix }}%
\providecommand \urlprefix  [0]{URL }%
\providecommand \Eprint [0]{\href }%
\providecommand \doibase [0]{http://dx.doi.org/}%
\providecommand \selectlanguage [0]{\@gobble}%
\providecommand \bibinfo  [0]{\@secondoftwo}%
\providecommand \bibfield  [0]{\@secondoftwo}%
\providecommand \translation [1]{[#1]}%
\providecommand \BibitemOpen [0]{}%
\providecommand \bibitemStop [0]{}%
\providecommand \bibitemNoStop [0]{.\EOS\space}%
\providecommand \EOS [0]{\spacefactor3000\relax}%
\providecommand \BibitemShut  [1]{\csname bibitem#1\endcsname}%
\let\auto@bib@innerbib\@empty
\bibitem [{\citenamefont {Newman}(2010)}]{NewmanBook}%
  \BibitemOpen
  \bibfield  {author} {\bibinfo {author} {\bibfnamefont {M.}~\bibnamefont
  {Newman}},\ }\href {https://books.google.com.br/books?id=LrFaU4XCsUoC} {\emph
  {\bibinfo {title} {Networks: An Introduction}}}\ (\bibinfo  {publisher} {OUP
  Oxford},\ \bibinfo {year} {2010})\BibitemShut {NoStop}%
\bibitem [{\citenamefont {Barrat}\ \emph {et~al.}(2008)\citenamefont {Barrat},
  \citenamefont {Barth{\'e}lemy},\ and\ \citenamefont
  {Vespignani}}]{BarratBook}%
  \BibitemOpen
  \bibfield  {author} {\bibinfo {author} {\bibfnamefont {A.}~\bibnamefont
  {Barrat}}, \bibinfo {author} {\bibfnamefont {M.}~\bibnamefont
  {Barth{\'e}lemy}}, \ and\ \bibinfo {author} {\bibfnamefont {A.}~\bibnamefont
  {Vespignani}},\ }\href {https://books.google.com.br/books?id=TmgePn9uQD4C}
  {\emph {\bibinfo {title} {Dynamical Processes on Complex Networks}}}\
  (\bibinfo  {publisher} {Cambridge University Press},\ \bibinfo {year}
  {2008})\BibitemShut {NoStop}%
\bibitem [{\citenamefont {Dorogovtsev}\ \emph {et~al.}(2008)\citenamefont
  {Dorogovtsev}, \citenamefont {Goltsev},\ and\ \citenamefont
  {Mendes}}]{Doro2008}%
  \BibitemOpen
  \bibfield  {author} {\bibinfo {author} {\bibfnamefont {S.~N.}\ \bibnamefont
  {Dorogovtsev}}, \bibinfo {author} {\bibfnamefont {A.~V.}\ \bibnamefont
  {Goltsev}}, \ and\ \bibinfo {author} {\bibfnamefont {J.~F.~F.}\ \bibnamefont
  {Mendes}},\ }\bibfield  {title} {\enquote {\bibinfo {title} {Critical
  phenomena in complex networks},}\ }\href {\doibase
  10.1103/RevModPhys.80.1275} {\bibfield  {journal} {\bibinfo  {journal} {Rev.
  Mod. Phys.}\ }\textbf {\bibinfo {volume} {80}},\ \bibinfo {pages}
  {1275--1335} (\bibinfo {year} {2008})}\BibitemShut {NoStop}%
\bibitem [{\citenamefont {Baxter}(2007)}]{BaxterBook}%
  \BibitemOpen
  \bibfield  {author} {\bibinfo {author} {\bibfnamefont {R.J.}\ \bibnamefont
  {Baxter}},\ }\href {https://books.google.com.br/books?id=G3owDULfBuEC} {\emph
  {\bibinfo {title} {Exactly Solved Models in Statistical Mechanics}}},\ Dover
  books on physics\ (\bibinfo  {publisher} {Dover Publications},\ \bibinfo
  {year} {2007})\BibitemShut {NoStop}%
\bibitem [{\citenamefont {Neri}(2010)}]{Izaakthesis}%
  \BibitemOpen
  \bibfield  {author} {\bibinfo {author} {\bibfnamefont {Izaak}\ \bibnamefont
  {Neri}},\ }\emph {\bibinfo {title} {Statistical mechanics of spin models on
  graphs}},\ \href@noop {} {Ph.D. thesis},\ \bibinfo  {school} {Catholic
  University of Leuven} (\bibinfo {year} {2010})\BibitemShut {NoStop}%
\bibitem [{\citenamefont {Dommers}(2013)}]{Dommersthesis}%
  \BibitemOpen
  \bibfield  {author} {\bibinfo {author} {\bibfnamefont {Sander}\ \bibnamefont
  {Dommers}},\ }\emph {\bibinfo {title} {Spin models on random graphs}},\
  \href@noop {} {Ph.D. thesis},\ \bibinfo  {school} {Eindhoven University of
  Technology} (\bibinfo {year} {2013})\BibitemShut {NoStop}%
\bibitem [{\citenamefont {Mezard}\ \emph {et~al.}(1987)\citenamefont {Mezard},
  \citenamefont {Parisi},\ and\ \citenamefont {Virasoro}}]{Mezard1987}%
  \BibitemOpen
  \bibfield  {author} {\bibinfo {author} {\bibfnamefont {M.}~\bibnamefont
  {Mezard}}, \bibinfo {author} {\bibfnamefont {G.}~\bibnamefont {Parisi}}, \
  and\ \bibinfo {author} {\bibfnamefont {M.A.}\ \bibnamefont {Virasoro}},\
  }\href {https://books.google.com.br/books?id=DwY8DQAAQBAJ} {\emph {\bibinfo
  {title} {Spin Glass Theory And Beyond: An Introduction To The Replica Method
  And Its Applications}}},\ World Scientific Lecture Notes In Physics\
  (\bibinfo  {publisher} {World Scientific Publishing Company},\ \bibinfo
  {year} {1987})\BibitemShut {NoStop}%
\bibitem [{\citenamefont {M{\'e}zard}\ and\ \citenamefont
  {Montanari}(2009)}]{MezardBook}%
  \BibitemOpen
  \bibfield  {author} {\bibinfo {author} {\bibfnamefont {M.}~\bibnamefont
  {M{\'e}zard}}\ and\ \bibinfo {author} {\bibfnamefont {A.}~\bibnamefont
  {Montanari}},\ }\href {https://books.google.com.br/books?id=jhCM7i0a6UUC}
  {\emph {\bibinfo {title} {Information, Physics, and Computation}}},\ Oxford
  Graduate Texts\ (\bibinfo  {publisher} {OUP Oxford},\ \bibinfo {year}
  {2009})\BibitemShut {NoStop}%
\bibitem [{\citenamefont {Castellano}\ \emph {et~al.}(2009)\citenamefont
  {Castellano}, \citenamefont {Fortunato},\ and\ \citenamefont
  {Loreto}}]{Castellano2009}%
  \BibitemOpen
  \bibfield  {author} {\bibinfo {author} {\bibfnamefont {Claudio}\ \bibnamefont
  {Castellano}}, \bibinfo {author} {\bibfnamefont {Santo}\ \bibnamefont
  {Fortunato}}, \ and\ \bibinfo {author} {\bibfnamefont {Vittorio}\
  \bibnamefont {Loreto}},\ }\bibfield  {title} {\enquote {\bibinfo {title}
  {Statistical physics of social dynamics},}\ }\href {\doibase
  10.1103/RevModPhys.81.591} {\bibfield  {journal} {\bibinfo  {journal} {Rev.
  Mod. Phys.}\ }\textbf {\bibinfo {volume} {81}},\ \bibinfo {pages} {591--646}
  (\bibinfo {year} {2009})}\BibitemShut {NoStop}%
\bibitem [{\citenamefont {Baumann}\ \emph {et~al.}(2021)\citenamefont
  {Baumann}, \citenamefont {Lorenz-Spreen}, \citenamefont {Sokolov},\ and\
  \citenamefont {Starnini}}]{Baumann2021}%
  \BibitemOpen
  \bibfield  {author} {\bibinfo {author} {\bibfnamefont {Fabian}\ \bibnamefont
  {Baumann}}, \bibinfo {author} {\bibfnamefont {Philipp}\ \bibnamefont
  {Lorenz-Spreen}}, \bibinfo {author} {\bibfnamefont {Igor~M.}\ \bibnamefont
  {Sokolov}}, \ and\ \bibinfo {author} {\bibfnamefont {Michele}\ \bibnamefont
  {Starnini}},\ }\bibfield  {title} {\enquote {\bibinfo {title} {Emergence of
  polarized ideological opinions in multidimensional topic spaces},}\ }\href
  {\doibase 10.1103/PhysRevX.11.011012} {\bibfield  {journal} {\bibinfo
  {journal} {Phys. Rev. X}\ }\textbf {\bibinfo {volume} {11}},\ \bibinfo
  {pages} {011012} (\bibinfo {year} {2021})}\BibitemShut {NoStop}%
\bibitem [{\citenamefont {Bouchaud}(2013)}]{Bouchaud2013}%
  \BibitemOpen
  \bibfield  {author} {\bibinfo {author} {\bibfnamefont {J.-P.}\ \bibnamefont
  {Bouchaud}},\ }\bibfield  {title} {\enquote {\bibinfo {title} {Crises and
  collective socio-economic phenomena: Simple models and challenges},}\
  }\href@noop {} {\bibfield  {journal} {\bibinfo  {journal} {J. Stat. Phys.}\
  }\textbf {\bibinfo {volume} {151}},\ \bibinfo {pages} {567–606} (\bibinfo
  {year} {2013})}\BibitemShut {NoStop}%
\bibitem [{\citenamefont {Hopfield}(1982)}]{Hopfield82}%
  \BibitemOpen
  \bibfield  {author} {\bibinfo {author} {\bibfnamefont {J.~J.}\ \bibnamefont
  {Hopfield}},\ }\bibfield  {title} {\enquote {\bibinfo {title} {Neural
  networks and physical systems with emergent collective computational
  abilities},}\ }\href@noop {} {\bibfield  {journal} {\bibinfo  {journal}
  {Proc. Natl. Acad. Sci. U S A}\ }\textbf {\bibinfo {volume} {79}},\ \bibinfo
  {pages} {2554} (\bibinfo {year} {1982})}\BibitemShut {NoStop}%
\bibitem [{\citenamefont {Wemmenhove}\ and\ \citenamefont
  {Coolen}(2003)}]{Wemmenhove2003}%
  \BibitemOpen
  \bibfield  {author} {\bibinfo {author} {\bibfnamefont {B}~\bibnamefont
  {Wemmenhove}}\ and\ \bibinfo {author} {\bibfnamefont {A~C~C}\ \bibnamefont
  {Coolen}},\ }\bibfield  {title} {\enquote {\bibinfo {title} {Finite
  connectivity attractor neural networks},}\ }\href {\doibase
  10.1088/0305-4470/36/37/302} {\bibfield  {journal} {\bibinfo  {journal}
  {Journal of Physics A: Mathematical and General}\ }\textbf {\bibinfo {volume}
  {36}},\ \bibinfo {pages} {9617--9633} (\bibinfo {year} {2003})}\BibitemShut
  {NoStop}%
\bibitem [{\citenamefont {Castillo}\ \emph {et~al.}(2004)\citenamefont
  {Castillo}, \citenamefont {Wemmenhove}, \citenamefont {Hatchett},
  \citenamefont {Coolen}, \citenamefont {Skantzos},\ and\ \citenamefont
  {Nikoletopoulos}}]{Castillo2004}%
  \BibitemOpen
  \bibfield  {author} {\bibinfo {author} {\bibfnamefont {I~P{\'{e}}rez}\
  \bibnamefont {Castillo}}, \bibinfo {author} {\bibfnamefont {B}~\bibnamefont
  {Wemmenhove}}, \bibinfo {author} {\bibfnamefont {J~P~L}\ \bibnamefont
  {Hatchett}}, \bibinfo {author} {\bibfnamefont {A~C~C}\ \bibnamefont
  {Coolen}}, \bibinfo {author} {\bibfnamefont {N~S}\ \bibnamefont {Skantzos}},
  \ and\ \bibinfo {author} {\bibfnamefont {T}~\bibnamefont {Nikoletopoulos}},\
  }\bibfield  {title} {\enquote {\bibinfo {title} {Analytic solution of
  attractor neural networks on scale-free graphs},}\ }\href {\doibase
  10.1088/0305-4470/37/37/002} {\bibfield  {journal} {\bibinfo  {journal}
  {Journal of Physics A: Mathematical and General}\ }\textbf {\bibinfo {volume}
  {37}},\ \bibinfo {pages} {8789--8799} (\bibinfo {year} {2004})}\BibitemShut
  {NoStop}%
\bibitem [{\citenamefont {Challet}\ and\ \citenamefont
  {Zhang}(1997)}]{Challet1997}%
  \BibitemOpen
  \bibfield  {author} {\bibinfo {author} {\bibfnamefont {D.}~\bibnamefont
  {Challet}}\ and\ \bibinfo {author} {\bibfnamefont {Y.-C.}\ \bibnamefont
  {Zhang}},\ }\bibfield  {title} {\enquote {\bibinfo {title} {Emergence of
  cooperation and organization in an evolutionary game},}\ }\href {\doibase
  https://doi.org/10.1016/S0378-4371(97)00419-6} {\bibfield  {journal}
  {\bibinfo  {journal} {Physica A: Statistical Mechanics and its Applications}\
  }\textbf {\bibinfo {volume} {246}},\ \bibinfo {pages} {407--418} (\bibinfo
  {year} {1997})}\BibitemShut {NoStop}%
\bibitem [{\citenamefont {Challet}\ \emph {et~al.}(2005)\citenamefont
  {Challet}, \citenamefont {Marsili},\ and\ \citenamefont
  {Zhang}}]{Challet2005}%
  \BibitemOpen
  \bibfield  {author} {\bibinfo {author} {\bibfnamefont {D.}~\bibnamefont
  {Challet}}, \bibinfo {author} {\bibfnamefont {M.}~\bibnamefont {Marsili}}, \
  and\ \bibinfo {author} {\bibfnamefont {Y.C.}\ \bibnamefont {Zhang}},\ }\href
  {https://books.google.com.br/books?id=3W758QJ16JkC} {\emph {\bibinfo {title}
  {Minority Games}}},\ Oxford finance\ (\bibinfo  {publisher} {Oxford
  University Press},\ \bibinfo {year} {2005})\BibitemShut {NoStop}%
\bibitem [{\citenamefont {Seoane}(2021)}]{Seoane2021}%
  \BibitemOpen
  \bibfield  {author} {\bibinfo {author} {\bibfnamefont {Lu\'{\i}s~F.}\
  \bibnamefont {Seoane}},\ }\bibfield  {title} {\enquote {\bibinfo {title}
  {Games in rigged economies},}\ }\href {\doibase 10.1103/PhysRevX.11.031058}
  {\bibfield  {journal} {\bibinfo  {journal} {Phys. Rev. X}\ }\textbf {\bibinfo
  {volume} {11}},\ \bibinfo {pages} {031058} (\bibinfo {year}
  {2021})}\BibitemShut {NoStop}%
\bibitem [{\citenamefont {Brunel}(2000)}]{Brunel2000}%
  \BibitemOpen
  \bibfield  {author} {\bibinfo {author} {\bibfnamefont {N.}~\bibnamefont
  {Brunel}},\ }\bibfield  {title} {\enquote {\bibinfo {title} {Dynamics of
  sparsely connected networks of excitatory and inhibitory spiking neurons},}\
  }\href@noop {} {\bibfield  {journal} {\bibinfo  {journal} {Comput.
  Neurosci.}\ }\textbf {\bibinfo {volume} {8}},\ \bibinfo {pages} {183}
  (\bibinfo {year} {2000})}\BibitemShut {NoStop}%
\bibitem [{\citenamefont {Ostojic}(2014)}]{Ostojic2014}%
  \BibitemOpen
  \bibfield  {author} {\bibinfo {author} {\bibfnamefont {S.}~\bibnamefont
  {Ostojic}},\ }\bibfield  {title} {\enquote {\bibinfo {title} {Two types of
  asynchronous activity in networks of excitatory and inhibitory spiking
  neurons},}\ }\href@noop {} {\bibfield  {journal} {\bibinfo  {journal} {Nat.
  Neurosci.}\ }\textbf {\bibinfo {volume} {17}},\ \bibinfo {pages} {594}
  (\bibinfo {year} {2014})}\BibitemShut {NoStop}%
\bibitem [{\citenamefont {Schuessler}\ \emph {et~al.}(2020)\citenamefont
  {Schuessler}, \citenamefont {Dubreuil}, \citenamefont {Mastrogiuseppe},
  \citenamefont {Ostojic},\ and\ \citenamefont {Barak}}]{Schuessler2020}%
  \BibitemOpen
  \bibfield  {author} {\bibinfo {author} {\bibfnamefont {Friedrich}\
  \bibnamefont {Schuessler}}, \bibinfo {author} {\bibfnamefont {Alexis}\
  \bibnamefont {Dubreuil}}, \bibinfo {author} {\bibfnamefont {Francesca}\
  \bibnamefont {Mastrogiuseppe}}, \bibinfo {author} {\bibfnamefont {Srdjan}\
  \bibnamefont {Ostojic}}, \ and\ \bibinfo {author} {\bibfnamefont {Omri}\
  \bibnamefont {Barak}},\ }\bibfield  {title} {\enquote {\bibinfo {title}
  {Dynamics of random recurrent networks with correlated low-rank structure},}\
  }\href {\doibase 10.1103/PhysRevResearch.2.013111} {\bibfield  {journal}
  {\bibinfo  {journal} {Phys. Rev. Research}\ }\textbf {\bibinfo {volume}
  {2}},\ \bibinfo {pages} {013111} (\bibinfo {year} {2020})}\BibitemShut
  {NoStop}%
\bibitem [{\citenamefont {Rogers}\ \emph {et~al.}(2008)\citenamefont {Rogers},
  \citenamefont {Castillo}, \citenamefont {K\"uhn},\ and\ \citenamefont
  {Takeda}}]{Rogers2008}%
  \BibitemOpen
  \bibfield  {author} {\bibinfo {author} {\bibfnamefont {Tim}\ \bibnamefont
  {Rogers}}, \bibinfo {author} {\bibfnamefont {Isaac~P\'erez}\ \bibnamefont
  {Castillo}}, \bibinfo {author} {\bibfnamefont {Reimer}\ \bibnamefont
  {K\"uhn}}, \ and\ \bibinfo {author} {\bibfnamefont {Koujin}\ \bibnamefont
  {Takeda}},\ }\bibfield  {title} {\enquote {\bibinfo {title} {Cavity approach
  to the spectral density of sparse symmetric random matrices},}\ }\href
  {\doibase 10.1103/PhysRevE.78.031116} {\bibfield  {journal} {\bibinfo
  {journal} {Phys. Rev. E}\ }\textbf {\bibinfo {volume} {78}},\ \bibinfo
  {pages} {031116} (\bibinfo {year} {2008})}\BibitemShut {NoStop}%
\bibitem [{\citenamefont {Metz}\ \emph {et~al.}(2019)\citenamefont {Metz},
  \citenamefont {Neri},\ and\ \citenamefont {Rogers}}]{Metz2019}%
  \BibitemOpen
  \bibfield  {author} {\bibinfo {author} {\bibfnamefont {Fernando~Lucas}\
  \bibnamefont {Metz}}, \bibinfo {author} {\bibfnamefont {Izaak}\ \bibnamefont
  {Neri}}, \ and\ \bibinfo {author} {\bibfnamefont {Tim}\ \bibnamefont
  {Rogers}},\ }\bibfield  {title} {\enquote {\bibinfo {title} {Spectral theory
  of sparse non-hermitian random matrices},}\ }\href {\doibase
  10.1088/1751-8121/ab1ce0} {\bibfield  {journal} {\bibinfo  {journal} {Journal
  of Physics A: Mathematical and Theoretical}\ }\textbf {\bibinfo {volume}
  {52}},\ \bibinfo {pages} {434003} (\bibinfo {year} {2019})}\BibitemShut
  {NoStop}%
\bibitem [{\citenamefont {Neri}\ and\ \citenamefont {Metz}(2020)}]{Neri2020}%
  \BibitemOpen
  \bibfield  {author} {\bibinfo {author} {\bibfnamefont {Izaak}\ \bibnamefont
  {Neri}}\ and\ \bibinfo {author} {\bibfnamefont {Fernando~Lucas}\ \bibnamefont
  {Metz}},\ }\bibfield  {title} {\enquote {\bibinfo {title} {Linear stability
  analysis of large dynamical systems on random directed graphs},}\ }\href
  {\doibase 10.1103/PhysRevResearch.2.033313} {\bibfield  {journal} {\bibinfo
  {journal} {Phys. Rev. Research}\ }\textbf {\bibinfo {volume} {2}},\ \bibinfo
  {pages} {033313} (\bibinfo {year} {2020})}\BibitemShut {NoStop}%
\bibitem [{\citenamefont {Krumbeck}\ \emph {et~al.}(2021)\citenamefont
  {Krumbeck}, \citenamefont {Yang}, \citenamefont {Constable},\ and\
  \citenamefont {Rogers}}]{Krumbeck2021}%
  \BibitemOpen
  \bibfield  {author} {\bibinfo {author} {\bibfnamefont {Yvonne}\ \bibnamefont
  {Krumbeck}}, \bibinfo {author} {\bibfnamefont {Qian}\ \bibnamefont {Yang}},
  \bibinfo {author} {\bibfnamefont {George W.~A.}\ \bibnamefont {Constable}}, \
  and\ \bibinfo {author} {\bibfnamefont {Tim}\ \bibnamefont {Rogers}},\
  }\bibfield  {title} {\enquote {\bibinfo {title} {Fluctuation spectra of large
  random dynamical systems reveal hidden structure in ecological networks},}\
  }\href@noop {} {\bibfield  {journal} {\bibinfo  {journal} {Nat. Commun.}\
  }\textbf {\bibinfo {volume} {12}},\ \bibinfo {pages} {3625} (\bibinfo {year}
  {2021})}\BibitemShut {NoStop}%
\bibitem [{\citenamefont {Kuramoto}(1975)}]{Kuramoto1975}%
  \BibitemOpen
  \bibfield  {author} {\bibinfo {author} {\bibfnamefont {Yoshiki}\ \bibnamefont
  {Kuramoto}},\ }\bibfield  {title} {\enquote {\bibinfo {title}
  {Self-entrainment of a population of coupled non-linear oscillators},}\ }in\
  \href@noop {} {\emph {\bibinfo {booktitle} {International Symposium on
  Mathematical Problems in Theoretical Physics}}},\ \bibinfo {editor} {edited
  by\ \bibinfo {editor} {\bibfnamefont {Huzihiro}\ \bibnamefont {Araki}}}\
  (\bibinfo  {publisher} {Springer Berlin Heidelberg},\ \bibinfo {address}
  {Berlin, Heidelberg},\ \bibinfo {year} {1975})\ pp.\ \bibinfo {pages}
  {420--422}\BibitemShut {NoStop}%
\bibitem [{\citenamefont {Strogatz}(2000)}]{strogatz2000kuramoto}%
  \BibitemOpen
  \bibfield  {author} {\bibinfo {author} {\bibfnamefont {Steven~H}\
  \bibnamefont {Strogatz}},\ }\bibfield  {title} {\enquote {\bibinfo {title}
  {From {K}uramoto to {C}rawford: exploring the onset of synchronization in
  populations of coupled oscillators},}\ }\href@noop {} {\bibfield  {journal}
  {\bibinfo  {journal} {Physica D: Nonlinear Phenomena}\ }\textbf {\bibinfo
  {volume} {143}},\ \bibinfo {pages} {1--20} (\bibinfo {year}
  {2000})}\BibitemShut {NoStop}%
\bibitem [{\citenamefont {Rodrigues}\ \emph {et~al.}(2016)\citenamefont
  {Rodrigues}, \citenamefont {Peron}, \citenamefont {Ji},\ and\ \citenamefont
  {Kurths}}]{Rodrigues2016}%
  \BibitemOpen
  \bibfield  {author} {\bibinfo {author} {\bibfnamefont {Francisco~A.}\
  \bibnamefont {Rodrigues}}, \bibinfo {author} {\bibfnamefont {Thomas K.~DM.}\
  \bibnamefont {Peron}}, \bibinfo {author} {\bibfnamefont {Peng}\ \bibnamefont
  {Ji}}, \ and\ \bibinfo {author} {\bibfnamefont {Jürgen}\ \bibnamefont
  {Kurths}},\ }\bibfield  {title} {\enquote {\bibinfo {title} {The {K}uramoto
  model in complex networks},}\ }\href {\doibase
  https://doi.org/10.1016/j.physrep.2015.10.008} {\bibfield  {journal}
  {\bibinfo  {journal} {Physics Reports}\ }\textbf {\bibinfo {volume} {610}},\
  \bibinfo {pages} {1--98} (\bibinfo {year} {2016})}\BibitemShut {NoStop}%
\bibitem [{\citenamefont {da~Fonseca}\ and\ \citenamefont
  {Abud}(2018)}]{fonseca2018kuramoto}%
  \BibitemOpen
  \bibfield  {author} {\bibinfo {author} {\bibfnamefont {JD}~\bibnamefont
  {da~Fonseca}}\ and\ \bibinfo {author} {\bibfnamefont {CV}~\bibnamefont
  {Abud}},\ }\bibfield  {title} {\enquote {\bibinfo {title} {The {K}uramoto
  model revisited},}\ }\href@noop {} {\bibfield  {journal} {\bibinfo  {journal}
  {Journal of Statistical Mechanics: Theory and Experiment}\ }\textbf {\bibinfo
  {volume} {2018}},\ \bibinfo {pages} {103204} (\bibinfo {year}
  {2018})}\BibitemShut {NoStop}%
\bibitem [{\citenamefont {Antenucci}\ \emph {et~al.}(2015)\citenamefont
  {Antenucci}, \citenamefont {Ib\'a\~nez Berganza},\ and\ \citenamefont
  {Leuzzi}}]{Antenucci2015}%
  \BibitemOpen
  \bibfield  {author} {\bibinfo {author} {\bibfnamefont {F.}~\bibnamefont
  {Antenucci}}, \bibinfo {author} {\bibfnamefont {M.}~\bibnamefont {Ib\'a\~nez
  Berganza}}, \ and\ \bibinfo {author} {\bibfnamefont {L.}~\bibnamefont
  {Leuzzi}},\ }\bibfield  {title} {\enquote {\bibinfo {title} {Statistical
  physics of nonlinear wave interaction},}\ }\href {\doibase
  10.1103/PhysRevB.92.014204} {\bibfield  {journal} {\bibinfo  {journal} {Phys.
  Rev. B}\ }\textbf {\bibinfo {volume} {92}},\ \bibinfo {pages} {014204}
  (\bibinfo {year} {2015})}\BibitemShut {NoStop}%
\bibitem [{\citenamefont {Marruzzo}\ and\ \citenamefont
  {Leuzzi}(2015)}]{Marruzzo2015}%
  \BibitemOpen
  \bibfield  {author} {\bibinfo {author} {\bibfnamefont {Alessia}\ \bibnamefont
  {Marruzzo}}\ and\ \bibinfo {author} {\bibfnamefont {Luca}\ \bibnamefont
  {Leuzzi}},\ }\bibfield  {title} {\enquote {\bibinfo {title} {Nonlinear $xy$
  and $p$-clock models on sparse random graphs: Mode-locking transition of
  localized waves},}\ }\href {\doibase 10.1103/PhysRevB.91.054201} {\bibfield
  {journal} {\bibinfo  {journal} {Phys. Rev. B}\ }\textbf {\bibinfo {volume}
  {91}},\ \bibinfo {pages} {054201} (\bibinfo {year} {2015})}\BibitemShut
  {NoStop}%
\bibitem [{\citenamefont {Antenucci}\ \emph {et~al.}(2021)\citenamefont
  {Antenucci}, \citenamefont {Lerario}, \citenamefont {Fernand\'ez},
  \citenamefont {De~Marco}, \citenamefont {De~Giorgi}, \citenamefont
  {Ballarini}, \citenamefont {Sanvitto},\ and\ \citenamefont
  {Leuzzi}}]{Antenucci2021}%
  \BibitemOpen
  \bibfield  {author} {\bibinfo {author} {\bibfnamefont {Fabrizio}\
  \bibnamefont {Antenucci}}, \bibinfo {author} {\bibfnamefont {Giovanni}\
  \bibnamefont {Lerario}}, \bibinfo {author} {\bibfnamefont {Blanca~Silva}\
  \bibnamefont {Fernand\'ez}}, \bibinfo {author} {\bibfnamefont {Luisa}\
  \bibnamefont {De~Marco}}, \bibinfo {author} {\bibfnamefont {Milena}\
  \bibnamefont {De~Giorgi}}, \bibinfo {author} {\bibfnamefont {Dario}\
  \bibnamefont {Ballarini}}, \bibinfo {author} {\bibfnamefont {Daniele}\
  \bibnamefont {Sanvitto}}, \ and\ \bibinfo {author} {\bibfnamefont {Luca}\
  \bibnamefont {Leuzzi}},\ }\bibfield  {title} {\enquote {\bibinfo {title}
  {Demonstration of self-starting nonlinear mode locking in random lasers},}\
  }\href {\doibase 10.1103/PhysRevLett.126.173901} {\bibfield  {journal}
  {\bibinfo  {journal} {Phys. Rev. Lett.}\ }\textbf {\bibinfo {volume} {126}},\
  \bibinfo {pages} {173901} (\bibinfo {year} {2021})}\BibitemShut {NoStop}%
\bibitem [{\citenamefont {Lupo}(2017)}]{Cosimothesis}%
  \BibitemOpen
  \bibfield  {author} {\bibinfo {author} {\bibfnamefont {Cosimo}\ \bibnamefont
  {Lupo}},\ }\emph {\bibinfo {title} {Critical properties of disordered XY
  model on sparse random graphs}},\ \href@noop {} {Ph.D. thesis},\ \bibinfo
  {school} {Sapienza University of Rome} (\bibinfo {year} {2017})\BibitemShut
  {NoStop}%
\bibitem [{\citenamefont {Lupo}\ and\ \citenamefont
  {Ricci-Tersenghi}(2018)}]{Lupo2018}%
  \BibitemOpen
  \bibfield  {author} {\bibinfo {author} {\bibfnamefont {Cosimo}\ \bibnamefont
  {Lupo}}\ and\ \bibinfo {author} {\bibfnamefont {Federico}\ \bibnamefont
  {Ricci-Tersenghi}},\ }\bibfield  {title} {\enquote {\bibinfo {title}
  {Comparison of gabay--toulouse and de almeida--thouless instabilities for the
  spin-glass xy model in a field on sparse random graphs},}\ }\href {\doibase
  10.1103/PhysRevB.97.014414} {\bibfield  {journal} {\bibinfo  {journal} {Phys.
  Rev. B}\ }\textbf {\bibinfo {volume} {97}},\ \bibinfo {pages} {014414}
  (\bibinfo {year} {2018})}\BibitemShut {NoStop}%
\bibitem [{\citenamefont {Lupo}\ \emph {et~al.}(2019)\citenamefont {Lupo},
  \citenamefont {Parisi},\ and\ \citenamefont {Ricci-Tersenghi}}]{Lupo2019}%
  \BibitemOpen
  \bibfield  {author} {\bibinfo {author} {\bibfnamefont {Cosimo}\ \bibnamefont
  {Lupo}}, \bibinfo {author} {\bibfnamefont {Giorgio}\ \bibnamefont {Parisi}},
  \ and\ \bibinfo {author} {\bibfnamefont {Federico}\ \bibnamefont
  {Ricci-Tersenghi}},\ }\bibfield  {title} {\enquote {\bibinfo {title} {The
  random field {XY} model on sparse random graphs shows replica symmetry
  breaking and marginally stable ferromagnetism},}\ }\href {\doibase
  10.1088/1751-8121/ab2287} {\bibfield  {journal} {\bibinfo  {journal} {Journal
  of Physics A: Mathematical and Theoretical}\ }\textbf {\bibinfo {volume}
  {52}},\ \bibinfo {pages} {284001} (\bibinfo {year} {2019})}\BibitemShut
  {NoStop}%
\bibitem [{\citenamefont {Olfati-Saber}(2006)}]{Olfati2006}%
  \BibitemOpen
  \bibfield  {author} {\bibinfo {author} {\bibfnamefont {Reza}\ \bibnamefont
  {Olfati-Saber}},\ }\bibfield  {title} {\enquote {\bibinfo {title} {Swarms on
  sphere: A programmable swarm with synchronous behaviors like oscillator
  networks},}\ }in\ \href {\doibase 10.1109/CDC.2006.376811} {\emph {\bibinfo
  {booktitle} {Proceedings of the 45th IEEE Conference on Decision and
  Control}}}\ (\bibinfo {year} {2006})\ pp.\ \bibinfo {pages}
  {5060--5066}\BibitemShut {NoStop}%
\bibitem [{\citenamefont {Lohe}(2009)}]{lohe2009non}%
  \BibitemOpen
  \bibfield  {author} {\bibinfo {author} {\bibfnamefont {MA}~\bibnamefont
  {Lohe}},\ }\bibfield  {title} {\enquote {\bibinfo {title} {Non-abelian
  kuramoto models and synchronization},}\ }\href@noop {} {\bibfield  {journal}
  {\bibinfo  {journal} {Journal of Physics A: Mathematical and Theoretical}\
  }\textbf {\bibinfo {volume} {42}},\ \bibinfo {pages} {395101} (\bibinfo
  {year} {2009})}\BibitemShut {NoStop}%
\bibitem [{\citenamefont {Zhu}(2013)}]{Zhu2013}%
  \BibitemOpen
  \bibfield  {author} {\bibinfo {author} {\bibfnamefont {Jiandong}\
  \bibnamefont {Zhu}},\ }\bibfield  {title} {\enquote {\bibinfo {title}
  {Synchronization of {K}uramoto model in a high-dimensional linear space},}\
  }\href {\doibase https://doi.org/10.1016/j.physleta.2013.09.010} {\bibfield
  {journal} {\bibinfo  {journal} {Physics Letters A}\ }\textbf {\bibinfo
  {volume} {377}},\ \bibinfo {pages} {2939--2943} (\bibinfo {year}
  {2013})}\BibitemShut {NoStop}%
\bibitem [{\citenamefont {Chandra}\ \emph {et~al.}(2019)\citenamefont
  {Chandra}, \citenamefont {Girvan},\ and\ \citenamefont {Ott}}]{Chandra2019}%
  \BibitemOpen
  \bibfield  {author} {\bibinfo {author} {\bibfnamefont {Sarthak}\ \bibnamefont
  {Chandra}}, \bibinfo {author} {\bibfnamefont {Michelle}\ \bibnamefont
  {Girvan}}, \ and\ \bibinfo {author} {\bibfnamefont {Edward}\ \bibnamefont
  {Ott}},\ }\bibfield  {title} {\enquote {\bibinfo {title} {Continuous versus
  discontinuous transitions in the $d$-dimensional generalized {K}uramoto
  {M}odel: Odd $d$ is different},}\ }\href {\doibase 10.1103/PhysRevX.9.011002}
  {\bibfield  {journal} {\bibinfo  {journal} {Phys. Rev. X}\ }\textbf {\bibinfo
  {volume} {9}},\ \bibinfo {pages} {011002} (\bibinfo {year}
  {2019})}\BibitemShut {NoStop}%
\bibitem [{\citenamefont {Kochma{\'{n}}ski}\ \emph {et~al.}(2013)\citenamefont
  {Kochma{\'{n}}ski}, \citenamefont {Paszkiewicz},\ and\ \citenamefont
  {Wolski}}]{Kochma2013}%
  \BibitemOpen
  \bibfield  {author} {\bibinfo {author} {\bibfnamefont {M}~\bibnamefont
  {Kochma{\'{n}}ski}}, \bibinfo {author} {\bibfnamefont {T}~\bibnamefont
  {Paszkiewicz}}, \ and\ \bibinfo {author} {\bibfnamefont {S}~\bibnamefont
  {Wolski}},\ }\bibfield  {title} {\enquote {\bibinfo {title}
  {Curie{\textendash}weiss magnet{\textemdash}a simple model of phase
  transition},}\ }\href {\doibase 10.1088/0143-0807/34/6/1555} {\bibfield
  {journal} {\bibinfo  {journal} {European Journal of Physics}\ }\textbf
  {\bibinfo {volume} {34}},\ \bibinfo {pages} {1555--1573} (\bibinfo {year}
  {2013})}\BibitemShut {NoStop}%
\bibitem [{\citenamefont {Sherrington}\ and\ \citenamefont
  {Kirkpatrick}(1975)}]{Sherrington1975}%
  \BibitemOpen
  \bibfield  {author} {\bibinfo {author} {\bibfnamefont {David}\ \bibnamefont
  {Sherrington}}\ and\ \bibinfo {author} {\bibfnamefont {Scott}\ \bibnamefont
  {Kirkpatrick}},\ }\bibfield  {title} {\enquote {\bibinfo {title} {Solvable
  model of a spin-glass},}\ }\href {\doibase 10.1103/PhysRevLett.35.1792}
  {\bibfield  {journal} {\bibinfo  {journal} {Phys. Rev. Lett.}\ }\textbf
  {\bibinfo {volume} {35}},\ \bibinfo {pages} {1792--1796} (\bibinfo {year}
  {1975})}\BibitemShut {NoStop}%
\bibitem [{\citenamefont {Kirkpatrick}\ and\ \citenamefont
  {Sherrington}(1978)}]{Sherrington1978}%
  \BibitemOpen
  \bibfield  {author} {\bibinfo {author} {\bibfnamefont {Scott}\ \bibnamefont
  {Kirkpatrick}}\ and\ \bibinfo {author} {\bibfnamefont {David}\ \bibnamefont
  {Sherrington}},\ }\bibfield  {title} {\enquote {\bibinfo {title}
  {Infinite-ranged models of spin-glasses},}\ }\href {\doibase
  10.1103/PhysRevB.17.4384} {\bibfield  {journal} {\bibinfo  {journal} {Phys.
  Rev. B}\ }\textbf {\bibinfo {volume} {17}},\ \bibinfo {pages} {4384--4403}
  (\bibinfo {year} {1978})}\BibitemShut {NoStop}%
\bibitem [{\citenamefont {Hatchett}\ and\ \citenamefont
  {Uezu}(2008)}]{Hatchett2008}%
  \BibitemOpen
  \bibfield  {author} {\bibinfo {author} {\bibfnamefont {Jonathan P.~L.}\
  \bibnamefont {Hatchett}}\ and\ \bibinfo {author} {\bibfnamefont {Tatsuya}\
  \bibnamefont {Uezu}},\ }\bibfield  {title} {\enquote {\bibinfo {title} {Mean
  field and cavity analysis for coupled oscillator networks},}\ }\href
  {\doibase 10.1103/PhysRevE.78.036106} {\bibfield  {journal} {\bibinfo
  {journal} {Phys. Rev. E}\ }\textbf {\bibinfo {volume} {78}},\ \bibinfo
  {pages} {036106} (\bibinfo {year} {2008})}\BibitemShut {NoStop}%
\bibitem [{\citenamefont {P.~L.~Hatchett}\ and\ \citenamefont
  {Uezu}(2009)}]{Hatchett2009}%
  \BibitemOpen
  \bibfield  {author} {\bibinfo {author} {\bibfnamefont {Jonathan}\
  \bibnamefont {P.~L.~Hatchett}}\ and\ \bibinfo {author} {\bibfnamefont
  {Tatsuya}\ \bibnamefont {Uezu}},\ }\bibfield  {title} {\enquote {\bibinfo
  {title} {Dynamical behavior of phase oscillator networks on the {B}ethe
  lattice},}\ }\href {\doibase 10.1143/JPSJ.78.024001} {\bibfield  {journal}
  {\bibinfo  {journal} {Journal of the Physical Society of Japan}\ }\textbf
  {\bibinfo {volume} {78}},\ \bibinfo {pages} {024001} (\bibinfo {year}
  {2009})},\ \Eprint
  {http://arxiv.org/abs/https://doi.org/10.1143/JPSJ.78.024001}
  {https://doi.org/10.1143/JPSJ.78.024001} \BibitemShut {NoStop}%
\bibitem [{\citenamefont {M\'ezard}\ and\ \citenamefont
  {Parisi}(2001)}]{Mezard2001}%
  \BibitemOpen
  \bibfield  {author} {\bibinfo {author} {\bibfnamefont {M.}~\bibnamefont
  {M\'ezard}}\ and\ \bibinfo {author} {\bibfnamefont {G.}~\bibnamefont
  {Parisi}},\ }\bibfield  {title} {\enquote {\bibinfo {title} {The {B}ethe
  lattice spin glass revisited},}\ }\href@noop {} {\bibfield  {journal}
  {\bibinfo  {journal} {Eur. Phys. J. B}\ }\textbf {\bibinfo {volume} {20}},\
  \bibinfo {pages} {217} (\bibinfo {year} {2001})}\BibitemShut {NoStop}%
\bibitem [{\citenamefont {M\'ezard}\ and\ \citenamefont
  {Parisi}(2003)}]{Mezard2003}%
  \BibitemOpen
  \bibfield  {author} {\bibinfo {author} {\bibfnamefont {M.}~\bibnamefont
  {M\'ezard}}\ and\ \bibinfo {author} {\bibfnamefont {G.}~\bibnamefont
  {Parisi}},\ }\bibfield  {title} {\enquote {\bibinfo {title} {The cavity
  method at zero temperature},}\ }\href@noop {} {\bibfield  {journal} {\bibinfo
   {journal} {J. Stat. Phys.}\ }\textbf {\bibinfo {volume} {111}},\ \bibinfo
  {pages} {1} (\bibinfo {year} {2003})}\BibitemShut {NoStop}%
\bibitem [{\citenamefont {Wald}(1944)}]{Wald1944}%
  \BibitemOpen
  \bibfield  {author} {\bibinfo {author} {\bibfnamefont {Abraham}\ \bibnamefont
  {Wald}},\ }\bibfield  {title} {\enquote {\bibinfo {title} {On cumulative sums
  of random variables},}\ }\href@noop {} {\bibfield  {journal} {\bibinfo
  {journal} {Ann. Math. Stat.}\ }\textbf {\bibinfo {volume} {15}},\ \bibinfo
  {pages} {283} (\bibinfo {year} {1944})}\BibitemShut {NoStop}%
\bibitem [{\citenamefont {Robbins}(1948)}]{Robbins1948}%
  \BibitemOpen
  \bibfield  {author} {\bibinfo {author} {\bibfnamefont {Herbert}\ \bibnamefont
  {Robbins}},\ }\bibfield  {title} {\enquote {\bibinfo {title} {The asymptotic
  distribution of the sum of a random number of random variables},}\
  }\href@noop {} {\bibfield  {journal} {\bibinfo  {journal} {Bull. Amer. Math.
  Soc.}\ }\textbf {\bibinfo {volume} {54}},\ \bibinfo {pages} {1151} (\bibinfo
  {year} {1948})}\BibitemShut {NoStop}%
\bibitem [{\citenamefont {Gnedenko}\ and\ \citenamefont
  {Korolev}(1996)}]{GnedenkoBook}%
  \BibitemOpen
  \bibfield  {author} {\bibinfo {author} {\bibfnamefont {B.V.}\ \bibnamefont
  {Gnedenko}}\ and\ \bibinfo {author} {\bibfnamefont {V.Y.}\ \bibnamefont
  {Korolev}},\ }\href {https://books.google.com.br/books?id=IkAfPWgSKIgC}
  {\emph {\bibinfo {title} {Random Summation: Limit Theorems and
  Applications}}}\ (\bibinfo  {publisher} {Taylor \& Francis},\ \bibinfo {year}
  {1996})\BibitemShut {NoStop}%
\bibitem [{\citenamefont {Leone}\ \emph {et~al.}(2002)\citenamefont {Leone},
  \citenamefont {V\'azquez}, \citenamefont {Vespignani},\ and\ \citenamefont
  {Zecchina}}]{Leone2002}%
  \BibitemOpen
  \bibfield  {author} {\bibinfo {author} {\bibfnamefont {M}~\bibnamefont
  {Leone}}, \bibinfo {author} {\bibfnamefont {A.}~\bibnamefont {V\'azquez}},
  \bibinfo {author} {\bibfnamefont {A.}~\bibnamefont {Vespignani}}, \ and\
  \bibinfo {author} {\bibfnamefont {R.}~\bibnamefont {Zecchina}},\ }\bibfield
  {title} {\enquote {\bibinfo {title} {Ferromagnetic ordering in graphs with
  arbitrary degree distribution},}\ }\href@noop {} {\bibfield  {journal}
  {\bibinfo  {journal} {Eur. Phys. J. B}\ }\textbf {\bibinfo {volume} {28}},\
  \bibinfo {pages} {191} (\bibinfo {year} {2002})}\BibitemShut {NoStop}%
\bibitem [{\citenamefont {de~Almeida}\ and\ \citenamefont
  {Thouless}(1978)}]{Almeida1978}%
  \BibitemOpen
  \bibfield  {author} {\bibinfo {author} {\bibfnamefont {J~R~L}\ \bibnamefont
  {de~Almeida}}\ and\ \bibinfo {author} {\bibfnamefont {D~J}\ \bibnamefont
  {Thouless}},\ }\bibfield  {title} {\enquote {\bibinfo {title} {Stability of
  the sherrington-kirkpatrick solution of a spin glass model},}\ }\href
  {\doibase 10.1088/0305-4470/11/5/028} {\bibfield  {journal} {\bibinfo
  {journal} {Journal of Physics A: Mathematical and General}\ }\textbf
  {\bibinfo {volume} {11}},\ \bibinfo {pages} {983--990} (\bibinfo {year}
  {1978})}\BibitemShut {NoStop}%
\bibitem [{\citenamefont {Skantzos}\ \emph {et~al.}(2005)\citenamefont
  {Skantzos}, \citenamefont {Castillo},\ and\ \citenamefont
  {Hatchett}}]{Skantzos2005}%
  \BibitemOpen
  \bibfield  {author} {\bibinfo {author} {\bibfnamefont {Nikos~S.}\
  \bibnamefont {Skantzos}}, \bibinfo {author} {\bibfnamefont {Isaac~P\'erez}\
  \bibnamefont {Castillo}}, \ and\ \bibinfo {author} {\bibfnamefont {Jonathan
  P.~L.}\ \bibnamefont {Hatchett}},\ }\bibfield  {title} {\enquote {\bibinfo
  {title} {Cavity approach for real variables on diluted graphs and application
  to synchronization in small-world lattices},}\ }\href {\doibase
  10.1103/PhysRevE.72.066127} {\bibfield  {journal} {\bibinfo  {journal} {Phys.
  Rev. E}\ }\textbf {\bibinfo {volume} {72}},\ \bibinfo {pages} {066127}
  (\bibinfo {year} {2005})}\BibitemShut {NoStop}%
\bibitem [{\citenamefont {Coolen}\ \emph {et~al.}(2005)\citenamefont {Coolen},
  \citenamefont {Skantzos}, \citenamefont {Castillo}, \citenamefont {Vicente},
  \citenamefont {Hatchett}, \citenamefont {Wemmenhove},\ and\ \citenamefont
  {Nikoletopoulos}}]{Coolen2005}%
  \BibitemOpen
  \bibfield  {author} {\bibinfo {author} {\bibfnamefont {A~C~C}\ \bibnamefont
  {Coolen}}, \bibinfo {author} {\bibfnamefont {N~S}\ \bibnamefont {Skantzos}},
  \bibinfo {author} {\bibfnamefont {I~P{\'{e}}rez}\ \bibnamefont {Castillo}},
  \bibinfo {author} {\bibfnamefont {C~J~P{\'{e}}rez}\ \bibnamefont {Vicente}},
  \bibinfo {author} {\bibfnamefont {J~P~L}\ \bibnamefont {Hatchett}}, \bibinfo
  {author} {\bibfnamefont {B}~\bibnamefont {Wemmenhove}}, \ and\ \bibinfo
  {author} {\bibfnamefont {T}~\bibnamefont {Nikoletopoulos}},\ }\bibfield
  {title} {\enquote {\bibinfo {title} {Finitely connected vector spin systems
  with random matrix interactions},}\ }\href {\doibase
  10.1088/0305-4470/38/39/001} {\bibfield  {journal} {\bibinfo  {journal}
  {Journal of Physics A: Mathematical and General}\ }\textbf {\bibinfo {volume}
  {38}},\ \bibinfo {pages} {8289--8317} (\bibinfo {year} {2005})}\BibitemShut
  {NoStop}%
\bibitem [{\citenamefont {Bollob{\'a}s}(2001)}]{BolloBook}%
  \BibitemOpen
  \bibfield  {author} {\bibinfo {author} {\bibfnamefont {B.}~\bibnamefont
  {Bollob{\'a}s}},\ }\href {https://books.google.com.br/books?id=EX6rQgAACAAJ}
  {\emph {\bibinfo {title} {Random Graphs}}},\ Cambridge Studies in Advanced
  Mathematics\ (\bibinfo  {publisher} {Cambridge University Press},\ \bibinfo
  {year} {2001})\BibitemShut {NoStop}%
\bibitem [{\citenamefont {Stanley}(1968)}]{Stanley1968}%
  \BibitemOpen
  \bibfield  {author} {\bibinfo {author} {\bibfnamefont {H.~E.}\ \bibnamefont
  {Stanley}},\ }\bibfield  {title} {\enquote {\bibinfo {title} {Dependence of
  critical properties on dimensionality of spins},}\ }\href {\doibase
  10.1103/PhysRevLett.20.589} {\bibfield  {journal} {\bibinfo  {journal} {Phys.
  Rev. Lett.}\ }\textbf {\bibinfo {volume} {20}},\ \bibinfo {pages} {589--592}
  (\bibinfo {year} {1968})}\BibitemShut {NoStop}%
\bibitem [{\citenamefont {Molloy}\ and\ \citenamefont {Reed}(1995)}]{Molloy95}%
  \BibitemOpen
  \bibfield  {author} {\bibinfo {author} {\bibfnamefont {Michael}\ \bibnamefont
  {Molloy}}\ and\ \bibinfo {author} {\bibfnamefont {Bruce}\ \bibnamefont
  {Reed}},\ }\bibfield  {title} {\enquote {\bibinfo {title} {A critical point
  for random graphs with a given degree sequence},}\ }\href {\doibase
  10.1002/rsa.3240060204} {\bibfield  {journal} {\bibinfo  {journal} {Random
  Structures \& Algorithms}\ }\textbf {\bibinfo {volume} {6}},\ \bibinfo
  {pages} {161--180} (\bibinfo {year} {1995})}\BibitemShut {NoStop}%
\bibitem [{\citenamefont {Newman}\ \emph {et~al.}(2001)\citenamefont {Newman},
  \citenamefont {Strogatz},\ and\ \citenamefont {Watts}}]{Newman2001}%
  \BibitemOpen
  \bibfield  {author} {\bibinfo {author} {\bibfnamefont {M.~E.~J.}\
  \bibnamefont {Newman}}, \bibinfo {author} {\bibfnamefont {S.~H.}\
  \bibnamefont {Strogatz}}, \ and\ \bibinfo {author} {\bibfnamefont {D.~J.}\
  \bibnamefont {Watts}},\ }\bibfield  {title} {\enquote {\bibinfo {title}
  {Random graphs with arbitrary degree distributions and their applications},}\
  }\href {\doibase 10.1103/PhysRevE.64.026118} {\bibfield  {journal} {\bibinfo
  {journal} {Phys. Rev. E}\ }\textbf {\bibinfo {volume} {64}},\ \bibinfo
  {pages} {026118} (\bibinfo {year} {2001})}\BibitemShut {NoStop}%
\bibitem [{\citenamefont {Fosdick}\ \emph {et~al.}(2018)\citenamefont
  {Fosdick}, \citenamefont {Larremore}, \citenamefont {Nishimura},\ and\
  \citenamefont {Ugander}}]{Fosdick2018}%
  \BibitemOpen
  \bibfield  {author} {\bibinfo {author} {\bibfnamefont {Bailey~K.}\
  \bibnamefont {Fosdick}}, \bibinfo {author} {\bibfnamefont {Daniel~B.}\
  \bibnamefont {Larremore}}, \bibinfo {author} {\bibfnamefont {Joel}\
  \bibnamefont {Nishimura}}, \ and\ \bibinfo {author} {\bibfnamefont {Johan}\
  \bibnamefont {Ugander}},\ }\bibfield  {title} {\enquote {\bibinfo {title}
  {Configuring random graph models with fixed degree sequences},}\ }\href
  {\doibase 10.1137/16M1087175} {\bibfield  {journal} {\bibinfo  {journal}
  {SIAM Review}\ }\textbf {\bibinfo {volume} {60}},\ \bibinfo {pages}
  {315--355} (\bibinfo {year} {2018})},\ \Eprint
  {http://arxiv.org/abs/https://doi.org/10.1137/16M1087175}
  {https://doi.org/10.1137/16M1087175} \BibitemShut {NoStop}%
\bibitem [{\citenamefont {Evans}\ \emph {et~al.}(2011)\citenamefont {Evans},
  \citenamefont {Hastings}, \citenamefont {Peacock},\ and\ \citenamefont
  {Forbes}}]{EvansBook}%
  \BibitemOpen
  \bibfield  {author} {\bibinfo {author} {\bibfnamefont {M.}~\bibnamefont
  {Evans}}, \bibinfo {author} {\bibfnamefont {N.}~\bibnamefont {Hastings}},
  \bibinfo {author} {\bibfnamefont {B.}~\bibnamefont {Peacock}}, \ and\
  \bibinfo {author} {\bibfnamefont {C.}~\bibnamefont {Forbes}},\ }\href
  {https://books.google.com.br/books?id=YhF1osrQ4psC} {\emph {\bibinfo {title}
  {Statistical Distributions}}}\ (\bibinfo  {publisher} {Wiley},\ \bibinfo
  {year} {2011})\BibitemShut {NoStop}%
\bibitem [{\citenamefont {Garlaschelli}(2009)}]{Garlaschelli2009}%
  \BibitemOpen
  \bibfield  {author} {\bibinfo {author} {\bibfnamefont {Diego}\ \bibnamefont
  {Garlaschelli}},\ }\bibfield  {title} {\enquote {\bibinfo {title} {The
  weighted random graph model},}\ }\href {\doibase
  10.1088/1367-2630/11/7/073005} {\bibfield  {journal} {\bibinfo  {journal}
  {New Journal of Physics}\ }\textbf {\bibinfo {volume} {11}},\ \bibinfo
  {pages} {073005} (\bibinfo {year} {2009})}\BibitemShut {NoStop}%
\bibitem [{\citenamefont {Volz}\ \emph {et~al.}(2011)\citenamefont {Volz},
  \citenamefont {Miller}, \citenamefont {Galvani},\ and\ \citenamefont
  {Ancel~Meyers}}]{Volz2011}%
  \BibitemOpen
  \bibfield  {author} {\bibinfo {author} {\bibfnamefont {Erik~M.}\ \bibnamefont
  {Volz}}, \bibinfo {author} {\bibfnamefont {Joel~C.}\ \bibnamefont {Miller}},
  \bibinfo {author} {\bibfnamefont {Alison}\ \bibnamefont {Galvani}}, \ and\
  \bibinfo {author} {\bibfnamefont {Lauren}\ \bibnamefont {Ancel~Meyers}},\
  }\bibfield  {title} {\enquote {\bibinfo {title} {Effects of heterogeneous and
  clustered contact patterns on infectious disease dynamics},}\ }\href
  {\doibase 10.1371/journal.pcbi.1002042} {\bibfield  {journal} {\bibinfo
  {journal} {PLOS Computational Biology}\ }\textbf {\bibinfo {volume} {7}},\
  \bibinfo {pages} {1--13} (\bibinfo {year} {2011})}\BibitemShut {NoStop}%
\bibitem [{\citenamefont {Miller}\ and\ \citenamefont
  {Volz}(2013)}]{Miller2013}%
  \BibitemOpen
  \bibfield  {author} {\bibinfo {author} {\bibfnamefont {Joel~C.}\ \bibnamefont
  {Miller}}\ and\ \bibinfo {author} {\bibfnamefont {Erik~M.}\ \bibnamefont
  {Volz}},\ }\bibfield  {title} {\enquote {\bibinfo {title} {Incorporating
  disease and population structure into models of sir disease in contact
  networks},}\ }\href {\doibase 10.1371/journal.pone.0069162} {\bibfield
  {journal} {\bibinfo  {journal} {PLOS ONE}\ }\textbf {\bibinfo {volume} {8}},\
  \bibinfo {pages} {1--14} (\bibinfo {year} {2013})}\BibitemShut {NoStop}%
\bibitem [{\citenamefont {Aleta}\ \emph {et~al.}(2020)\citenamefont {Aleta},
  \citenamefont {Ferraz~de Arruda},\ and\ \citenamefont {Moreno}}]{Aleta2020}%
  \BibitemOpen
  \bibfield  {author} {\bibinfo {author} {\bibfnamefont {Alberto}\ \bibnamefont
  {Aleta}}, \bibinfo {author} {\bibfnamefont {Guilherme}\ \bibnamefont
  {Ferraz~de Arruda}}, \ and\ \bibinfo {author} {\bibfnamefont {Yamir}\
  \bibnamefont {Moreno}},\ }\bibfield  {title} {\enquote {\bibinfo {title}
  {Data-driven contact structures: From homogeneous mixing to multilayer
  networks},}\ }\href {\doibase 10.1371/journal.pcbi.1008035} {\bibfield
  {journal} {\bibinfo  {journal} {PLOS Computational Biology}\ }\textbf
  {\bibinfo {volume} {16}},\ \bibinfo {pages} {1--16} (\bibinfo {year}
  {2020})}\BibitemShut {NoStop}%
\bibitem [{\citenamefont {Bordenave}\ and\ \citenamefont
  {Lelarge}(2010)}]{Bordenave2010}%
  \BibitemOpen
  \bibfield  {author} {\bibinfo {author} {\bibfnamefont {Charles}\ \bibnamefont
  {Bordenave}}\ and\ \bibinfo {author} {\bibfnamefont {Marc}\ \bibnamefont
  {Lelarge}},\ }\bibfield  {title} {\enquote {\bibinfo {title} {Resolvent of
  large random graphs},}\ }\href {\doibase 10.1002/rsa.20313} {\bibfield
  {journal} {\bibinfo  {journal} {Random Structures \& Algorithms}\ }\textbf
  {\bibinfo {volume} {37}},\ \bibinfo {pages} {332--352} (\bibinfo {year}
  {2010})},\ \Eprint
  {http://arxiv.org/abs/https://onlinelibrary.wiley.com/doi/pdf/10.1002/rsa.20313}
  {https://onlinelibrary.wiley.com/doi/pdf/10.1002/rsa.20313} \BibitemShut
  {NoStop}%
\bibitem [{\citenamefont {Monasson}(1998)}]{Monasson1998}%
  \BibitemOpen
  \bibfield  {author} {\bibinfo {author} {\bibfnamefont {R{\'{e}}mi}\
  \bibnamefont {Monasson}},\ }\bibfield  {title} {\enquote {\bibinfo {title}
  {Optimization problems and replica symmetry breaking in finite connectivity
  spin glasses},}\ }\href {\doibase 10.1088/0305-4470/31/2/012} {\bibfield
  {journal} {\bibinfo  {journal} {Journal of Physics A: Mathematical and
  General}\ }\textbf {\bibinfo {volume} {31}},\ \bibinfo {pages} {513--529}
  (\bibinfo {year} {1998})}\BibitemShut {NoStop}%
\bibitem [{\citenamefont {Yook}\ and\ \citenamefont {Kim}(2018)}]{yook2018two}%
  \BibitemOpen
  \bibfield  {author} {\bibinfo {author} {\bibfnamefont {Soon-Hyung}\
  \bibnamefont {Yook}}\ and\ \bibinfo {author} {\bibfnamefont {Yup}\
  \bibnamefont {Kim}},\ }\bibfield  {title} {\enquote {\bibinfo {title} {Two
  order parameters for the kuramoto model on complex networks},}\ }\href@noop
  {} {\bibfield  {journal} {\bibinfo  {journal} {Physical Review E}\ }\textbf
  {\bibinfo {volume} {97}},\ \bibinfo {pages} {042317} (\bibinfo {year}
  {2018})}\BibitemShut {NoStop}%
\bibitem [{\citenamefont {Dorogovtsev}\ \emph {et~al.}(2002)\citenamefont
  {Dorogovtsev}, \citenamefont {Goltsev},\ and\ \citenamefont
  {Mendes}}]{Doro2002}%
  \BibitemOpen
  \bibfield  {author} {\bibinfo {author} {\bibfnamefont {S.~N.}\ \bibnamefont
  {Dorogovtsev}}, \bibinfo {author} {\bibfnamefont {A.~V.}\ \bibnamefont
  {Goltsev}}, \ and\ \bibinfo {author} {\bibfnamefont {J.~F.~F.}\ \bibnamefont
  {Mendes}},\ }\bibfield  {title} {\enquote {\bibinfo {title} {Ising model on
  networks with an arbitrary distribution of connections},}\ }\href {\doibase
  10.1103/PhysRevE.66.016104} {\bibfield  {journal} {\bibinfo  {journal} {Phys.
  Rev. E}\ }\textbf {\bibinfo {volume} {66}},\ \bibinfo {pages} {016104}
  (\bibinfo {year} {2002})}\BibitemShut {NoStop}%
\bibitem [{\citenamefont {Moreno}\ and\ \citenamefont
  {Pacheco}(2004)}]{moreno2004synchronization}%
  \BibitemOpen
  \bibfield  {author} {\bibinfo {author} {\bibfnamefont {Yamir}\ \bibnamefont
  {Moreno}}\ and\ \bibinfo {author} {\bibfnamefont {Amalio~F}\ \bibnamefont
  {Pacheco}},\ }\bibfield  {title} {\enquote {\bibinfo {title} {Synchronization
  of {K}uramoto oscillators in scale-free networks},}\ }\href@noop {}
  {\bibfield  {journal} {\bibinfo  {journal} {EPL (Europhysics Letters)}\
  }\textbf {\bibinfo {volume} {68}},\ \bibinfo {pages} {603} (\bibinfo {year}
  {2004})}\BibitemShut {NoStop}%
\bibitem [{\citenamefont {Ichinomiya}(2004)}]{ichinomiya2004frequency}%
  \BibitemOpen
  \bibfield  {author} {\bibinfo {author} {\bibfnamefont {Takashi}\ \bibnamefont
  {Ichinomiya}},\ }\bibfield  {title} {\enquote {\bibinfo {title} {Frequency
  synchronization in a random oscillator network},}\ }\href@noop {} {\bibfield
  {journal} {\bibinfo  {journal} {Physical Review E}\ }\textbf {\bibinfo
  {volume} {70}},\ \bibinfo {pages} {026116} (\bibinfo {year}
  {2004})}\BibitemShut {NoStop}%
\bibitem [{\citenamefont {Restrepo}\ \emph {et~al.}(2005)\citenamefont
  {Restrepo}, \citenamefont {Ott},\ and\ \citenamefont
  {Hunt}}]{restrepo2005onset}%
  \BibitemOpen
  \bibfield  {author} {\bibinfo {author} {\bibfnamefont {Juan~G}\ \bibnamefont
  {Restrepo}}, \bibinfo {author} {\bibfnamefont {Edward}\ \bibnamefont {Ott}},
  \ and\ \bibinfo {author} {\bibfnamefont {Brian~R}\ \bibnamefont {Hunt}},\
  }\bibfield  {title} {\enquote {\bibinfo {title} {Onset of synchronization in
  large networks of coupled oscillators},}\ }\href@noop {} {\bibfield
  {journal} {\bibinfo  {journal} {Physical Review E}\ }\textbf {\bibinfo
  {volume} {71}},\ \bibinfo {pages} {036151} (\bibinfo {year}
  {2005})}\BibitemShut {NoStop}%
\bibitem [{\citenamefont {Arenas}\ \emph {et~al.}(2008)\citenamefont {Arenas},
  \citenamefont {D{\'\i}az-Guilera}, \citenamefont {Kurths}, \citenamefont
  {Moreno},\ and\ \citenamefont {Zhou}}]{arenas2008synchronization}%
  \BibitemOpen
  \bibfield  {author} {\bibinfo {author} {\bibfnamefont {Alex}\ \bibnamefont
  {Arenas}}, \bibinfo {author} {\bibfnamefont {Albert}\ \bibnamefont
  {D{\'\i}az-Guilera}}, \bibinfo {author} {\bibfnamefont {Jurgen}\ \bibnamefont
  {Kurths}}, \bibinfo {author} {\bibfnamefont {Yamir}\ \bibnamefont {Moreno}},
  \ and\ \bibinfo {author} {\bibfnamefont {Changsong}\ \bibnamefont {Zhou}},\
  }\bibfield  {title} {\enquote {\bibinfo {title} {Synchronization in complex
  networks},}\ }\href@noop {} {\bibfield  {journal} {\bibinfo  {journal}
  {Physics reports}\ }\textbf {\bibinfo {volume} {469}},\ \bibinfo {pages}
  {93--153} (\bibinfo {year} {2008})}\BibitemShut {NoStop}%
\bibitem [{\citenamefont {Peron}\ \emph {et~al.}(2019)\citenamefont {Peron},
  \citenamefont {de~Resende}, \citenamefont {Mata}, \citenamefont {Rodrigues},\
  and\ \citenamefont {Moreno}}]{peron2019onset}%
  \BibitemOpen
  \bibfield  {author} {\bibinfo {author} {\bibfnamefont {Thomas}\ \bibnamefont
  {Peron}}, \bibinfo {author} {\bibfnamefont {Bruno Messias~F}\ \bibnamefont
  {de~Resende}}, \bibinfo {author} {\bibfnamefont {Ang{\'e}lica~S}\
  \bibnamefont {Mata}}, \bibinfo {author} {\bibfnamefont {Francisco~A}\
  \bibnamefont {Rodrigues}}, \ and\ \bibinfo {author} {\bibfnamefont {Yamir}\
  \bibnamefont {Moreno}},\ }\bibfield  {title} {\enquote {\bibinfo {title}
  {Onset of synchronization of {K}uramoto oscillators in scale-free
  networks},}\ }\href@noop {} {\bibfield  {journal} {\bibinfo  {journal}
  {Physical Review E}\ }\textbf {\bibinfo {volume} {100}},\ \bibinfo {pages}
  {042302} (\bibinfo {year} {2019})}\BibitemShut {NoStop}%
\bibitem [{\citenamefont {Sakaguchi}(1988)}]{sakaguchi1988cooperative}%
  \BibitemOpen
  \bibfield  {author} {\bibinfo {author} {\bibfnamefont {Hidetsugu}\
  \bibnamefont {Sakaguchi}},\ }\bibfield  {title} {\enquote {\bibinfo {title}
  {Cooperative phenomena in coupled oscillator systems under external
  fields},}\ }\href@noop {} {\bibfield  {journal} {\bibinfo  {journal}
  {Progress of theoretical physics}\ }\textbf {\bibinfo {volume} {79}},\
  \bibinfo {pages} {39--46} (\bibinfo {year} {1988})}\BibitemShut {NoStop}%
\bibitem [{\citenamefont {Sonnenschein}\ and\ \citenamefont
  {Schimansky-Geier}(2013)}]{sonnenschein2013approximate}%
  \BibitemOpen
  \bibfield  {author} {\bibinfo {author} {\bibfnamefont {Bernard}\ \bibnamefont
  {Sonnenschein}}\ and\ \bibinfo {author} {\bibfnamefont {Lutz}\ \bibnamefont
  {Schimansky-Geier}},\ }\bibfield  {title} {\enquote {\bibinfo {title}
  {Approximate solution to the stochastic {K}uramoto model},}\ }\href@noop {}
  {\bibfield  {journal} {\bibinfo  {journal} {Physical Review E}\ }\textbf
  {\bibinfo {volume} {88}},\ \bibinfo {pages} {052111} (\bibinfo {year}
  {2013})}\BibitemShut {NoStop}%
\bibitem [{\citenamefont {Sonnenschein}\ and\ \citenamefont
  {Schimansky-Geier}(2012)}]{sonnenschein2012onset}%
  \BibitemOpen
  \bibfield  {author} {\bibinfo {author} {\bibfnamefont {Bernard}\ \bibnamefont
  {Sonnenschein}}\ and\ \bibinfo {author} {\bibfnamefont {Lutz}\ \bibnamefont
  {Schimansky-Geier}},\ }\bibfield  {title} {\enquote {\bibinfo {title} {Onset
  of synchronization in complex networks of noisy oscillators},}\ }\href@noop
  {} {\bibfield  {journal} {\bibinfo  {journal} {Physical Review E}\ }\textbf
  {\bibinfo {volume} {85}},\ \bibinfo {pages} {051116} (\bibinfo {year}
  {2012})}\BibitemShut {NoStop}%
\bibitem [{\citenamefont {Kwon}\ and\ \citenamefont
  {Thouless}(1991)}]{Kwon1991}%
  \BibitemOpen
  \bibfield  {author} {\bibinfo {author} {\bibfnamefont {C.}~\bibnamefont
  {Kwon}}\ and\ \bibinfo {author} {\bibfnamefont {D.~J.}\ \bibnamefont
  {Thouless}},\ }\bibfield  {title} {\enquote {\bibinfo {title} {Spin glass
  with two replicas on a {B}ethe lattice},}\ }\href {\doibase
  10.1103/PhysRevB.43.8379} {\bibfield  {journal} {\bibinfo  {journal} {Phys.
  Rev. B}\ }\textbf {\bibinfo {volume} {43}},\ \bibinfo {pages} {8379--8390}
  (\bibinfo {year} {1991})}\BibitemShut {NoStop}%
\bibitem [{\citenamefont {Pagnani}\ \emph {et~al.}(2003)\citenamefont
  {Pagnani}, \citenamefont {Parisi},\ and\ \citenamefont
  {Rati\'eville}}]{Pagnani2003}%
  \BibitemOpen
  \bibfield  {author} {\bibinfo {author} {\bibfnamefont {A.}~\bibnamefont
  {Pagnani}}, \bibinfo {author} {\bibfnamefont {G.}~\bibnamefont {Parisi}}, \
  and\ \bibinfo {author} {\bibfnamefont {M.}~\bibnamefont {Rati\'eville}},\
  }\bibfield  {title} {\enquote {\bibinfo {title} {Near-optimal configurations
  in mean-field disordered systems},}\ }\href {\doibase
  10.1103/PhysRevE.68.046706} {\bibfield  {journal} {\bibinfo  {journal} {Phys.
  Rev. E}\ }\textbf {\bibinfo {volume} {68}},\ \bibinfo {pages} {046706}
  (\bibinfo {year} {2003})}\BibitemShut {NoStop}%
\bibitem [{\citenamefont {Neri}\ \emph {et~al.}(2010)\citenamefont {Neri},
  \citenamefont {Metz},\ and\ \citenamefont {Boll{\'{e}}}}]{Neri2010}%
  \BibitemOpen
  \bibfield  {author} {\bibinfo {author} {\bibfnamefont {I}~\bibnamefont
  {Neri}}, \bibinfo {author} {\bibfnamefont {F~L}\ \bibnamefont {Metz}}, \ and\
  \bibinfo {author} {\bibfnamefont {D}~\bibnamefont {Boll{\'{e}}}},\ }\bibfield
   {title} {\enquote {\bibinfo {title} {The phase diagram of {L}{\'{e}}vy spin
  glasses},}\ }\href {\doibase 10.1088/1742-5468/2010/01/p01010} {\bibfield
  {journal} {\bibinfo  {journal} {Journal of Statistical Mechanics: Theory and
  Experiment}\ }\textbf {\bibinfo {volume} {2010}},\ \bibinfo {pages} {P01010}
  (\bibinfo {year} {2010})}\BibitemShut {NoStop}%
\bibitem [{\citenamefont {Zwillinger}\ and\ \citenamefont
  {Jeffrey}(2000)}]{Gradstein}%
  \BibitemOpen
  \bibfield  {author} {\bibinfo {author} {\bibfnamefont {D.}~\bibnamefont
  {Zwillinger}}\ and\ \bibinfo {author} {\bibfnamefont {A.}~\bibnamefont
  {Jeffrey}},\ }\href {https://books.google.com.br/books?id=h4y-36vKIZgC}
  {\emph {\bibinfo {title} {Table of Integrals, Series, and Products}}}\
  (\bibinfo  {publisher} {Elsevier Science},\ \bibinfo {year}
  {2000})\BibitemShut {NoStop}%
\bibitem [{\citenamefont {Metz}\ and\ \citenamefont {Silva}(2020)}]{Metz2020}%
  \BibitemOpen
  \bibfield  {author} {\bibinfo {author} {\bibfnamefont {Fernando~L.}\
  \bibnamefont {Metz}}\ and\ \bibinfo {author} {\bibfnamefont {Jeferson~D.}\
  \bibnamefont {Silva}},\ }\bibfield  {title} {\enquote {\bibinfo {title}
  {Spectral density of dense random networks and the breakdown of the wigner
  semicircle law},}\ }\href {\doibase 10.1103/PhysRevResearch.2.043116}
  {\bibfield  {journal} {\bibinfo  {journal} {Phys. Rev. Research}\ }\textbf
  {\bibinfo {volume} {2}},\ \bibinfo {pages} {043116} (\bibinfo {year}
  {2020})}\BibitemShut {NoStop}%
\bibitem [{\citenamefont {Sompolinsky}(1986)}]{Sompolinsky1986}%
  \BibitemOpen
  \bibfield  {author} {\bibinfo {author} {\bibfnamefont {H.}~\bibnamefont
  {Sompolinsky}},\ }\bibfield  {title} {\enquote {\bibinfo {title} {Neural
  networks with nonlinear synapses and a static noise},}\ }\href {\doibase
  10.1103/PhysRevA.34.2571} {\bibfield  {journal} {\bibinfo  {journal} {Phys.
  Rev. A}\ }\textbf {\bibinfo {volume} {34}},\ \bibinfo {pages} {2571--2574}
  (\bibinfo {year} {1986})}\BibitemShut {NoStop}%
\bibitem [{\citenamefont {Canning}\ and\ \citenamefont
  {Naef}(1992)}]{Canning1992}%
  \BibitemOpen
  \bibfield  {author} {\bibinfo {author} {\bibfnamefont {A.}~\bibnamefont
  {Canning}}\ and\ \bibinfo {author} {\bibfnamefont {J.-P.}\ \bibnamefont
  {Naef}},\ }\bibfield  {title} {\enquote {\bibinfo {title} {Phase diagrams and
  the instability of the spin glass states for the diluted hopfield neural
  network model},}\ }\href {\doibase 10.1051/jp1:1992245} {\bibfield  {journal}
  {\bibinfo  {journal} {J. Phys. I France}\ }\textbf {\bibinfo {volume} {2}},\
  \bibinfo {pages} {1791--1801} (\bibinfo {year} {1992})}\BibitemShut {NoStop}%
\bibitem [{\citenamefont {Metz}\ and\ \citenamefont
  {Theumann}(2006)}]{Metz2006}%
  \BibitemOpen
  \bibfield  {author} {\bibinfo {author} {\bibfnamefont {F.L.}\ \bibnamefont
  {Metz}}\ and\ \bibinfo {author} {\bibfnamefont {W.K.}\ \bibnamefont
  {Theumann}},\ }\bibfield  {title} {\enquote {\bibinfo {title} {Feed-forward
  chains of recurrent attractor neural networks with finite dilution near
  saturation},}\ }\href {\doibase https://doi.org/10.1016/j.physa.2005.11.049}
  {\bibfield  {journal} {\bibinfo  {journal} {Physica A: Statistical Mechanics
  and its Applications}\ }\textbf {\bibinfo {volume} {368}},\ \bibinfo {pages}
  {273--286} (\bibinfo {year} {2006})}\BibitemShut {NoStop}%
\bibitem [{\citenamefont {Boettcher}(2020)}]{Boettcher2020}%
  \BibitemOpen
  \bibfield  {author} {\bibinfo {author} {\bibfnamefont {Stefan}\ \bibnamefont
  {Boettcher}},\ }\bibfield  {title} {\enquote {\bibinfo {title} {Ground state
  properties of the diluted sherrington-kirkpatrick spin glass},}\ }\href
  {\doibase 10.1103/PhysRevLett.124.177202} {\bibfield  {journal} {\bibinfo
  {journal} {Phys. Rev. Lett.}\ }\textbf {\bibinfo {volume} {124}},\ \bibinfo
  {pages} {177202} (\bibinfo {year} {2020})}\BibitemShut {NoStop}%
\bibitem [{\citenamefont {Jakeman}\ and\ \citenamefont
  {Pusey}(1978)}]{Jakeman1978}%
  \BibitemOpen
  \bibfield  {author} {\bibinfo {author} {\bibfnamefont {E.}~\bibnamefont
  {Jakeman}}\ and\ \bibinfo {author} {\bibfnamefont {P.~N.}\ \bibnamefont
  {Pusey}},\ }\bibfield  {title} {\enquote {\bibinfo {title} {Significance of
  $k$ distributions in scattering experiments},}\ }\href {\doibase
  10.1103/PhysRevLett.40.546} {\bibfield  {journal} {\bibinfo  {journal} {Phys.
  Rev. Lett.}\ }\textbf {\bibinfo {volume} {40}},\ \bibinfo {pages} {546--550}
  (\bibinfo {year} {1978})}\BibitemShut {NoStop}%
\bibitem [{\citenamefont {Godr\'eche}(2021)}]{Godreche2021}%
  \BibitemOpen
  \bibfield  {author} {\bibinfo {author} {\bibfnamefont {Claude}\ \bibnamefont
  {Godr\'eche}},\ }\bibfield  {title} {\enquote {\bibinfo {title} {Condensation
  and extremes for a fluctuating number of independent random variables},}\
  }\href@noop {} {\bibfield  {journal} {\bibinfo  {journal} {J. Stat. Phys.}\
  }\textbf {\bibinfo {volume} {182}},\ \bibinfo {pages} {13} (\bibinfo {year}
  {2021})}\BibitemShut {NoStop}%
\bibitem [{\citenamefont {Mimura}\ and\ \citenamefont
  {Coolen}(2009)}]{Mimura2009}%
  \BibitemOpen
  \bibfield  {author} {\bibinfo {author} {\bibfnamefont {Kazushi}\ \bibnamefont
  {Mimura}}\ and\ \bibinfo {author} {\bibfnamefont {A~C~C}\ \bibnamefont
  {Coolen}},\ }\bibfield  {title} {\enquote {\bibinfo {title} {Parallel
  dynamics of disordered ising spin systems on finitely connected directed
  random graphs with arbitrary degree distributions},}\ }\href {\doibase
  10.1088/1751-8113/42/41/415001} {\bibfield  {journal} {\bibinfo  {journal}
  {Journal of Physics A: Mathematical and Theoretical}\ }\textbf {\bibinfo
  {volume} {42}},\ \bibinfo {pages} {415001} (\bibinfo {year}
  {2009})}\BibitemShut {NoStop}%
\bibitem [{\citenamefont {Neri}\ and\ \citenamefont
  {Boll{\'{e}}}(2009)}]{Neri2009}%
  \BibitemOpen
  \bibfield  {author} {\bibinfo {author} {\bibfnamefont {I}~\bibnamefont
  {Neri}}\ and\ \bibinfo {author} {\bibfnamefont {D}~\bibnamefont
  {Boll{\'{e}}}},\ }\bibfield  {title} {\enquote {\bibinfo {title} {The cavity
  approach to parallel dynamics of ising spins on a graph},}\ }\href {\doibase
  10.1088/1742-5468/2009/08/p08009} {\bibfield  {journal} {\bibinfo  {journal}
  {Journal of Statistical Mechanics: Theory and Experiment}\ }\textbf {\bibinfo
  {volume} {2009}},\ \bibinfo {pages} {P08009} (\bibinfo {year}
  {2009})}\BibitemShut {NoStop}%
\bibitem [{\citenamefont {Concetti}(2018)}]{Concetti2018}%
  \BibitemOpen
  \bibfield  {author} {\bibinfo {author} {\bibfnamefont {F.}~\bibnamefont
  {Concetti}},\ }\bibfield  {title} {\enquote {\bibinfo {title} {The full
  replica symmetry breaking in the ising spin glass on random regular graph},}\
  }\href@noop {} {\bibfield  {journal} {\bibinfo  {journal} {J. Stat. Phys.}\
  }\textbf {\bibinfo {volume} {173}},\ \bibinfo {pages} {1459} (\bibinfo {year}
  {2018})}\BibitemShut {NoStop}%
\bibitem [{\citenamefont {Parisi}(1979)}]{Parisi1979}%
  \BibitemOpen
  \bibfield  {author} {\bibinfo {author} {\bibfnamefont {G.}~\bibnamefont
  {Parisi}},\ }\bibfield  {title} {\enquote {\bibinfo {title} {Infinite number
  of order parameters for spin-glasses},}\ }\href {\doibase
  10.1103/PhysRevLett.43.1754} {\bibfield  {journal} {\bibinfo  {journal}
  {Phys. Rev. Lett.}\ }\textbf {\bibinfo {volume} {43}},\ \bibinfo {pages}
  {1754--1756} (\bibinfo {year} {1979})}\BibitemShut {NoStop}%
\bibitem [{\citenamefont {Parisi}(1983)}]{Parisi1983}%
  \BibitemOpen
  \bibfield  {author} {\bibinfo {author} {\bibfnamefont {Giorgio}\ \bibnamefont
  {Parisi}},\ }\bibfield  {title} {\enquote {\bibinfo {title} {Order parameter
  for spin-glasses},}\ }\href {\doibase 10.1103/PhysRevLett.50.1946} {\bibfield
   {journal} {\bibinfo  {journal} {Phys. Rev. Lett.}\ }\textbf {\bibinfo
  {volume} {50}},\ \bibinfo {pages} {1946--1948} (\bibinfo {year}
  {1983})}\BibitemShut {NoStop}%
\bibitem [{\citenamefont {Derrida}\ \emph {et~al.}(1987)\citenamefont
  {Derrida}, \citenamefont {Gardner},\ and\ \citenamefont
  {Zippelius}}]{Derrida1987}%
  \BibitemOpen
  \bibfield  {author} {\bibinfo {author} {\bibfnamefont {B}~\bibnamefont
  {Derrida}}, \bibinfo {author} {\bibfnamefont {E}~\bibnamefont {Gardner}}, \
  and\ \bibinfo {author} {\bibfnamefont {A}~\bibnamefont {Zippelius}},\
  }\bibfield  {title} {\enquote {\bibinfo {title} {An exactly solvable
  asymmetric neural network model},}\ }\href {\doibase
  10.1209/0295-5075/4/2/007} {\bibfield  {journal} {\bibinfo  {journal}
  {Europhysics Letters ({EPL})}\ }\textbf {\bibinfo {volume} {4}},\ \bibinfo
  {pages} {167--173} (\bibinfo {year} {1987})}\BibitemShut {NoStop}%
\bibitem [{\citenamefont {Watkin}\ and\ \citenamefont
  {Sherrington}(1991)}]{Watkin1991}%
  \BibitemOpen
  \bibfield  {author} {\bibinfo {author} {\bibfnamefont {T.~L.~H}\ \bibnamefont
  {Watkin}}\ and\ \bibinfo {author} {\bibfnamefont {D}~\bibnamefont
  {Sherrington}},\ }\bibfield  {title} {\enquote {\bibinfo {title} {A neural
  network with low symmetric connectivity},}\ }\href {\doibase
  10.1209/0295-5075/14/8/012} {\bibfield  {journal} {\bibinfo  {journal}
  {Europhysics Letters ({EPL})}\ }\textbf {\bibinfo {volume} {14}},\ \bibinfo
  {pages} {791--796} (\bibinfo {year} {1991})}\BibitemShut {NoStop}%
\bibitem [{\citenamefont {Newman}(2009)}]{Newman2009}%
  \BibitemOpen
  \bibfield  {author} {\bibinfo {author} {\bibfnamefont {M.~E.~J.}\
  \bibnamefont {Newman}},\ }\bibfield  {title} {\enquote {\bibinfo {title}
  {Random graphs with clustering},}\ }\href {\doibase
  10.1103/PhysRevLett.103.058701} {\bibfield  {journal} {\bibinfo  {journal}
  {Phys. Rev. Lett.}\ }\textbf {\bibinfo {volume} {103}},\ \bibinfo {pages}
  {058701} (\bibinfo {year} {2009})}\BibitemShut {NoStop}%
\bibitem [{\citenamefont {Metz}\ \emph {et~al.}(2011)\citenamefont {Metz},
  \citenamefont {Neri},\ and\ \citenamefont {Boll\'e}}]{Metz2011}%
  \BibitemOpen
  \bibfield  {author} {\bibinfo {author} {\bibfnamefont {F.~L.}\ \bibnamefont
  {Metz}}, \bibinfo {author} {\bibfnamefont {I.}~\bibnamefont {Neri}}, \ and\
  \bibinfo {author} {\bibfnamefont {D.}~\bibnamefont {Boll\'e}},\ }\bibfield
  {title} {\enquote {\bibinfo {title} {Spectra of sparse regular graphs with
  loops},}\ }\href {\doibase 10.1103/PhysRevE.84.055101} {\bibfield  {journal}
  {\bibinfo  {journal} {Phys. Rev. E}\ }\textbf {\bibinfo {volume} {84}},\
  \bibinfo {pages} {055101} (\bibinfo {year} {2011})}\BibitemShut {NoStop}%
\bibitem [{\citenamefont {Kirkley}\ \emph {et~al.}(2021)\citenamefont
  {Kirkley}, \citenamefont {Cantwell},\ and\ \citenamefont
  {Newman}}]{Kirkley2021}%
  \BibitemOpen
  \bibfield  {author} {\bibinfo {author} {\bibfnamefont {Alec}\ \bibnamefont
  {Kirkley}}, \bibinfo {author} {\bibfnamefont {George~T.}\ \bibnamefont
  {Cantwell}}, \ and\ \bibinfo {author} {\bibfnamefont {M.~E.~J.}\ \bibnamefont
  {Newman}},\ }\bibfield  {title} {\enquote {\bibinfo {title} {Belief
  propagation for networks with loops},}\ }\href {\doibase
  10.1126/sciadv.abf1211} {\bibfield  {journal} {\bibinfo  {journal} {Science
  Advances}\ }\textbf {\bibinfo {volume} {7}},\ \bibinfo {pages} {eabf1211}
  (\bibinfo {year} {2021})}\BibitemShut {NoStop}%
\bibitem [{\citenamefont {Coolen}(2001)}]{coolen2001statistical}%
  \BibitemOpen
  \bibfield  {author} {\bibinfo {author} {\bibfnamefont {ACC}\ \bibnamefont
  {Coolen}},\ }\bibfield  {title} {\enquote {\bibinfo {title} {Statistical
  mechanics of recurrent neural networks i—statics},}\ }in\ \href@noop {}
  {\emph {\bibinfo {booktitle} {Handbook of biological physics}}},\
  Vol.~\bibinfo {volume} {4}\ (\bibinfo  {publisher} {Elsevier},\ \bibinfo
  {year} {2001})\ pp.\ \bibinfo {pages} {553--618}\BibitemShut {NoStop}%
\end{thebibliography}%

\end{document}